\documentclass[ALICE,manyauthors]{cernphprep}
\usepackage[comma,square,numbers,sort&compress]{natbib}
\usepackage{hyperref}
\usepackage{lineno}
\usepackage{xspace}
\usepackage{color,soul}
\usepackage{comment}
\usepackage{scalerel}
\usepackage[dvipsnames]{xcolor}
\usepackage{booktabs, multirow}
\usepackage{adjustbox}
\usepackage{rotating}
\usepackage[T1]{fontenc}
\usepackage{orcidlink}

\begin{document}
%

\newcommand{\pp}           {pp\xspace}
\newcommand{\ppbar}        {\mbox{$\mathrm {p\overline{p}}$}\xspace}
\newcommand{\XeXe}         {\mbox{Xe--Xe}\xspace}
\newcommand{\PbPb}         {\mbox{Pb--Pb}\xspace}
\newcommand{\pA}           {\mbox{pA}\xspace}
\newcommand{\pPb}          {\mbox{p--Pb}\xspace}
\newcommand{\AuAu}         {\mbox{Au--Au}\xspace}
\newcommand{\dAu}          {\mbox{d--Au}\xspace}

\newcommand{\sbold}        {\ensuremath{\sqrt{\mathbf{s}}}\xspace}
\newcommand{\s}            {\ensuremath{\sqrt{s}}\xspace}
\newcommand{\snn}          {\ensuremath{\sqrt{s_{\mathrm{NN}}}}\xspace}
\newcommand{\pt}           {\ensuremath{p_{\rm T}}\xspace}
\newcommand{\pthat}        {\ensuremath{\hat{p}_{\rm T}}\xspace}

\newcommand{\mpt}           {\ensuremath{\left<p_{\rm T}\right>}\xspace}
\newcommand{\meanpt}       {$\langle p_{\mathrm{T}}\rangle$\xspace}
\newcommand{\ycms}         {\ensuremath{y_{\rm CMS}}\xspace}
\newcommand{\ylab}         {\ensuremath{y_{\rm lab}}\xspace}
\newcommand{\etarange}[1]  {\mbox{$\left | \eta \right |~<~#1$}}
\newcommand{\yrange}[1]    {\mbox{$\left | y \right |~<~#1$}}
\newcommand{\dndy}         {\ensuremath{\mathrm{d}N_\mathrm{ch}/\mathrm{d}y}\xspace}
\newcommand{\dndeta}       {\ensuremath{\mathrm{d}N_\mathrm{ch}/\mathrm{d}\eta}\xspace}
\newcommand{\myield}       {\ensuremath{\left<\mathrm{d}N/\mathrm{d}y\right>}\xspace}
\newcommand{\myieldpi}       {\ensuremath{\left<\mathrm{d}N_\mathrm{\pi}/\mathrm{d}y\right>}\xspace}

\newcommand{\avdndeta}     {\ensuremath{\langle\dndeta\rangle}\xspace}
\newcommand{\dNdy}         {\ensuremath{\mathrm{d}N_\mathrm{ch}/\mathrm{d}y}\xspace}
\newcommand{\Npart}        {\ensuremath{N_\mathrm{part}}\xspace}
\newcommand{\Ncoll}        {\ensuremath{N_\mathrm{coll}}\xspace}
\newcommand{\dEdx}         {\ensuremath{\textrm{d}E/\textrm{d}x}\xspace}
\newcommand{\RpPb}         {\ensuremath{R_{\rm pPb}}\xspace}
\newcommand{\INELgtO}            {\ensuremath{\text{INEL} > 0}\xspace}

\newcommand{\nineH}        {$\sqrt{s}~=~0.9$~Te\kern-.1emV\xspace}
\newcommand{\seven}        {$\sqrt{s}~=~7$~Te\kern-.1emV\xspace}
\newcommand{\twoH}         {$\sqrt{s}~=~0.2$~Te\kern-.1emV\xspace}
\newcommand{\twosevensix}  {$\sqrt{s}~=~2.76$~Te\kern-.1emV\xspace}
\newcommand{\five}         {$\sqrt{s}~=~5.02$~Te\kern-.1emV\xspace}
\newcommand{\twosevensixnn}{$\sqrt{s_{\mathrm{NN}}}~=~2.76$~Te\kern-.1emV\xspace}
\newcommand{\fivenn}       {$\sqrt{s_{\mathrm{NN}}}~=~5.02$~Te\kern-.1emV\xspace}
\newcommand{\LT}           {L{\'e}vy-Tsallis\xspace}
\newcommand{\GeVc}         {Ge\kern-.1emV/$c$\xspace}
\newcommand{\MeVc}         {Me\kern-.1emV/$c$\xspace}
\newcommand{\TeV}          {Te\kern-.1emV\xspace}
\newcommand{\GeV}          {Ge\kern-.1emV\xspace}
\newcommand{\MeV}          {Me\kern-.1emV\xspace}
\newcommand{\GeVmass}      {Ge\kern-.2emV/$c^2$\xspace}
\newcommand{\MeVmass}      {Me\kern-.2emV/$c^2$\xspace}
\newcommand{\lumi}         {\ensuremath{\mathcal{L}}\xspace}

\newcommand{\ITS}          {\rm{ITS}\xspace}
\newcommand{\TOF}          {\rm{TOF}\xspace}
\newcommand{\ZDC}          {\rm{ZDC}\xspace}
\newcommand{\ZDCs}         {\rm{ZDCs}\xspace}
\newcommand{\ZNA}          {\rm{ZNA}\xspace}
\newcommand{\ZNC}          {\rm{ZNC}\xspace}
\newcommand{\SPD}          {\rm{SPD}\xspace}
\newcommand{\SDD}          {\rm{SDD}\xspace}
\newcommand{\SSD}          {\rm{SSD}\xspace}
\newcommand{\TPC}          {\rm{TPC}\xspace}
\newcommand{\TRD}          {\rm{TRD}\xspace}
\newcommand{\VZERO}        {\rm{V0}\xspace}
\newcommand{\VZEROM}       {\rm{V0M}\xspace}
\newcommand{\VZEROA}       {\rm{V0A}\xspace}
\newcommand{\VZEROC}       {\rm{V0C}\xspace}
\newcommand{\Vdecay} 	   {\ensuremath{V^{0}}\xspace}

\newcommand{\ee}           {\ensuremath{e^{+}e^{-}}} 
\newcommand{\pip}          {\ensuremath{\pi^{+}}\xspace}
\newcommand{\pim}          {\ensuremath{\pi^{-}}\xspace}
\newcommand{\kap}          {\ensuremath{\rm{K}^{+}}\xspace}
\newcommand{\kam}          {\ensuremath{\rm{K}^{-}}\xspace}
\newcommand{\pbar}         {\ensuremath{\rm\overline{p}}\xspace}
\newcommand{\kzero}        {\ensuremath{{\rm K}^{0}_{\rm{S}}}\xspace}
\newcommand{\lmb}          {\ensuremath{\Lambda}\xspace}
\newcommand{\almb}         {\ensuremath{\overline{\Lambda}}\xspace}
\newcommand{\Om}           {\ensuremath{\Omega^-}\xspace}
\newcommand{\Mo}           {\ensuremath{\overline{\Omega}^+}\xspace}
\newcommand{\X}            {\ensuremath{\Xi^-}\xspace}
\newcommand{\Ix}           {\ensuremath{\Xi^+}\xspace}
\newcommand{\Xis}          {\ensuremath{\Xi^{\pm}}\xspace}
\newcommand{\Oms}          {\ensuremath{\Omega^{\pm}}\xspace}
\newcommand{\degree}       {\ensuremath{^{\rm o}}\xspace}
\newcommand{\PHI}          {\ensuremath{\phi}\xspace}
\newcommand{\KSTAR}        {\ensuremath{\rm K^{*0}}\xspace}
\newcommand{\KSTARb}        {\ensuremath{\rm{\overline{K^{*0}}}}\xspace}

\newcommand{\pion}         {\ensuremath{\pi}\xspace}
\newcommand{\kaon}         {\ensuremath{\textrm{K}}\xspace}
\newcommand{\pr}           {\ensuremath{\textrm{p}}\xspace}

\newcommand{\RT}{\ensuremath{R_{\rm T}}\xspace}

\newcommand{\Raa}{\ensuremath{\rm R_{AA}}\xspace}
\newcommand{\Rpa}{\ensuremath{\rm R_{pA}}\xspace}
\newcommand{\Rppb}{\ensuremath{R_{\rm pPb}}\xspace}

\newcommand{\cme}{\ensuremath{\sqrt{s}}\xspace}
\newcommand{\gev}{\ensuremath{{\rm GeV}/c}\xspace}
\newcommand{\auau}{\ensuremath{\rm Au\!-\!Au}\xspace}
\newcommand{\pikp}{\ensuremath{\pi,{\rm K},{\rm p}}\xspace}
\newcommand{\ptopi}{\ensuremath{{\rm p } / \pi}\xspace}
\newcommand{\ktopi}{\ensuremath{({\rm K}^{+}+{\rm K}^{-}) / (\pi^{+}+\pi^{-})}\xspace}
\newcommand{\twotwo}{\ensuremath{2\rightarrow 2}\xspace}
\newcommand{\ltok}{\ensuremath{({\rm \Lambda}^{0}+\bar {\rm \Lambda}^{0})/(\rm 2 K^{0}_{s} )} \xspace}

\newcommand{\snnt}[1]{\ensuremath{\sqrt{s_{\rm NN}} = #1 \text{\,TeV}}\xspace}
\newcommand{\snnnotext}[1]{\ensuremath{\sqrt{s_{NN}} = #1}\xspace}
\newcommand{\sppt}[1]{\ensuremath{\sqrt{s} = #1 \text{\,TeV}}\xspace}
\newcommand{\sppg}[1]{\ensuremath{\sqrt{s} = #1 \text{\,GeV}}\xspace}
\newcommand{\gevc}[1]{\ensuremath{#1 \text{\,GeV/$c$}}\xspace}
\newcommand{\eff}[1]{\ensuremath{\epsilon_{#1}}\xspace}
\newcommand{\bareyield}{\ensuremath{Y}\xspace}
\newcommand{\yield}[1]{\ensuremath{Y_{#1}}\xspace}
\newcommand{\ch}{\ensuremath{\text{ch}}\xspace}
\newcommand{\etalab}{\ensuremath{\eta_{lab}}\xspace}

\newcommand{\VZ}{\ensuremath{V^{0}}\xspace}
\newcommand{\VZs}{\ensuremath{V^{0}\text{s}}\xspace}
\newcommand{\RAA}{\ensuremath{R_{\text{AA}}}\xspace}
\newcommand{\MRAA}{\ensuremath{\mathbf{R_{\text{\textbf{AA}}}}}\xspace}
\newcommand{\ppb}{p--Pb\xspace}
\newcommand{\pbpb}{Pb--Pb\xspace}
\newcommand{\km}{\ensuremath{K^{-}}\xspace}
\newcommand{\kp}{\ensuremath{K^{+}}\xspace}
\newcommand{\p}{\ensuremath{p}\xspace}
\newcommand{\mathpt}{\ensuremath{\mathbf{p_{\text{T}}}}\xspace}
\newcommand{\dpi}{\ensuremath{\Delta_{\pi}}\xspace}
\newcommand{\dkaon}{\ensuremath{\Delta_{K}}\xspace}
\newcommand{\dproton}{\ensuremath{\Delta_{p}}\xspace}
\newcommand{\mdpi}{\ensuremath{\mathbf{\Delta_{\pi}}}\xspace}
\newcommand{\mdkaon}{\ensuremath{\mathbf{\Delta_{K}}}\xspace}
\newcommand{\mdproton}{\ensuremath{\mathbf{\Delta_{p}}}\xspace}
\newcommand{\rpi}{\ensuremath{R_{\pi}}\xspace}
\newcommand{\mathdedx}{\ensuremath{\mathbf{\text{d}E/\text{d}x}}\xspace}
\newcommand{\dedx}{\ensuremath{\text{d}E/\text{d}x}\xspace}
\newcommand{\mdedx}{\ensuremath{\left <\text{d}E/\text{d}x \right>}\xspace}
\newcommand{\mathmdedx}{\ensuremath{\mathbf{\left <\text{d}E/\text{d}x \right>}}\xspace}
\newcommand{\mdedxpi}{\ensuremath{\left <\text{d}E/\text{d}x \right>_{\pi}}\xspace}
\newcommand{\meanp}{\ensuremath{\langle p \rangle}\xspace}
\newcommand{\sdedx}{\ensuremath{\sigma_{\text{d}E/\text{d}x}}\xspace}
\newcommand{\relres}{\ensuremath{\sigma/\left <\text{d}E/\text{d}x \right>}\xspace}
\newcommand{\res}{\ensuremath{\sigma_{\text{d}E/\text{d}x}}\xspace}
\newcommand{\ncl}{\ensuremath{\text{Ncl}}\xspace}
\newcommand{\mncl}{\ensuremath{\langle \text{Ncl} \rangle}\xspace}

\newcommand{\chpi}{\ensuremath{\pi^{+}+\pi^{-}}\xspace}
\newcommand{\chk}{\ensuremath{{\rm K}^{+}+{\rm K}^{-}}\xspace}
\newcommand{\chp}{\ensuremath{{\rm p}+{\rm \bar{p}}}\xspace}

\newcommand{\bg}{\ensuremath{\beta\gamma}\xspace}

\newcommand{\LA}{\ensuremath{\Lambda}\xspace}
\newcommand{\AL}{\ensuremath{\bar{\Lambda}}\xspace}
\newcommand{\KOs}{\ensuremath{\rm K^0_S}\xspace}
\newcommand{\dMassG}{\ensuremath{\Delta m_\gamma}\xspace}
\newcommand{\dMassL}{\ensuremath{\Delta m_\Lambda}\xspace}
\newcommand{\dMassAL}{\ensuremath{\Delta m_{\bar{\Lambda}}}\xspace}
\newcommand{\dMassKO}{\ensuremath{\Delta m_{K^0_s}}\xspace}

\newcommand{\Ks}{\ensuremath{\rm K^0_S}\xspace}
\newcommand{\dMassKs}{\ensuremath{\Delta m_{K^0_s}}\xspace}

\newcommand{\XI}{\ensuremath{\Xi}\xspace}
\newcommand{\PI}{\ensuremath{\pi}\xspace}
\newcommand{\XIM}{\ensuremath{\Xi^-}\xspace}
\newcommand{\XIP}{\ensuremath{\overline{\Xi}^+}\xspace}
\newcommand{\XISUM}{\ensuremath{\Xi^-+\bar{\Xi}^+}\xspace}
\newcommand{\Nch}{\ensuremath{N_{\text{ch}}}\xspace}
\newcommand{\Ntracks}{\ensuremath{N_{\text{tracks}}}\xspace}
\newcommand{\tracklet}{\ensuremath{N_{\text{tracklets}}^{\text{$|\eta| < 0.8$}}}\xspace}
\newcommand{\Minv}{\ensuremath{M_{\text{inv}}}\xspace}
\newcommand{\SO}{\ensuremath{S_{\text{O}}}\xspace}
\newcommand{\TrkRec}{\ensuremath{\textrm{Trk}_{\textrm{Rec}}}\xspace}
\newcommand{\TrkGen}{\ensuremath{\textrm{Trk}_{\textrm{Gen}}}\xspace}
\newcommand{\SORec}{\ensuremath{S_{\text{O}_{\textrm{Rec}}}^{(\pt = 1.0)}}\xspace}
\newcommand{\SOGen}{\ensuremath{S_{\text{O}_{\textrm{Gen}}}^{(\pt = 1.0)}}\xspace}

\newcommand{\SOPT}{\ensuremath{S_{\text{O}}^{\pt=1}}\xspace}
\newcommand{\nsigma}{\ensuremath{n\sigma}\xspace}
\newcommand{\NSPD}{\ensuremath{N_{{\textrm{SPD}}_{\textrm{Trklts}}}}\xspace} 
\newcommand{\NSIGMAKTPC}{\ensuremath{n\sigma_{\rm K}^{\rm TPC}}\xspace}
\newcommand{\NSIGMAKTOF}{\ensuremath{n\sigma_{\rm K}^{\rm TOF}}\xspace}
\newcommand{\topone}{0--1\%}
\newcommand{\topten}{0--10\%}

\begin{titlepage}
\PHyear{2023}       
\PHnumber{215}      
\PHdate{27 September}  

\title{Light-flavor particle production in high-multiplicity pp collisions at $\mathbf{\sqrt{\textit{s}} = 13}$ \TeV as a function of transverse spherocity}
\ShortTitle{Particle production in pp at \s = 13 \TeV vs. transverse spherocity}   

\Collaboration{ALICE Collaboration\thanks{See Appendix~\ref{app:collab} for the list of collaboration members}}
\ShortAuthor{ALICE Collaboration} 

\begin{abstract}
Results on the transverse spherocity dependence of light-flavor particle production ($\pi$, K, p, $\phi$, ${\rm K^{*0}}$, ${\rm K}^{0}_{\rm{S}}$, $\Lambda$, $\Xi$) at midrapidity in high-multiplicity pp collisions at $\sqrt{s} = 13$~TeV were obtained with the ALICE apparatus. The transverse spherocity estimator (\SOPT) categorizes events by their azimuthal topology. Utilizing narrow selections on \SOPT, it is possible to contrast particle production in collisions dominated by many soft initial interactions with that observed in collisions dominated by one or more hard scatterings. 
Results are reported for two multiplicity estimators covering different pseudorapidity regions.  The \SOPT estimator is found to effectively constrain the hardness of the events when the midrapidity ($\left | \eta \right |< 0.8$) estimator is used. 

The production rates of strange particles are found to be slightly higher for soft isotropic topologies, and severely suppressed in hard jet-like topologies. These effects are more pronounced for hadrons with larger mass and strangeness content, and observed when the topological selection is done within a narrow multiplicity interval. This demonstrates that an important aspect of the universal scaling of strangeness enhancement with final-state multiplicity is that high-multiplicity collisions are dominated by soft, isotropic processes. On the contrary, strangeness production in events with jet-like processes is significantly reduced.

The results presented in this article are compared with several QCD-inspired Monte Carlo event generators. Models that incorporate a two-component phenomenology, either through mechanisms accounting for string density, or thermal production, are able to describe the observed strangeness enhancement as a function of \SOPT. 

\end{abstract}
\end{titlepage}

\setcounter{page}{2} 


\section{Introduction}
Studies of high-multiplicity proton--proton (\pp) and proton--lead (\pPb) collisions have revealed that small collision systems exhibit signatures previously considered unique features of heavy-ion collisions. Some of these signatures, such as the enhanced production of strange hadrons~\cite{Naturepaper}, and collective flow~\cite{Ref:Flow1}~\cite{Ref:Flow2}, can be explained by the formation of a strongly interacting medium. Strangeness enhancement was one of the first proposed quark--gluon plasma (QGP) signatures~\cite{Koch:1986ud}, as the QGP-transition temperature is typically expected of being in the order of the strange quark mass, allowing for thermal production of strange quarks in a QGP. Collectivity was historically also expected to require thermalization and provide information on the equation of state~\cite{Teaney:2000cw}, while more recently it has been understood how collectivity can build up from kinetic equilibrium alone~\cite{Kurkela:2018xxd}. However, the formation of a medium in these small systems challenges current theoretical frameworks, because their initial small volumes imply lifetimes so short that it is unclear to what degree the systems can equilibrate (see Ref.~\cite{Kurkela:2018xxd} and references therein).

The observation of collective flow, as well as strangeness enhancement in particular, implies that pp collisions at LHC and RHIC energies can no longer be described as semi-incoherent sums of parton--parton collisions, an idea that has been central to most general-purpose quantum chromodynamics (QCD)-inspired Monte-Carlo event generators, such as PYTHIA~\cite{Sjostrand:2007gs} and Herwig~\cite{Herwig}. “Jet Universality'' is another long-standing idea for the phenomenological understanding of QCD assuming that, while the partonic processes vary with system and beam energy, the produced color fields and their hadronization are universal. In the context of Lund strings, this implies that the string tension and string-fragmentation parameters are the same regardless of collision system. For a recent discussion, we refer to Ref.~\cite{Torbjorn}. The discovery of strangeness enhancement scaling with the multiplicity~\cite{alice_newstrangeness}\cite{ALICE:2020nkc} violates the assumptions of jet universality, and for this reason QCD-inspired generators have incorporated additional phenomenological final-state pre-hadronization mechanisms, such as string percolation~\cite{Ref:SPerc}, color ropes~\cite{Ropes}, baryon junctions~\cite{Junctions} and/or new types of baryon-favored color reconnection~\cite{HerwigStrange}. Furthermore, charm fragmentation fractions in pp collisions also differ significantly from the values measured in $\textrm{e}^+\textrm{e}^-$ collisions~\cite{ALICE:2021dhb}.

In contrast, QGP-inspired models include a system evolution with volume and multiplicity and so the qualitative features of both collective flow and strangeness enhancement are expected. EPOS-LHC is an event generator where the initial interactions lead to a two-phase state (core--corona) consisting of a dense core of QGP, and a diluted corona ~\cite{EPOS-LHC}. Strangeness enhancement in EPOS-LHC is due to a change in the relative contribution of the corona (low strangeness production from pomerons) and core (high thermal strangeness production from QGP) with multiplicity.

As can be seen from the previous discussion, strange-particle production
appears to be a powerful probe for QGP-like effects in small systems, and the origin of the
strangeness enhancement constitutes an important open question. Therefore, results on strangeness production in small systems can facilitate progress in the understanding of both QCD dynamics and hadronization, in addition to putting constraints on phenomenological models.

ALICE has previously reported that strangeness enhancement as a function of the average multiplicity at midrapidity does not depend on the collision system nor center-of-mass energy per nucleon--nucleon collision, ranging from pp to p--Pb and \snn = 2.76 TeV to \s =13 TeV~\cite{13tevvsmultpapers}. This universality implies a strong correlation between the underlying physics processes that drive both the enhancement and the multiplicity. In this article, we have tested if the observed universality can be broken by contrasting high-multiplicity events dominated by one or more hard scatterings, with events dominated by multiple softer interactions.

At high transverse momentum (\pt), strangeness production in hadronic collisions tends to originate from “hard'', perturbative QCD processes. These hadrons are either produced directly through flavor creation ($\rm XX \rightarrow \rm s\bar{s}$) or flavor excitation ($\rm{s}X \rightarrow \rm{s}X$), or indirectly as a result of radiation and/or hadronization associated with hard processes, e.g., through gluon splitting following the partonic evolution ($\rm g \rightarrow s\bar{s}$). In contrast, the production of “soft'', low-\pt strange hadrons ($\pt \leq 2~\gev$) is dominated by non-perturbative QCD processes, where novel QCD dynamics could be found. The final-state azimuthal topology is expected to reflect which of these QCD processes are primarily driving particle production for a given event. Events dominated by one or more hard scatterings will presumably lead to pronounced back-to-back jet structures, while events that contain several softer scatterings will result in isotropic distributions. We note that previous results obtained by ALICE, both utilizing the same event shape estimator~\cite{ALICE:2019dfi} and a similar one, the transverse sphericity~\cite{ALICE:2012cor}, found evidence for this behavior in the way that \meanpt would depend on the event topology. 

To identify the final-state azimuthal topology, we will here use a modified variant of the transverse spherocity (\SO) estimator proposed in Ref.~\cite{ALICE:2019dfi}, for being more sensitive to the underlying processes. The lower bound of \SO aims to distinguish “Jet-like'' events, which in this study are events characterized by an azimuthal topology similar to a pair of back-to-back jets (implying a tight clustering of particles with a difference in azimuthal angle $\Delta\phi \approx 0$ or $\pi$~\cite{ALICE:2019dfi}), which on average produces a larger number of high-\pt hadrons compared with the \SO--integrated distribution, thereby “hardening'' the \pt-differential spectra. Conversely, the upper bound of \SO selects “Isotropic'' events, defined by an azimuthal topology which is close to symmetric, with an absence of preferred direction. The hypothesis is that \SO, by contrasting events with these different topologies, can be used to control the degree of QGP-like effects, like strangeness enhancement and radial flow, in high-multiplicity \pp collisions~\cite{Antonio2}~\cite{Antonio3}.

In this article, we present the first results on strangeness production in high-multiplicity pp collisions at \sppt{13} as a function of the transverse spherocity. The results obtained for the light-flavor hadrons are presented as the sum of particles and anti-particles, explicitly as \pip+ \pim, \kap+ \kam, \KSTAR + \KSTARb, p + \pbar, \lmb+ \almb, \X +  \Ix where the exceptions are \kzero and \PHI. Hereinafter, the sum of particle and anti-particles will be referred to as $\pi$, K, \kzero, \KSTAR, p, \PHI,  \lmb, and $\Xi$, unless otherwise explicitly mentioned. 

The article is organized as follows. The ALICE main detectors used in this analysis are detailed in Section~\ref{Sec:ALICE}. Section~\ref{Sec:Mult} describes the high-multiplicity definitions used throughout this article. The transverse spherocity observable, along with caveats, is defined in Section~\ref{Sec:SOPT}. The details concerning particle identification (PID), yield extraction, and correction procedures are discussed in Section~\ref{Sec:Details}. The details regarding experimental corrections and systematic uncertainties are discussed in Sections~\ref{Sec:Corr} and~\ref{Sec:Sys}, respectively. The results are reported in Section~\ref{Sec:Results}, and finally the summary and conclusions of our study are discussed in Section~\ref{Sec:Conc}.

\section{Experimental setup and event selection}\label{Sec:ALICE}

A detailed description of the ALICE apparatus in its Run 1 and 2 configuration, as well as its performance, can be found in Refs.~\cite{ALICE1,ALICE2}. This section will briefly describe the main ALICE detectors used for the event selection, \SOPT determination, and extraction of particle spectra. All radii in the following are given as distances from the beam axis.

The detectors of the ALICE apparatus can be grouped as follows: the central barrel at midrapidity, the muon arm at forward rapidity, and the forward global detectors. To be able to efficiently trigger on inelastic \pp collisions, ALICE employs two forward scintillator arrays, V0A and V0C, with a pseudorapidity coverage of $2.8 < \eta < 5.1$ and $-3.7 < \eta < -1.7$,
respectively.

Both \SOPT and particle yields are determined by using tracks in the central barrel. The main detectors for tracking in the central barrel are the Inner Tracking System (ITS) and the Time Projection Chamber (TPC). The ITS detector is composed of six layers of cylindrical silicon detectors with full azimuthal acceptance, where the radius of the innermost (outermost) layer is $3.9$\,cm ($43$\,cm). The innermost layers consist of two arrays of hybrid silicon pixel detectors (SPD), whose fine granularity provides high precision tracking closest to the primary vertex (PV). The SPD is also used to reconstruct tracklets, short two-point track segments covering the pseudorapidity region $\etarange{1.4}$. The tracklets provide both efficient primary vertex determination and a precise estimate of the charged-particle multiplicity. The remaining layers of the ITS are only used for tracking in the analyses presented here. The TPC is the primary tracking device in ALICE that provides a full three-dimensional trajectory for  each track. It is a large cylindrical detector that surrounds the ITS detector with an inner and outer radii of 85\,cm and 250\,cm, respectively.  It has full azimuthal acceptance and a pseudorapidity coverage of $-0.9< \eta <0.9$ for full-length tracks. The ITS and TPC are situated inside the solenoidal L3 magnet with a uniform magnetic field of 0.5\,T. For global tracks, where the full information of the ITS and TPC are used together, a momentum resolution of 1--10\,\% is achieved for momenta ranging from 0.05 to \gevc{100}.

The ALICE apparatus utilizes a broad range of different particle identification techniques to identify the mass of each analyzed particle. For weakly-decaying $V^0$ (\kzero, \lmb) and Cascades ($\Xi$), the TPC tracking is able to provide topological track matching of the decay products, where particles are then identified through peaks in invariant mass distributions. For the other particles, the PID is provided by the specific energy loss, d$E$/d$x$, measured in the TPC and the particle's velocity (given a measured momentum in the TPC) is provided by the Time-of-flight (TOF) detector. The TOF consists of Multi-Gap Resistive Plate Chambers (MRPC), which are used for the particle identification by measuring the total time of flight of the identified hadrons. It is located at about 3.7\,m from the interaction point, with a full azimuthal acceptance and has a pseudorapidity coverage of $-0.9< \eta <0.9$. Details of the particle identification procedure are given in Sec.~\ref{Sec:Details}.
 
\subsection{Event selection}\label{Sec:Mult}

Data used in this analysis were collected with a minimum bias trigger, which requires one or more hits in both V0 scintillator arrays in coincidence with proton beams from both directions. The contamination from beam-induced background is removed offline by using the timing information in the V0 detectors and taking into account the correlation between tracklets and clusters in the SPD detector, as discussed in detail in~\cite{ALICE2}. The primary vertex is reconstructed by correlating hits in the two SPD layers and only events with a primary vertex within $\pm$10\,cm of the nominal interaction point along the beam direction are accepted for this analysis. Due to the fast readout time of the SPD, any contamination from out-of-bunch pile-up is rejected. The contamination from in-bunch pile-up events is removed offline by excluding events with multiple vertices reconstructed in the SPD. Any remaining pile-up will be from collisions which produce little or no particles. Due to the required high-multiplicity event-selection, this will have a negligible impact on the final results presented in this article,

The measurements reported here were performed on minimum-bias triggered events that additionally have at least one charged particle measured in the pseudorapidity interval $|\eta| < 1$ (\INELgtO)~\cite{INEL}, corresponding to about 75\,\% of the total inelastic cross section. Two different multiplicity estimators are used, the total charge deposited in the full coverage of both V0 detectors (V0M), and the number of SPD tracklets within $\etarange 0.8$ (\tracklet). For each of these estimators, the multiplicity is classified as a percentile, where 0\% corresponds to the highest and 100\% to the lowest multiplicity. The high-multiplicity events used throughout this article include the top-1\% (0--1\%) and the top-10\% (0--10\%) multiplicity percentiles, with a minimum of 10 reconstructed charged primary tracks at midrapidity (\etarange 0.8).

\section{Unweighted transverse spherocity \SOPT}
\label{Sec:SOPT}

In this analysis, the unweighted transverse spherocity \SOPT is used to
quantify the topology in the azimuthal plane. It is calculated as
\begin{equation}\label{Eq:SOPT}
\centering
    \SOPT = \frac{\pi^2}{4} \min_{\hat{n}} \left(\frac{\Sigma_i
      |\hat{p_{\textrm{T},i}} \times \hat{n}|}{N_{\textrm{trks}}}  \right)^{2}.
\end{equation}
The sum is calculated over all charged particles with $\pt > \gevc{0.15}$, where $\hat{p_{\textrm{T}}}$ represents the transverse momentum unit vector, $N_{\textrm{trks}}$ is the number of charged particles in a given event and $\hat{n}$ is the unit vector that minimizes \SOPT. 

Loose selection criteria were chosen for primary tracks to ensure a high efficiency and uniform azimuthal acceptance over the full TPC volume. At least 50 or more measured clusters are required for a track in the TPC. Furthermore, the TPC tracks must be matched to hits in the ITS, but, to ensure homogeneous azimuthal acceptance, the tracks are not explicitly required to have hits in the SPD. These requirements improve the tracking precision, and reject tracks from out-of-bunch pile-up. Finally, selections of distance of closest approach (DCA) along the beam axis ($|\rm DCA_{\it{z}}|<3.2 $ cm), and in the xy-plane ($|\rm DCA_{\it{xy}}|<2.4 $ cm) are applied, ensuring that the reconstructed TPC track points to the primary vertex.

Unlike the traditional transverse spherocity estimator discussed in~\cite{ALICE:2019dfi}, the transverse momentum ($\pt$) of each track in this article is normalized to 1 ($\pt = 1$) when measuring the \SOPT. This modification was required to have similar sensitivity between neutral and charged particles. Otherwise, there would be significant differences between the production of neutral and charged kaons in jet-like events. This can be understood considering that for the traditional spherocity, a single high-\pt track will have a large weight in the spherocity calculation, which can occur for a charged kaon but never for a neutral kaon. By assigning the same weight to all measured primary charged tracks, one reduces the possible charged-vs-neutral biases.

However, this also means that results obtained with the two different spherocity estimators (traditional and unweighted) can only be qualitatively compared. Furthermore, as \SOPT is independent of the \pt of the particles, it requires a substantial amount of particles to be a good measure of the event topology. For this reason, the number of charged tracks is required to be greater or equal to 10 for events included in this analysis, which limits the applicability of the unweighted spherocity estimator to only apply for high-multiplicity pp collisions, where the average \dndeta is greater than 10.

\begin{figure}[htbp!]
    \centering
    \includegraphics[scale=0.5]{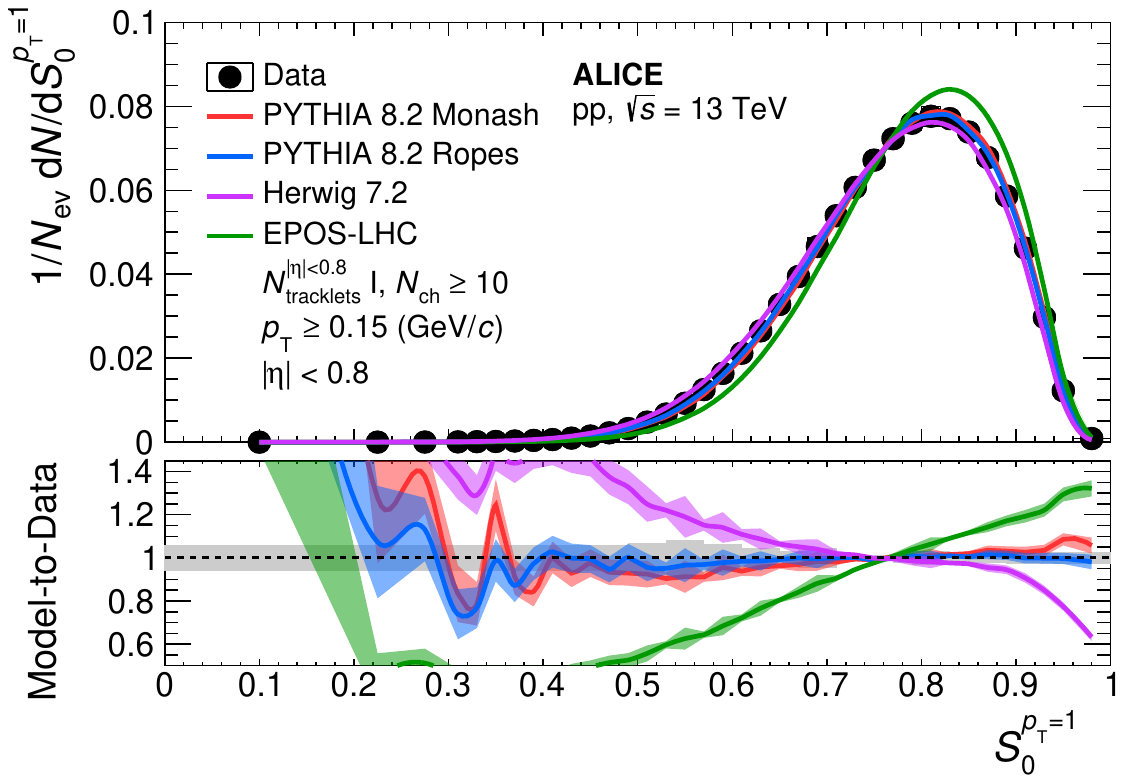}
    \includegraphics[scale=0.5]{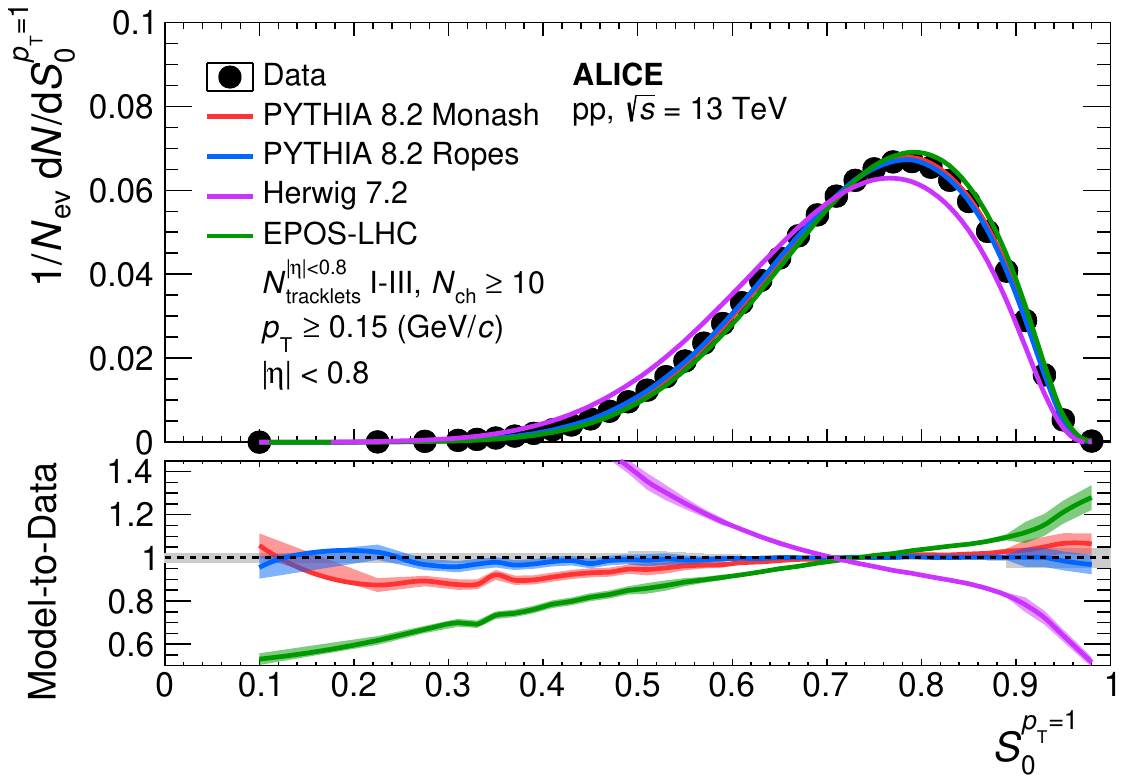}
    \includegraphics[scale=0.5]{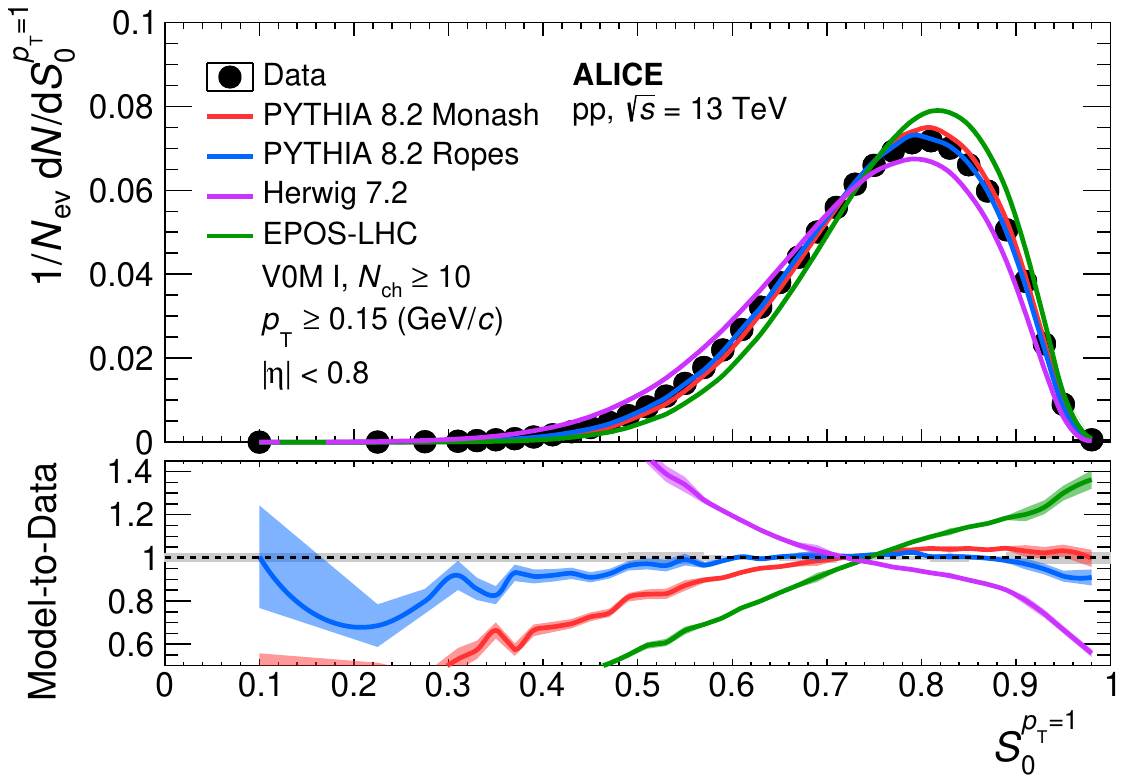}
    \caption{Upper panels: The measured and fully corrected \SOPT distributions. Lower panels: Ratio between model calculations and experimental data. These are presented for \tracklet I (top), I-III (middle) and V0M I (bottom). The roman numerals I (I-III) correspond the top 0--1\% (0--10\%) multiplicity for each respective estimator. The curves represent different model predictions, where the shaded area represents the statistical uncertainty of the models. The relative systematic uncertainty is shown as a gray area around unity in the lower panels.}
    \label{Fig:SO}
\end{figure}

The values of \SOPT will, by construction, lie between 0 and 1. Having events with $\SOPT \approx 0$ imply that $|\hat{p_{\textrm{T}}} \times \hat{n}| \approx 0$ for all tracks, which indicates that all tracks are parallel in the azimuthal plane, suggesting that the event is dominated by a single back-to-back dijet. Events where \SOPT $\rightarrow$ 1 imply that all particles are uniformly distributed azimuthally, suggesting the absence of a preferred direction. Note that part of the isotropy of the event can be affected by anisotropic flow~\cite{Ref:Flow1}. The unfolded \SOPT distributions are presented in Fig.~\ref{Fig:SO}, along with model predictions, for the three different multiplicity selections used in this article. Particle candidates that enter into the \SOPT calculation for the model predictions are required to meet the ALICE definition of a primary charged particle~\cite{alice_primary}.  The correction of the \SOPT distributions is based on the Bayesian unfolding technique~\cite{DAgostini:1994fjx}. The same unfolding method used in previous ALICE publications~\cite{ALICE:2023yuk} is applied here.

It is crucial to emphasize that, unlike the distributions presented in Fig.~\ref{Fig:SO}, the individual particle \pt-spectra are not corrected with Bayesian unfolding.
A key aspect of the analysis presented in this article is the capability to compare the measured results with MC generator predictions. Extensive studies were performed to understand and mitigate any experimental biases due to the \SOPT selection. The experimental bias is evaluated by generating PYTHIA 8~\footnote{The fully simulated PYTHIA events are the same used to obtain the tracking efficiency correction discussed in Sec.~\ref{Sec:Details}. We refer to this section for technical details on the simulation.} events, that are propagated through a full ALICE simulation of the experimental apparatus. The \SOPT definition has been constructed to require that the generated events agree with the reconstructed simulated events, where the \pt spectra have been corrected for the minimum bias reconstruction efficiency\footnote{Secondary particles were rejected using MC information.}. This ensures that the measured and generated results are directly comparable, where the measured results do not require further unfolding. This is achieved by adopting the following criteria, which minimizes the experimental uncertainty and makes \SOPT a robust and model-independent observable:

\begin{itemize}
    \item \textbf{\SOPT selection in quantiles.}\\
    One major source of experimental bias is the smearing of the measured \SOPT distribution by detector effects. It was verified through MC studies that one can minimize the effect of the smearing by measuring \SOPT in quantiles, similar to how multiplicity classes are defined in ALICE~\cite{ALICE:2019dfi}. The stability of the quantiles is due to the fact that spherocity distribution does not have large gradients. A robust model-to-data comparison is achieved by calculating the model comparison in measured percentiles, rather than for the numerical \SOPT ranges we report in this article.
\end{itemize}

\begin{itemize}
    \item \textbf{Exclusion of neutral decay modes for resonance particles}\\
    The decay daughters of both \PHI and \KSTAR meet the standard ALICE definition of primary particles~\cite{alice_primary}, subsequently leading to the resonance daughters entering the measurement of \SOPT for a given event. These daughters will contribute to the multiplicity estimate if measured at midrapidity, and as both the \PHI and \KSTAR resonances have charged and neutral decay modes, the branching ratio for the charged decay mode will artificially inflate in high-multiplicity events. This bias can be completely avoided by only including (and correcting for) the following charged decay modes for resonance particles in the resulting spectra: \PHI $\rightarrow$ $K^+K^-$ and \KSTAR $\rightarrow$ K\PI. 
\end{itemize}

\begin{itemize}
    \item \textbf{Contamination of secondary particles}\\
    There is an effect that comes from the loose selection criteria for primary tracks used in this analysis. The loose DCA selections, which are there to maintain a full azimuthal acceptance, consequently lead to some decay daughters from $V^0$s and cascades to be incorrectly reconstructed as primary particles. To minimize the experimental bias from this effect, we include the primary \LA, \KOs, and \XI directly into the calculation of \SOPT for the generated MC predictions. The inclusion of these particles at the generator level makes the results presented in this article directly comparable to model predictions. 
\end{itemize}

 Note that the \SOPT distribution is the same across all presented particle species in this article. The implementation of these corrections reduce the experimental bias to a few percent, presented in Table~\ref{Tab:S0bias}, which lists the remaining fractional systematic uncertainties due to the \SOPT selection. The only exception is for low \pt \LA(\AL) and \KOs, where we do not present results for $\pt < \gevc{1}$. Our understanding is that for these very low \pt tracks, the azimuthal angles of the decay daughters that enter the \SOPT measurement are too different from their mother to give a precise result. Charged pions, kaons and (anti)protons are the most abundant particles, as well as the particles that enter directly into the \SOPT measurement, ensuring that these particle species are more robust against experimental biases. 

\begin{table}[h!]
\centering
\caption{The fractional systematic uncertainties used to estimate the experimental bias introduced when triggering on \SOPT. These uncertainties are a conservative estimate, obtained from the narrow \SOPT selections utilized in this analysis. The uncertainties are \pt-independent, and are applied to both jet-like and isotropic event selections, unless otherwise specified.}\label{Tab:S0bias}

\begin{tabular}{|l|cccccc|}
\multicolumn{1}{r}{\parbox[b][1.4em]{2em}{} } & \PI & K & p & $\mathrm{\phi}$ and $\mathrm{K}^{*}$ & \KOs and \LA & \multicolumn{1}{c}{$\Xi$} \\ \hline
\parbox[b][1.2em]{2em}{} &  &  &  &  & \pt$>1$ \GeVc &   \\
\SOPT \pt spectra & 0\% & 0\% & 0\% & 3\% & Isotropic: 1\% & 3\%  \\
  &  &  &  &  & jet-like: 4\% &   \\ \hline
\parbox[b][1.2em]{2em}{} &  &  &  &  & \pt$>1$ \GeVc &   \\
\SOPT h/\PI ratio & N/A & 0\% & 0\% & 3\% & Isotropic: 1\% & 3\%  \\
  &  &  &  &  & jet-like: 4\% &   \\ \hline

\end{tabular}
\end{table}

The unfolded numerical ranges for the selected multiplicity and \SOPT quantiles are listed in
Table~\ref{Tab:SOPT}. The roman numerals represent the multiplicity labels for these percentiles reported in earlier ALICE publications~\cite{ALICE:2019dfi,ALICE:2023yuk}. It is important to stress that the \SOPT selections for the model comparisons presented throughout this article are done on their corresponding \SOPT quantiles, not the numerical quantities listed in Table~\ref{Tab:SOPT}.

\begin{table}[h!]
\centering
\caption{Values corresponding to the \SOPT ranges of different quantiles used for the event selections in this analysis, obtained from the corrected \SOPT distribution.}\label{Tab:SOPT}

\begin{tabular}{|cc|ccc|}
\hline
\multicolumn{2}{|r|}{\parbox[b][1.2em]{2em}{} Multiplicity:} & \tracklet I--III (0--10\%) & \tracklet I (0--1\%) & V0M I (0--1\%) \\ \hline
\multicolumn{5}{l}{\parbox[b][1.4em]{1em}{Jet$\text{-}$like}} \\ \hline
\multicolumn{2}{|l|}{\parbox[b][1.1em]{1em}{}\SOPT 0--1\%} & $<0.423$ & $<0.503$ & $<0.453$ \\
\multicolumn{2}{|l|}{\SOPT 0--5\%} & $<0.523$ & $<0.588$ & $<0.553$ \\
\multicolumn{2}{|l|}{\SOPT 0--10\%} & $<0.573$ & $<0.633$ & $<0.603$ \\
\multicolumn{2}{|l|}{\SOPT 0--20\%} & $<0.638$ & $<0.689$ & $<0.663$ \\
\hline
\multicolumn{5}{l}{\parbox[b][1.2em]{1em}{Isotropic}} \\ \hline
\multicolumn{2}{|l|}{\parbox[b][1.1em]{1em}{}\SOPT 80--100\%} & $>0.838$ & $>0.863$ & $>0.858$ \\
\multicolumn{2}{|l|}{\SOPT 90--100\%} & $>0.878$ & $>0.898$ & $>0.888$ \\
\multicolumn{2}{|l|}{\SOPT 95--100\%} & $>0.903$ & $>0.918$ & $>0.913$ \\
\multicolumn{2}{|l|}{\SOPT 99--100\%} & $>0.933$ & $>0.943$ & $>0.943$ \\ \hline
\end{tabular}
\end{table}

\newpage

\section{Measurements of transverse momentum spectra}\label{Sec:Details}

In this section, the procedure for the \pt spectra measurements will be presented. First, the analyses utilizing only primary charged tracks are discussed in Section~\ref{Sec:Prim}, followed in Section~\ref{Sec:Sec} by the analyses which utilize the weak decay topology, reconstructed via secondary tracks. All particle spectra are measured in the exact same pseudorapidity interval ($|\eta|<0.8$) for which the \SOPT is defined, to ensure maximal correlation between event topology and particle production. This allows for a clear, one-to-one comparison between MC generators and data. 

\subsection{Particle identification utilizing primary charged tracks}\label{Sec:Prim}
This subsection discusses analyses regarding the production of \pikp, and the resonances \PHI and
\KSTAR. For these analyses, global tracks are reconstructed using the combined information from the ITS and TPC to achieve high precision.  Track selection criteria are applied to limit the contamination from secondary particles, to maximize tracking efficiency and to improve the \dedx and momentum resolution for primary charged particles. The number of crossed pad rows in the TPC is
required to be at least 70 (out of a maximum of 159); the ratio of
the number of crossed pad rows to the number of findable clusters $N_{\mathrm{cl}}$ (the number of geometrically possible clusters which can be assigned to a track) is restricted to be greater than 0.8, see Ref.~\cite{ALICE2} for details. The
goodness-of-fit values $\chi^2$ per cluster ($\chi^2/N_{\rm cl}$) of the track fits in the TPC must be less than 4. Tracks must be associated with at least one cluster in the SPD, and the $\chi^2$ values per cluster in the ITS are restricted in order to select high-quality tracks. The DCA to the primary vertex in the plane perpendicular to the beam axis (DCA$_{xy}$) is required to be less than seven times the resolution of this quantity; this selection is \pt dependent, i.e. DCA$_{xy}$ $< 7\times(0.0015 \pm 0.05\pt) $ cm. A loose selection criterion is also applied for the DCA in the beam direction (DCA$_{z}$), by rejecting tracks with DCA$_{z}>$ 2 cm, to remove tracks from possible residual pileup events. The transverse momentum of each track must be greater than 0.15
GeV/$c$ and the pseudorapidity is restricted to the range $|\eta|<$ 0.8 to avoid edge effects in the TPC acceptance. Additionally, tracks produced by the reconstructed weak decays of pions and kaons (the “kink" decay topology) are rejected.

The Particle Identification (PID) technique adopted for these analyses is based on a selection on the $n\sigma$ distributions, which are defined as
\begin{align*}
    n\sigma = \frac{\text{Signal}_{\textrm{measured}} - \langle
      \text{Signal}_{\textrm{expected}} \rangle}{\sigma},
\end{align*}
where the Signal can be either the measured \dedx in the TPC, or the extracted 1/$\beta$ in the TOF, and $\sigma$ is the corresponding resolution.

\subsubsection{Charged pions, kaons and (anti)protons}

Primary \pikp are measured in the range of 0.3--\gevc{20.0}, and yields are extracted following standard particle identification techniques reported in previous ALICE publications~\cite{ALICE:2023yuk,ALICE:2019hno}. At low \pt the particles are identified with high purity through the TPC \dedx by fitting the $n\sigma$ distributions~\cite{ALICE:2019hno}. In this \pt range,  the large separation power between \PI--K and p--K allows for a particle identification on a track-by-track basis. The relative particle abundances are obtained by fitting the $n\sigma$ distributions in narrow intervals of transverse momentum. For the kaons a two-Gaussian parameterization is used to correct for contamination of electrons.

At intermediate \pt, the TOF detector is used for identification by fitting the $\beta$ distribution~\cite{ALICE:2020nkc}. Particle identification in this \pt range is also performed on a track-by-track basis, by fitting the convolution of a Gaussian parameterization and an exponential function to the $\beta$ distribution. Similarly to low \pt, the identified particle yield is extracted through fitting the $n\sigma$ distributions. 

Finally, the TPC \dedx is used to identify the relativistic high-\pt particles (rTPC), with a technique described in Ref.~\cite{pikp_rtpc}. Unlike the methods used for lower \pt intervals, this technique does not allow for track-by-track identification of \pikp. Instead, the relative particle abundances are measured by fitting a four-Gaussian parameterization to the \dedx distributions, in $\eta$ and momentum intervals, where each Gaussian corresponds to the signal of \pikp and electrons. The fractional yields in each Gaussian are then used to extract the identified particle abundances from the full charged primary particle \pt spectra in $\eta$ and momentum intervals. The explicit momentum intervals for each particle identification technique are found in Table~\ref{Tab:pikpPID}.

\begin{table}[h!]
\centering
\caption{A breakdown of the momentum intervals for the different PID techniques used in the \pikp analysis.}\label{Tab:pikpPID}

\begin{tabular}{ccc}
\hline
\multicolumn{1}{|c|}{Analysis} & \multicolumn{1}{c|}{PID technique} & \multicolumn{1}{c|}{\pt ranges (\gev)}               \\ \hline
\multicolumn{1}{l}{}           & \multicolumn{1}{l}{}               & \multicolumn{1}{l}{\hspace{0.4cm}$\pi$\hspace{1.4cm}   K\hspace{1.6cm}   p}                       \\ \hline
\multicolumn{1}{|c|}{TPC}      & \multicolumn{1}{c|}{n$\sigma$ fits}   & \multicolumn{1}{c|}{0.3--0.7\hspace{0.5cm}   0.3--0.6\hspace{0.5cm}    0.45--1.0} \\
\multicolumn{1}{|c|}{TOF}      & \multicolumn{1}{c|}{$\beta$ fits}     & \multicolumn{1}{c|}{\hspace{-0.2cm}0.7--3.0\hspace{0.5cm}   0.6--3.0\hspace{0.5cm}    1.0--3.0}                                \\
\multicolumn{1}{|c|}{rTPC}     & \multicolumn{1}{c|}{\dedx fits}    & \multicolumn{1}{c|}{\hspace{-0.2cm}3.0--20\hspace{0.5cm}   3.0--20\hspace{0.5cm}    3.0--20}                                \\ \hline
\end{tabular}
\end{table}

\subsubsection{Resonance particles $\phi$ and \KSTAR}

Resonance particles cannot be directly detected by the experimental apparatus, but one can extract their yield by analyzing the invariant mass \Minv distribution, obtained by correlating all pairs of possible decay daughters. For the analyses presented here, the relevant decays are:

\begin{align}
\label{Eq:Channels1}
\nonumber\PHI &\rightarrow \rm K^++K^-  & \rm B.R.~= (49.2 \pm 0.5)\%, \\ 
\KSTAR(\bar{\KSTAR}) &\rightarrow \rm K^+(K^-)+\PI^-(\PI^+)  & \rm B.R.~= (66.503 \pm 0.014)\%.
\end{align}

The extraction technique summarized in the following is similar to previously published procedures, see for example Ref.~\cite{alice_resonance} for further details.

The \Minv distribution is constructed utilizing \PI and $K$ hadrons,  identified via selections on the TPC and TOF n$\sigma$ distributions. This is critical to reduce the contamination of pions in the kaons sample. The combinatorial background is estimated by creating a mixed-event \Minv sample for each resonance decay channel. For each new event, a continuous interval of the last 9 previous events are utilized for the event-mixing. The \Minv is then calculated for pairs of daughter candidates originating from different events, which is then subtracted from the signal distribution. The precision with which the mixed-event distribution can describe the combinatorial background depends significantly on the event topology. For events with jet-like topologies, this description is very poor, as the mixed events have a completely different event topology if the ``jet axes'' are not aligned. One can easily understand this by expressing the \Minv (with masses $m_{1,2}$ and
momentum $p_{1,2}$, and energies $E_{1,2}$) in terms of the angle between the two momentum vectors, $\theta$,
\begin{equation}
M_{\textrm{inv}}^2 = m_1^2 + m_2^2 + 2E_1E_2 - 2|\vec{p_1}||\vec{p_2}|\cos{(\theta)}.
\label{Eq1}
\end{equation} By similar reasoning one can deduce that the description of the combinatorial background in isotropic events is very good. Therefore, one needs significant statistics to precisely estimate the remaining background, so that the yield extraction is accurate. Relative to the other analyses presented in this article, the resonance analyses are consequently limited in terms of how narrow one is able to make \SOPT event selection. This also results in a significant \SOPT dependence of the systematic uncertainty.

The reconstructed resonance particle yield is extracted, in \pt-differential intervals, using a Voigtian peak function for the signal and a 2nd-degree polynomial for the remaining combinatorial background. The yield extractions for \PHI and \KSTAR are then performed in two separate steps: First, by counting the entries after subtraction of the 2nd-degree polynomial in the invariant mass ranges 0.995--1.07\,\gev and 0.76--1.12\,\gev, respectively. Second, the Voigtian tails are integrated outside the aforementioned counted mass intervals to correct for the missing tails, which represents a minor fraction of the total yield. The final yields are obtained by the summation of the bin counted estimate and the integrated Voigtian tails.                                             
\subsection{Particle identification utilizing weak decay topology}\label{Sec:Sec}

The details regarding the
reconstruction of the weakly decaying particles, \kzero, \LA, and \XI,
which utilize secondary tracks, are reported here. The relevant decays are:

\begin{align}
\label{Eq:Channel2}
\nonumber\kzero &\rightarrow \rm \PI^+\PI^- & \rm B.R.~= (69.20\pm0.05)\%,\\ 
\nonumber\LA(\AL) &\rightarrow \rm p(\bar{p}) + \PI^-(\PI^+ ) & \rm B.R.~= (63.9 \pm 0.5)\%,\\
\XI^-(\overline{\XI}^+) &\rightarrow \rm \LA(\AL)+\PI^-(\PI^+)  & \rm B.R.~= (99.887 \pm 0.035)\%.
\end{align}

As a substantial fraction of secondary particles from the decays are
produced outside the ITS, the track criteria are different for these
secondary tracks than for primary particles:  no ITS information is required, and the tracks are required to \emph{not} point directly back to the collision vertex. The data sample used in these analyses contained a significant contribution of out-of-bunch pile up that had to be accounted for.

The topological constraints, as well as the track quality
requirements, are summarized in Table~\ref{tab:v0cuts} and Table~\ref{tab:xicuts} for \VZ and \XI candidates, respectively. Topological constraints are applied on the reconstructed decay geometry to reduce background, specifically the distances of the daughters to the PV, the DCA of the daughters to the secondary vertex, and the pointing angle of the \VZ candidate with respect to the PV. The topological identification starts with the formation of \VZ candidates, consisting of two secondary tracks that:
\begin{itemize}
\item do not point to the PV (DCA of daughters to PV),
\item appear to share the same secondary production vertex (\VZ DCA),
\item do point back to the PV (from the secondary vertex) when
  the momentum vectors are summed up (pointing angle).
\end{itemize}
To calculate the \Minv, the masses of the secondary tracks are assigned based on Eq.~\ref{Eq:Channel2} after testing the PID consistency. This yields up to three different hypotheses, \kzero, \LA, and \AL, for the \Minv of each \VZ. If a \VZ candidate has its invariant mass compatible to other \VZ masses under the respective competing hypotheses, it is rejected. Additionally, for \LA and \AL, a selection on an experimental estimate of the particle's proper lifetime is also used to reduce the background. All selection criteria used are summarized in Table.~\ref{tab:v0cuts}.

\begin{table}[h!]
\begin{center}
\caption{Topological selection criteria used in the identification of the \Ks , \LA , and \AL particles.}
\label{tab:v0cuts}
\begin{tabular}{|l|c|}
\hline
 \parbox[b][1.1em]{1em}{}Selection variable & Selection criteria for \Ks (\LA , \AL ) \\ \hline
 
 \multicolumn{2}{l}{\parbox[b][1.2em]{1em}{}Topology} \\ \hline
DCA between daughters  & $< 1.0~\mathrm{cm}$ \\
Cosine of pointing angle & $> 0.97 (0.995)$ \\
Transverse decay radius  $R_{xy}$ & $> 0.5~\mathrm{cm}$ \\ \hline

\multicolumn{2}{l}{\parbox[b][1.2em]{1em}{}Daughter tracks selection} \\ \hline
\parbox[b][1.1em]{1em}{}DCA of daughters to PV &  $> 0.06$ cm \\
TPC PID of daughters & $< 5 \, \sigma$\\
Track pseudorapidity & \etarange 0.8 \\
TPC crossed rows & $N_\mathrm{cr} > 70$ \\
TPC crossed rows to findable ratio & $N_\mathrm{cr}/N_\mathrm{f} > 0.8$ \\ \hline

\multicolumn{2}{l}{\parbox[b][1.2em]{1em}{}Candidate selection} \\ \hline
\parbox[b][1.1em]{1em}{}\VZ pseudorapidity & \etarange 0.8 \\
Transverse momentum & $1.0 < \pt < 25.0$ \gev \\
Proper lifetime (transverse) & $( \, R_{xy} \times m_{(\LA , \AL) } / p_\mathrm{T} \,) < 30$~cm  \\
Competing mass & $ > 4\, \sigma$ \\ \hline

\end{tabular}
\end{center}
\end{table}

For cascades (\XI), secondary \VZ candidates are matched to secondary bachelor tracks and by assuming the relevant masses one can test the cascade hypothesis for a \XI. Secondary \VZ are identified by means of a selection on the distance of the PV from the pointing direction of the candidate.

\begin{table}[h!]
  \begin{center}
    \caption{Summary of the topological selection
      values used for the \XI candidate selection. All impact parameter
      requirements on tracks are 2D ($xy$).}\label{tab:xicuts}
    \begin{tabular}{|l|c|}
      \hline
\parbox[b][1.1em]{1em}{}Selection variable & Selection criteria\\ \hline
\multicolumn{2}{l}{\parbox[b][1.2em]{1em}{}Topology of \XI} \\ \hline
	\parbox[b][1.1em]{1em}{}DCA between daughters (\VZ and \PI) & $< 1.6$ cm \\
      Cosine of pointing angle & $ > 0.97$ \\
      Cascade transverse decay radius & $>0.8$ cm \\
      DCA of bachelor to PV & $> 0.05$ cm \\
      \hline 
\multicolumn{2}{l}{\parbox[b][1.2em]{1em}{}Topology of secondary \LA} \\ \hline
      \parbox[b][1.1em]{1em}{}DCA between daughters (p and \PI) & $< 1.6$ cm \\
      \VZ impact parameter & $>0.07$ cm \\
      \VZ transverse decay radius & $>1.4$ cm \\
	Window around $\Lambda$ mass & $< 0.006$ (GeV/$c^2$) \\  
        DCA of daughters to PV & $>0.04$ cm \\     
      \hline
      \multicolumn{2}{l}{\parbox[b][1.2em]{1em}{}Common bachelor/daughter track selection} \\ \hline
      \parbox[b][1.1em]{1em}{}Track pseudorapidity & \etarange 0.8\\
      TPC clusters & $>$ 70\\
      TPC PID of daughters & $< 5 \, \sigma$ \\ \hline
    \end{tabular}
  \end{center}
\end{table}

As the TPC is a relatively slow detector with a readout time of
${\approx}100\,\mu$s, it is possible to accidentally accept secondary tracks from \VZ or cascades produced in earlier or later events. To reject these candidates, we require that at least one decay daughter is either associated to a cluster in the ITS {\em or} is matched to a hit in the TOF detector.

\section{Corrections}\label{Sec:Corr}

The \pt spectra of \PHI, \KSTAR, and \XI are only corrected for acceptance and tracking efficiencies, while the study of \pikp, \LA, and \kzero requires more sophisticated corrections to account for the contamination from secondary particles. The \pt spectra of \pion, \kaon and \pr are corrected for acceptance, reconstruction inefficiency, TPC--TOF matching efficiency (only in the TOF analysis) and secondary particle contamination. The reconstruction efficiencies are obtained from event simulations using the PYTHIA8 (Monash-2013 tune) Monte Carlo event generator~\cite{pythia8}. Then, the propagation of the simulated events through the ALICE apparatus is described with GEANT 3.21~\cite{Geant3}. Finally, the simulated events are reconstructed and analyzed using the same procedure as for data. This study found that the reconstruction efficiencies are independent of the multiplicity and spherocity selections. Thus, the values from the minimum-bias sample were applied. In addition, since low-momentum interactions of $\mathrm{K}^{-}$ and $\mathrm{\overline{p}}$ with the detector material are not completely described by GEANT 3, an additional correction to the reconstruction efficiencies of these two particles is estimated with GEANT 4~\cite{Geant4} and FLUKA~\cite{Fluka}. 

The secondary particle contamination correction takes into account the contamination to the \pt spectra of $\pi$ and $\mathrm{p}$ from weak decays: $\kzero \rightarrow \pi^{+}+\pi^{-}, \lmb \rightarrow \mathrm{p}+\pi^{+},  \Sigma^{+} \rightarrow \mathrm{p}+\pi^{0}$ (and charge conjugates) and material interactions. This correction is estimated from a multi-template fit to data distributions of the transverse distance-of-closest
approach $(\mathrm{DCA}_{xy})$~\cite{piKp_PbPb_276}. Three templates derived from reconstructed PYTHIA 8 events, which represent the expected shapes of
$\mathrm{DCA}_{xy}$ distributions of primary and secondary (from weak decays) particles, and of particles from interactions with the detector material are used. The fits are constrained in the interval $\pm 3~\mathrm{cm}$. Variations of this constraint are accounted for as a systematic uncertainty. Since the TPC and TOF analyses employ different track selections, these corrections are estimated separately for the TPC and TOF analyses. At $\pt = 0.45 ~\rm{GeV}/\it{c}$ the contribution from non-primary $\pi^{+} (\rm{p})$ was found to be about $4\% (20\%)$, while at $\pt = 2.0 ~\rm{GeV}/\it{c}$ it decreases to about $1\% (4\%)$. Since this correction decreases asymptotically, an extrapolation from the TOF measurement is applied in the rTPC analysis. At $\pt=10~\mathrm{GeV}/c$ the correction to the pion and proton spectra is about $0.1\%$.
 
Secondary production of \LA(\AL) from $\Xi^\pm$ and $\Xi^0$ baryons were determined from measured $\Xi^\pm$ spectra in data and the feed-down matrix obtained from MC simulations, which gives the probabilities of $\Xi^\pm$ and $\Xi^0$ decaying into \LA(\AL) at given transverse momenta. The feed-down matrix is calculated in minimum bias events and the $\Xi^\pm$ \pt spectra used for the final correction come from the same high-multiplicity and spherocity event selections. The secondary yields do not exceed 25\% of the total yields across the entire \pt-range.

\section{Systematic uncertainties}\label{Sec:Sys}
The estimation of the total systematic uncertainties for each \SOPT distribution is performed using the methods described in Ref.~\cite{ALICE:2023yuk}. Two sources of systematic uncertainty are considered, and the total uncertainty is given as the sum in quadrature of the two components:

\begin{itemize}
    \item Monte Carlo non-closure: PYTHIA8 with Monash tune is the default model used to generate the spherocity response matrix and \SOPT distributions with and without the detector efficiency losses. The generated \SOPT distributions are defined as previously described in Sec.~\ref{Sec:SOPT}. The unfolded \SOPT spectrum from the simulation is compared to the generated one. Thus, any statistically significant difference between the generated and unfolded distributions is referred to as MC non-closure and is added in quadrature to the total systematic uncertainty. This uncertainty is within $2\%$. For \PHI, \KSTAR, \kzero, \LA and \XI, the systematic uncertainty comes for variations in how the decay products enters the \SOPT calculation, particle-by-particle, and is driven by the effect that the same particle can decay in different modes (e.g., shorter or longer lifetime for the weakly decaying particles). For this reason, it is expected that, even if the relative abundances of particle species in the MC are not the same as in data, the systematic effect will be similar.
    
    \item Dependence on the choice of the MC model: EPOS-LHC is used as the alternative model to generate a different spherocity response matrix. This response matrix is used to unfold the \SOPT distributions. The ratio between the final unfolded distributions using PYTHIA8 and EPOS-LHC was quantified and added to the total systematic uncertainty. This uncertainty has a spherocity dependence. While in the interval $\SOPT > 0.8$ this uncertainty is about $3\%$, for the interval \\$\SOPT < 0.6$ the uncertainty is of about $5.8\%$.
\end{itemize}

Similar to prior sections, the systematic uncertainties for analyses depending on primary charged particles  are discussed in Section~\ref{Sec:SysPrim}. Likewise, the systematic uncertainties for longer-lived particles are discussed in Section~\ref{Sec:SysSec}. We note that in all cases, checks in the spirit of Ref.~\cite{Barlow:2002yb} were performed, comparing measurements of default and alternate criteria in quadrature. This is done in order to remove possible statistical effects, where the default and alternate selections are statistically correlated, from systematic uncertainties. 

\subsection{Systematic uncertainties for analyses utilizing primary charged particles}\label{Sec:SysPrim}

The total systematic uncertainty on the \pt spectra of \pikp is divided into two categories. The first class includes the uncertainties  that are common among the different PID techniques: the track selection criteria, the ITS--TPC and TPC--TOF matching efficiencies. These uncertainties are species and \pt dependent. The second class includes systematic uncertainties that are technique dependent: signal extraction method and the estimation of the secondary particle correction. This study utilizes the same methods used in previous ALICE analyses ~\cite{ALICE:2023yuk,ALICE:2020nkc,ALICE:2019hno,pikp_rtpc}. Most of the systematic uncertainties cancel in the \pt-differential particle ratios $(\mathrm{K}/\pi$ and $\mathrm{p}/\pi)$ except the ones attributed to the signal extraction and feed-down correction. Moreover, at high transverse momentum (rTPC analysis) the procedure described in~\cite{pikp_rtpc} is used to extract the signal extraction systematic uncertainty on the $\mathrm{K}/\pi$ and $\mathrm{p}/\pi$ ratios directly from fits to the \dedx distributions. Table~\ref{tab:systematics_pikp} shows a summary of the relative systematic uncertainties on the \pt spectra of \pion, \kaon, \pr, and the particle ratios. The results are shown for the spherocity classes: Jet-like and Isotropic [0--1]\%, and the \SOPT unbiased case. The multiplicity class corresponds to $N_{\mathrm{tracklets}}^{|\eta|<0.8}\, \mathrm{I{-}III}$. Similar results are obtained for the other multiplicity classes. 

\begin{sidewaystable}[p!]
\centering
\caption{Summary of the relative systematic uncertainties on the \pt spectra of \pion, \kaon, \pr, and the particle ratios. The uncertainties are reported as percentage values. The intervals represent the minimum and maximum values in the respective interval of identification. They are given for the spherocity classes: Jet-like and Isotropic [0-1]\%, and the \SOPT unbiased case. The multiplicity class corresponds to $N_{\mathrm{tracklets}}^{|\eta|<0.8}\, \mathrm{I{-}III}$.}
\begin{adjustbox}{width=\textwidth}
\begin{tabular}{c ccc ccc ccc}
    \toprule
\multirow{2}{*}{Source} 
        & \multicolumn{3}{c}{$\pi$} 
        & \multicolumn{3}{c}{$\mathrm{K}$} 
        & \multicolumn{3}{c}{$\mathrm{p}$}  \\
    \cmidrule(lr){2-4} \cmidrule(lr){5-7} \cmidrule(lr){8-10}
        & Jet-like & Isotropic & HM
        & Jet-like & Isotropic & HM
        & Jet-like & Isotropic & HM \\
    \midrule
    \multirow{1}{*}{Common}  
    &  &  &  &  &  &  &  &  & \\ 
    \midrule
    \addlinespace
    \multirow{1}{*}{ITS--TPC matching efficiency}  
    & 0.7 -- 3 & 0.7 -- 3 & 0.7 -- 3 & 0.7 -- 3 & 0.7 -- 3 & 0.7 -- 3 & 0.7 -- 3 & 0.7 -- 3 & 0.7 -- 3 \\
    \addlinespace
    \multirow{1}{*}{Track selection}  
    & 0.13 -- 1 & 0.13 -- 1 & 0.13 -- 1 & 0.13 -- 2 & 0.13 -- 2 & 0.13 -- 2 & 0.13 -- 2.7 & 0.13 -- 2.7 & 0.13 -- 3 \\
    \midrule
    \multirow{1}{*}{\textbf{TPC}}
    &  &  &  &  &  &  &  &  & \\
    \addlinespace
    \multirow{1}{*}{PID}  
    & 0.2 -- 1.8 & 0.2 -- 1.9 & 0.2 -- 1.8 & 2.8 -- 6 & 2.8 -- 6 & 2.8 -- 6 & 1.6 -- 9.8 & 1.6 -- 10 & 1.6 -- 9.9 \\
    \addlinespace
    \multirow{1}{*}{$N_{\mathrm{cl}}$}  
    & 0.5 -- 0.7 & 0.6 -- 0.7 & 0.6 -- 0.7 & 0.02 -- 0.8 & 0.03 -- 0.7 & 0 -- 0.5 & 0.7 -- 1.5 & 0.7 -- 1.4 & 0.7 -- 1.3 \\
    \addlinespace
    \multirow{1}{*}{Feed-Down}  
    & 0.3 -- 0.9 & 0.3 -- 0.9 & 0.3 -- 0.9 &  --  &  --  &  --  & 1.1 -- 10 & 1.1 -- 10 & 1.1 -- 10 \\
    \addlinespace
    \multirow{1}{*}{\textbf{TOF}}
    &  &  &  &  &  &  &  &  & \\
    \addlinespace
    \multirow{1}{*}{PID}  
    & 0.03 -- 2.9 & 0.03 -- 1.8 & 0.02 -- 2.1 & 0.3 -- 8.7 & 0.3 -- 4.2 & 0.2 -- 4.8 & 0.08 -- 4.8 & 0.08 -- 6.9 & 0.09 -- 5.4 \\
    \addlinespace
    \multirow{1}{*}{Feed-Down}  
    & 0 -- 0.2 & 0 -- 0.2 & 0 -- 0.2 &  --  &  --  &  --  & 0.2 -- 0.9 & 0.2 -- 0.9 & 0.2 -- 0.9 \\
    \addlinespace
    \multirow{1}{*}{TPC--TOF matching efficiency}  
    &  3  & 3 &  3 & 6 & 6 & 6 & 4 & 4 & 4 \\
    \addlinespace
    \multirow{1}{*}{\textbf{rTPC}}
    &  &  &  &  &  &  &  &  & \\
    \addlinespace
    \multirow{1}{*}{PID}  
    & 0.65 -- 2.8 & 0.68 -- 2.9 & 0.67 -- 2.8 & 2.7 -- 10.7 & 2.5 -- 10.7 & 2.5 -- 10.3 & 4.6 -- 15.0 & 4.8 -- 14.9 & 4.7 -- 14.1 \\
    \addlinespace
    \multirow{1}{*}{$N_{\mathrm{cl}}$}  
    & 0.5 -- 1.5 & 0.4 -- 1.6 & 0.6 -- 1.5 & 0.04 -- 0.7 & 0.05 -- 0.5 & 0.03 -- 0.7 & 0.08 -- 1.4 & 0.5 -- 1.7 & 0.1 -- 1.4 \\
    \addlinespace
    \multirow{1}{*}{Feed-Down}  
    & Negl. & Negl. & Negl. &  --  &  --  &  --  & 0 -- 0.2 & 0 -- 0.2 & 0 -- 0.2 \\
    \midrule
    \addlinespace
    \multirow{1}{*}{Total}  
    & 1.2 -- 4.4 & 1.2 -- 4.5 & 1.5 -- 5 & 3.1 -- 10.9 & 3.1 -- 10.9 & 3 -- 11 & 3.1 -- 15.7 & 3 -- 15.3 & 3 -- 15 \\
    \midrule
    \addlinespace
    \multirow{1}{*}{Particle ratios}  
    &  &  &  &  & $\mathrm{K}/\pi$ &  &  & $\mathrm{p}/\pi$ &  \\
    \midrule
    \addlinespace
    \multirow{1}{*}{Total}  
    & -- & -- & -- & 0.3 -- 11 & 0.3 -- 11.4 & 0.4 -- 11 & 0.8 -- 15 & 0.7 -- 14.6 & 1 -- 14 \\
    \bottomrule
\end{tabular}
\label{tab:systematics_pikp}
\end{adjustbox}
\end{sidewaystable}

The main contributions to the systematic uncertainty on the spectra of the resonance particles are listed in Table~\ref{Tab:SysRes}. The uncertainties are evaluated in groups, where each group contains sources of systematic uncertainties that cannot be evaluated individually. The systematic uncertainties are \pt-dependent, and the ranges listed in the table represent the minimum and maximum values. The maximum uncertainties for \PHI and \KSTAR are obtained at low \pt ($\pt < 1.5 \gevc$), with the uncertainties reaching local minima at intermediate \pt ($1.5 < \pt < 4.0 \gevc$), and then approaching towards the maximum value at high \pt ($\pt > 4.0 \gevc$). Both the correlated and uncorrelated sources defined in Table~\ref{Tab:SysRes} are used to account for the uncertainty in the \pt-differential particle spectra, but only the uncorrelated sources are considered for the \SOPT-dependent-to-\SOPT-integrated ratios.

The uncertainties related to signal extraction are estimated through variations of the fit range, constraints to the peak fit, and variations of the residual background function. ``Track selection \& PID'' consists of varying the $n\sigma$ requirements for valid resonance daughter candidates, as well as variations in the track quality criteria. This includes variations of the required crossed rows in the TPC, the initial vertex position along the beam axis, and the DCA along the beam axis (DCA$_z$). These variations are performed both in data and simulations, simultaneously.  Finally, the background estimation includes changes in how the combinatorial background is evaluated (event-mixing, reflected mass hypothesis, like-charge pairs).

\begin{table}
\centering
\caption{The most relevant systematic uncertainties for the resonance analysis as a function of \SOPT. ``HM'' in this table represents the \SOPT-integrated spectra. The uncertainties are reported as percentage values. Uncertainties are \pt-dependent, and ranges listed represent the minimum and maximum values presented in the final spectra (see text for details). }\label{Tab:SysRes}

\begin{tabular}{|l|lll|lll|} 
\hline
\multicolumn{1}{|r|}{\parbox[b][1.1em]{1em}{}Hadron:}                  &       &~\PHI &                           &       & \KSTAR &                       \\
\multicolumn{1}{|r|}{Topology:}                                 & Jet-like & HM                  & Iso                       & Jet-like & HM                    & Iso                   \\ 
\hline
\multicolumn{1}{l}{\parbox[b][1.2em]{1em}{}Uncorrelated sources}   &       &                     & \multicolumn{1}{l}{}      &       &                       & \multicolumn{1}{l}{}  \\ 
\hline
\parbox[b][1.1em]{1em}{}Signal extraction                        & 1--3 & 1--2           & 1.5--2.5                   & 3--7 & 2--4                 & 1--5                 \\
Track selection \& PID                      & 2--6 & 1--5             & 1--5                     & 1--5 & 1--4                 & 1--4                 \\
Background estimation                    & 1--3 & 0--1             & 1--2                   & 1--3 & 1--4                 & 1--4                 \\ 
\hline
\multicolumn{1}{l}{\parbox[b][1.2em]{1em}{}Correlated sources} &       &                     & \multicolumn{1}{l}{}      &       &                       & \multicolumn{1}{l}{}  \\ 
\hline
\parbox[b][1.1em]{1em}{}Tracking efficiency                      &    & 2                 &                        &    & 2                   &                   \\
Branching ratio                          &    & 1                 &                        &    & 2                   &                    \\
Hadronic interaction                     &  & 2--3               &                     &  & 0--2                 &                  \\
Material budget                          &  & 0--5               &                      &  & 0--5                 &                  \\ 
\hline
\multicolumn{7}{l}{\parbox[b][0.2em]{1em}{}} \\ \hline
\parbox[b][1.1em]{1em}{}Total uncertainty                        & 5--9 & 5--8               & 5--8 & 5--9 & 4--8                 & 4--8                 \\
\hline
\end{tabular}
\end{table}

\begin{table}[h!]
\centering
\caption{The most relevant systematic uncertainties for the long-lived particles \kzero, \LA(\AL), and \XI, as a function of \SOPT. ``HM'' in this table represents the \SOPT-unbiased spectra. The uncertainties are reported as percentage values. Uncertainties are \pt-dependent, and ranges listed represent the minimum and maximum values presented in the final spectra (see text for details). }\label{Tab:SysLong}

\begin{tabular}{|l|lllll|}
\hline
\multicolumn{1}{|r|}{Topology:}              & Jet-like     & Iso       & HM        & Jet-like/HM       & Iso/HM      \\ \hline

\multicolumn{6}{l}{\parbox[b][1.2em]{1em}{K$^0_\mathrm{S}$}} \\
\hline
Selection cuts        & 3    & 3--4     & 3--4     & Negl.          & 1          \\
Track pile--up        & 1 & 1--3 & 1 & 0--2 & 0--2   \\
Signal extraction     & 1--3     & 1--3     & 1--3     & Negl. & Negl. \\
Efficiency            & 2    & 2    & 2    & 2         & 2         \\
Material budget       & 4    & 4    & 4    & --              & --              \\
Experimental bias     & 4    & 1    & --         & 4         & 1         \\ \hline
Total uncertainty     & 7    & 6--7   & 5--6     & 5         & 2--3          \\ \hline
\multicolumn{6}{l}{\parbox[b][1.2em]{1em}{$\Lambda$($\overline{\Lambda})$}} \\
\hline
Selection cuts & 1--5     & 2--6     & 4--5     & 0--1          & 0--3          \\
Track pile--up        & 4--5 & 5 & 3--5 & 0--1.5 & 0--1   \\
Signal extraction     & 2--6     & 2--6     & 2--6     & 0--2          & 0--1          \\
Feed-down correction   & 1.0--1.5 & 1.0--1.5 & 1.0--1.5 & Negl. & Negl. \\
Efficiency            & 2    & 2    & 2    & 2         & 2         \\
Material budget       & 4    & 4    & 4    & --              & --              \\
Experimental bias     & 4    & 1    & --         & 4         & 1         \\ \hline
Total uncertainty     & 8--10    & 8--9     & 7--9     & 5         & 3--4          \\ \hline
\multicolumn{6}{l}{\parbox[b][1.2em]{1em}{$\Xi^\pm$}} \\
\hline
Selection cuts  & 0--1         & 0--1         & 0--1     & Negl.          & Negl.  \\
Track pile--up         & 2--3         & 2--3         & 2--3     & 2         & Negl.    \\
Signal extraction     & 0--1        & 0--1     & 0--1     & 1         & Negl. \\
Efficiency             & 2         & 2         & 2    & 2         & 2         \\
Material budget       & 1--9         & 1--9         & 1--9     & --              & --              \\
Experimental bias     & 3         & 3         & --    & 3         & 3         \\ \hline
Total uncertainty     & 5--10         & 5--10         & 4--10    & 4         & 3.5         \\ \hline
\end{tabular}
\end{table}

The correlated sources apply equally across the different \SOPT selections, and cancel in the ratio. The hadronic interaction and material budget represent the uncertainties in interactions between particles and the ALICE detector, and the uncertainty in the hadronic interaction cross section of particles traversing material in ALICE.

\subsection{Systematic uncertainties on analyses of long-lived particles}\label{Sec:SysSec}

The systematic uncertainties on long-lived weakly decaying particles, \Ks, \LA(\AL), and \XI, are reported in Table~\ref{Tab:SysLong} and have similar components for all particle species:
\begin{itemize}
\item Selection criteria: estimated by carrying out the analysis using looser and tighter variations of the selections in Tables~\ref{tab:v0cuts} and~\ref{tab:xicuts}. 
\item Track pile-up: assigned due to the requirement that at least one daughter (or bachelor track for \XI) has a fast-detector signal (ITS or TOF), see Sec.~\ref{Sec:Sec}. This systematic uncertainty was found by varying the number of required tracks and fast signals, and reflects how well these conditions are modelled in the MC simulation.
\item Signal extraction: estimated by varying the range in \Minv for signal and background and the shape of the background.
\item Efficiency: accounts for possible variations of the tracking
efficiency with multiplicity. The same uncertainty, 2\%, as used in a previous detailed study of multiplicity-dependent strangeness production in \s = 13 TeV pp collisions~\cite{alice_newstrangeness} is assigned here.
\item Material budget: estimated by varying parameters in the Monte Carlo description of the ALICE apparatus, see Ref.~\cite{ALICE2} for details.
\item Experimental bias: taken from Table~\ref{Tab:S0bias}. See discussion in Sec.~\ref{Sec:SOPT} for details.
\end{itemize}

In addition to the above-mentioned uncertainties, there is also a contribution from evaluating the secondary yields for \LA(\AL) particles. It was determined by varying \XI yields within their uncertainties, using an alternative method of constructing the feed-down matrix only from charged \XI baryons, as well as a flat systematic uncertainty to account for the possible multiplicity dependence of the matrix (shown in Table~\ref{Tab:SysLong} as ``Feed-down correction").

Finally, we note that some of the systematic uncertainties cancel when we compare results in the same multiplicity class but with various \SOPT selections. This uncertainty is shown in Table~\ref{Tab:SysLong} as ``jet-like/HM" and ``Iso/HM" and was estimated by doing the systematic variations for a jet-like selection and an unbiased (same multiplicity) selection in parallel and comparing the ratios of \pt spectra with and without the variation. The studies were performed for various \SOPT and multiplicity selections. No strong dependence was found inside the relevant selections (e.g. for tighter or looser jet-like selections). However, one should note that such a dependence is hard to pin down with good precision due to the limited number of candidates for very tight selections.

\section{Results and discussion}\label{Sec:Results}
 
The results presented in this section include the transverse momentum spectra, integrated yields, \myield, and mean transverse momentum, \mpt, as a function of \SOPT, as well as \pt-differential ratios between particle species (mainly with respect to pions), and the \pt-differential ratios-to-\PI relative to the \SOPT unbiased baseline (referred to as the ``double ratio''). Finally, the strangeness enhancement as a function of \SOPT is investigated.

In the following, the \myield and \mpt are calculated in the measured kinematic range, and corrected for the limited range by extrapolating the spectra to the unmeasured \pt regions using Levy--Tsallis fits down to \pt = 0. To account for the additional systematic uncertainties arising from this procedure, modified fit functions such as Boltzmann, $m_{\rm T}$-exponential, \pt-exponential, Fermi-Dirac (only for fermions) and Bose--Einstein statistics (only for bosons), and Boltzmann--Gibbs blast-wave functions are used as variations. The systematic uncertainties arising due to extrapolations are evaluated from the RMS of the variations with respect to the Levy--Tsallis fits. The extrapolation uncertainty is added in quadrature to the systematic uncertainties obtained from the measured \pt ranges to get the final systematic uncertainties for the \myield and \mpt. More details about this procedure can be found in Ref.~\cite{alice_newstrangeness}.

The experimental measurements will be compared with a broad selection of \pp MC models, namely PYTHIA 8.2~\cite{Torbjorn} (both the default Monash tune and with the added rope hadronization framework), Herwig 7.2~\cite{Herwig}, and EPOS-LHC~\cite{EPOS-LHC}. PYTHIA 8.2 is a QCD-inspired model that is built around the Lund-string model for hadronization~\cite{pythia8}. The default Monash tune is unable to describe the strangeness enhancement in small systems, while the rope extension (layers of overlapping strings that increase the string tension) of the Lund-string model~\cite{Ropes} has been introduced to accommodate this. For the PYTHIA 8.2 Ropes, note that this article only utilizes the "Flavour Ropes" for the model predictions, without the string-shoving mechanism~\cite{Bierlich:2016vgw} incorporated in the rope hadronization framework. Similarly, Herwig is a QCD-inspired \pp generator centered around a cluster hadronization model~\cite{Herwig}, which has recently been extended to be able to describe the strangeness enhancement~\cite{ALICE:2020nkc}. On the contrary, EPOS-LHC is a two-component core-corona model, which incorporates QGP features in the core to explain the strangeness enhancement.

The effect of the \SOPT selection is discussed in Sec.~\ref{Sec:ResHM} for the two multiplicity estimators. It is shown that the spherocity selection for the events with the highest midrapidity multiplicity, 0--1\% \tracklet, gives the best control of the event ``hardness'', i.e. select events with a large \mpt. Results obtained using this estimator are therefore first presented in Secs.~\ref{Sec:ResSpec}--~\ref{Sec:IntYield}, followed by results obtained using other multiplicity estimators in Sec.~\ref{Sec:ResExpMult}.

\subsection{High-multiplicity estimators and \SOPT}\label{Sec:ResHM}

\begin{figure}[h!]
    \centering
    \includegraphics[scale=0.6]{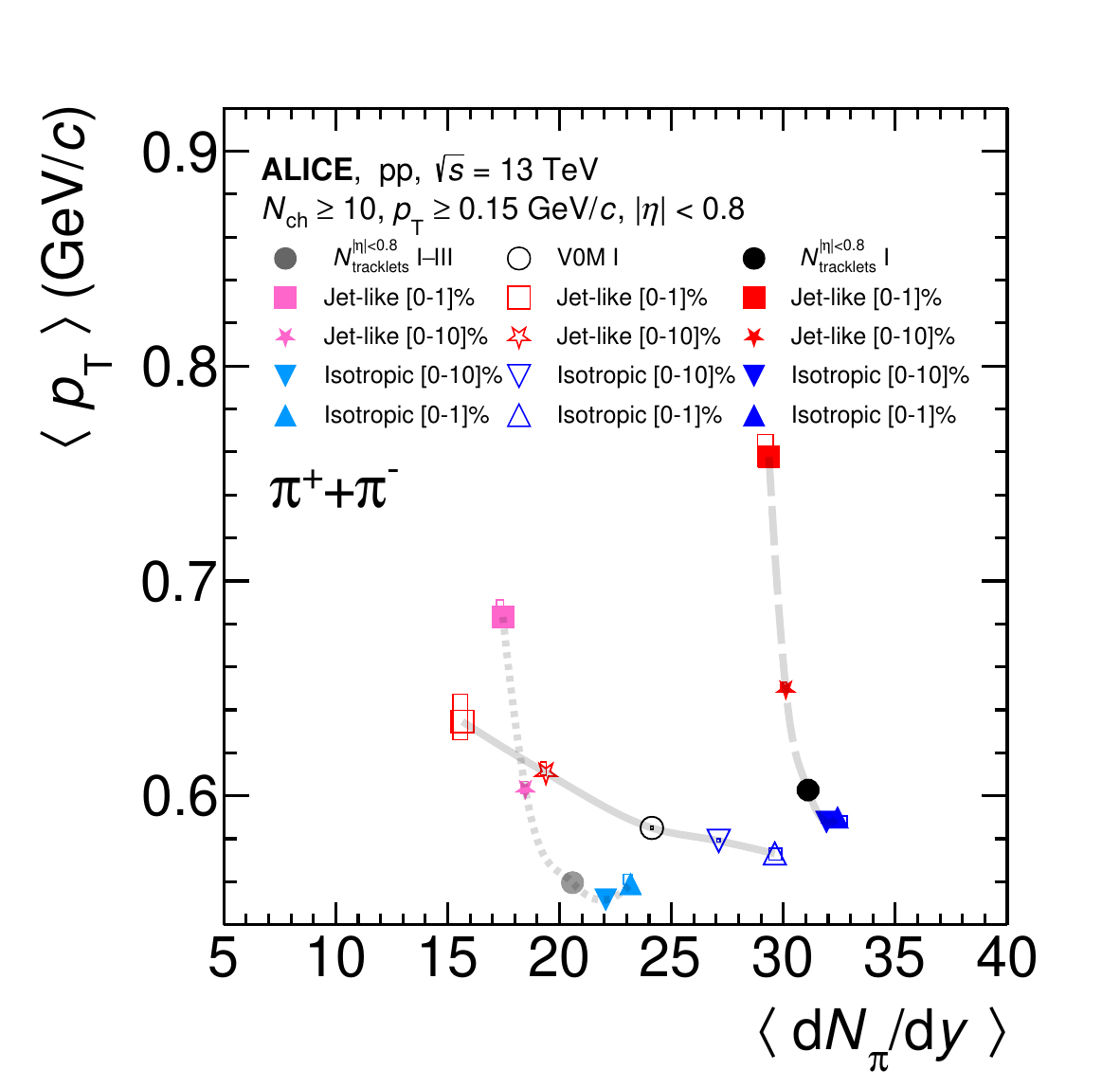}
    \caption{Correlation between $\langle p_{\rm{T}}\rangle$ and $\langle \rm{d}N_{\pi}/\rm{d}y \rangle$ as a function of \SOPT, in the 0--10\% and 0--1\% V0M and \tracklet multiplicity classes. The total systematic uncertainties are represented by empty boxes. The statistical uncertainty is smaller than the reported marker sizes.}
    \label{Fig:CL1vsV0M}
\end{figure}

If one considers a high-multiplicity \pp collision to be built up from independent subcollisions (a traditional MPI picture~\cite{MPI}), then there will be a trivial isotropization with increasing multiplicity. As previous ALICE measurements indicate a strong correlation between multiplicity and QGP-like effects, such as strangeness production~\cite{Naturepaper}, it is important for the study presented here to disentangle this possible trivial bias from the underlying physical properties of interest. To understand the impact of this on the \SOPT selection, the \mpt and the average pion yield $\langle \rm{d}\it{N}_{\pi}/\rm{d}\it{y} \rangle$, with different multiplicity and spherocity selection criteria are shown in Fig.~\ref{Fig:CL1vsV0M}. Results are shown for both the forward (V0M) and midrapidity (\tracklet) multiplicity estimators (described in Sec.~\ref{Sec:Mult}). A clear distinction is observed with respect to how the different multiplicity estimators relate to the \SOPT selection. This effect is solely driven by the rapidity region where the multiplicity is estimated, and not by properties of the ALICE apparatus.

It is observed that the V0M multiplicity selection maintains a similar \mpt, but contains large variations in \myieldpi for the different \SOPT selections. In contrast, the \tracklet selected events are characterized by large differences in \mpt among the event classes, selecting events according to their hardness. The implicit multiplicity dependence of \SOPT is minimized by sharply constraining the multiplicity using a midrapidity multiplicity estimation. This implies that the \tracklet multiplicity estimator, in tandem with a \SOPT selection, is best at separating events based on their hardness.

\begin{figure}[h!]
    \centering
    \includegraphics[scale=0.6]{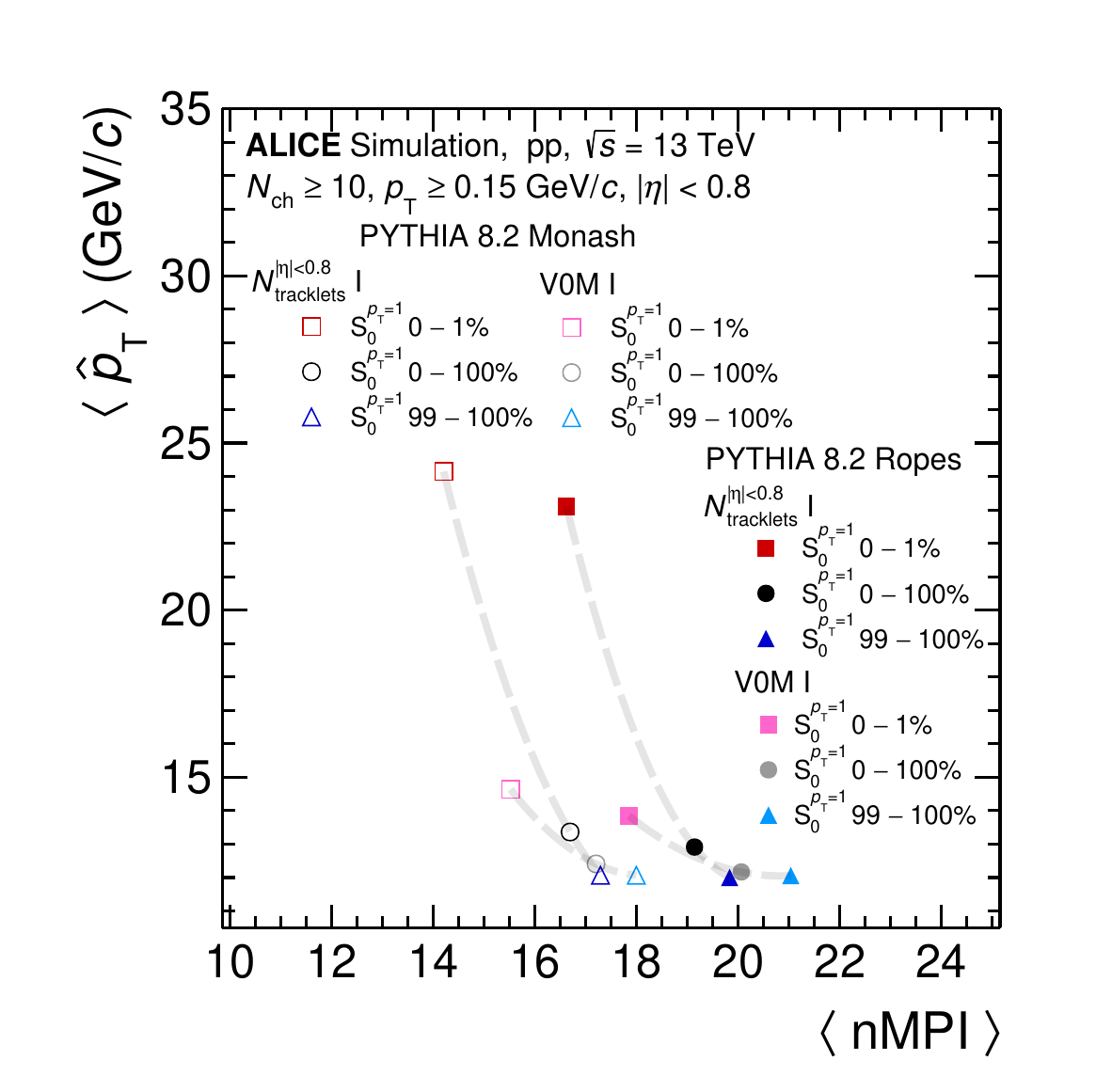}
    \caption{PYTHIA 8.2 correlation study between $\left<\pthat\right>$ and $\left<n\textrm{MPI}\right>$ as a function of \SOPT, in 0--1\% V0M and \tracklet multiplicity classes. The default PYTHIA 8.2 Monash variation is compared to PYTHIA 8.2 with color rope hadronization. The total systematic and statistical uncertainties are smaller than the marker sizes. The grey band is an interpolation between the points, to more clearly illustrate the trend of each multiplicity and model variation.}
    \label{Fig:pthatvsmpi}
\end{figure}

Moreover, PYTHIA 8.2 model studies of the correlation between the average transverse momentum transfer of the hardest parton--parton interaction, $\left<\pthat\right>$, and the average number of multi-parton interactions $\left<n\textrm{MPI}\right>$ is presented in Fig.~\ref{Fig:pthatvsmpi} for multiplicity and \SOPT intervals of 0--1\%. For each estimator, the multiplicity is calculated as the number of primary charged particles in the pseudorapidity interval(s) covered by that estimator. Figure~\ref{Fig:pthatvsmpi} shows that the hardest scattering for the events in the jet-like \SOPT 0--1\% event category is significantly harder than for the \SOPT-integrated, high-multiplicity reference. This is observed both for the default PYTHIA 8.2 Monash, and with the rope hadronization framework enabled. This feature is present when the multiplicity is estimated both in the \tracklet and V0M pseudorapidity regions, but the effect is particularly strong when the multiplicity is estimated at midrapidity. Furthermore, the isotropic \SOPT 99--100\% events are slightly softer than overall high-multiplicity events. In conjunction with the softer \mpt presented in Fig.~\ref{Fig:CL1vsV0M}, these findings suggest that the isotropic topologies are formed by multiple softer interactions, while the jet-like topologies have at least one hard scattering that is significantly harder than for the \SOPT-integrated selection.

In the following, a complete set of results will be presented using \tracklet 0--1\% to highlight the impact on the QCD dynamics of the extreme event topologies, while minimizing the effects of any trivial multiplicity (system size) dependence. The results are first presented for jet-like and isotropic events, utilizing a 0--10\% and 90--100\% \SOPT selection, respectively, to have a complete set of particle spectra. Furthermore, selected results are also presented for the most extreme 1\% percentiles of \SOPT. 

\subsection{Results of \SOPT-differential \pt spectra at \tracklet 0--1\%}\label{Sec:ResSpec}

\begin{figure}[h!]
    \centering
    \includegraphics[scale=0.8]{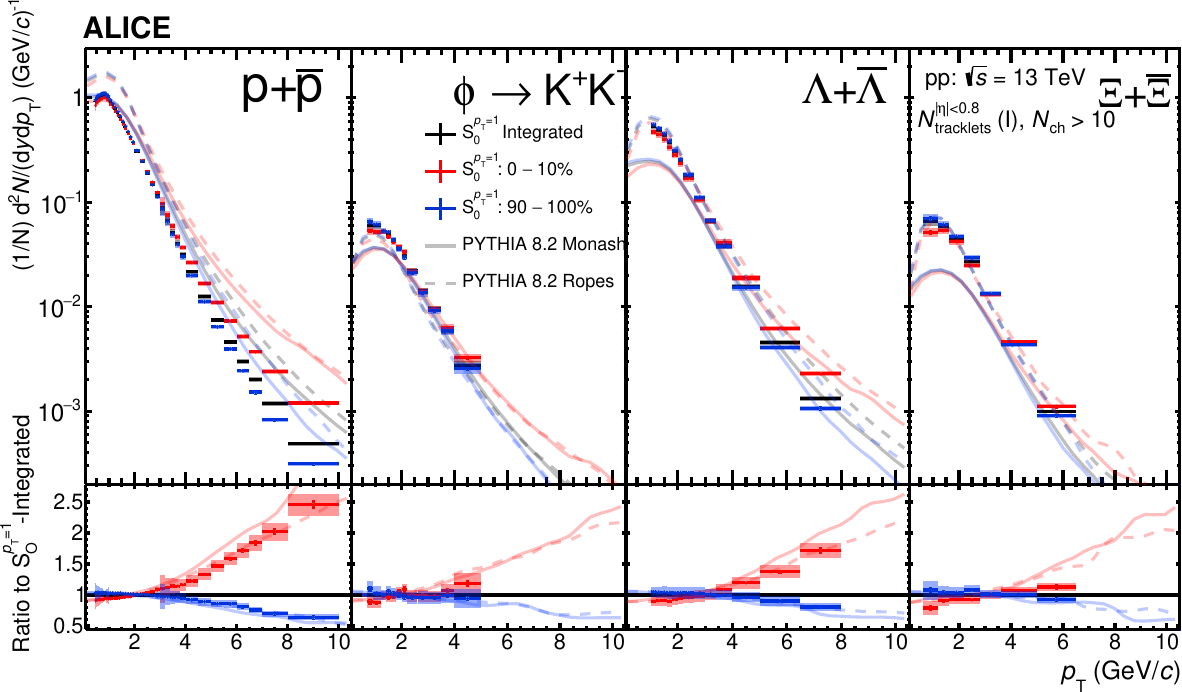}
    \includegraphics[scale=0.8]{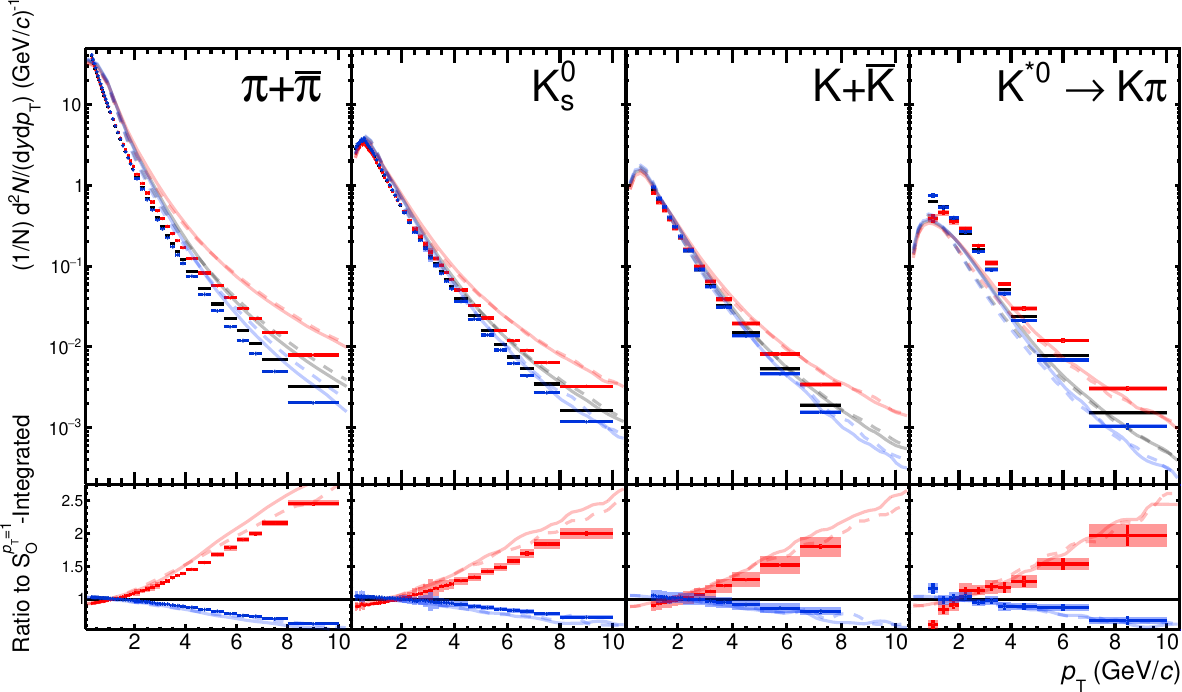}
    \caption{Transverse momentum distribution of \pikp, \KSTAR, \PHI, \KOs, \LA and \XI for \SOPT classes selected for events at high-multiplicity, determined by events in the top 1\% of \tracklet. The lower panels present the ratio between the \SOPT-integrated and \SOPT-differential events. Statistical and total systematic uncertainties are shown by error bars and boxes, respectively. The curves represent PYTHIA 8.2 model predictions of the same measurement. The average statistical uncertainties from the predictions across all particle species range in the order of 1--15 \% from low-to-high \pt.} 
    \label{Fig:CombSpectraA}
\end{figure}

\begin{figure}[h!]
    \centering
    \includegraphics[scale=0.8]{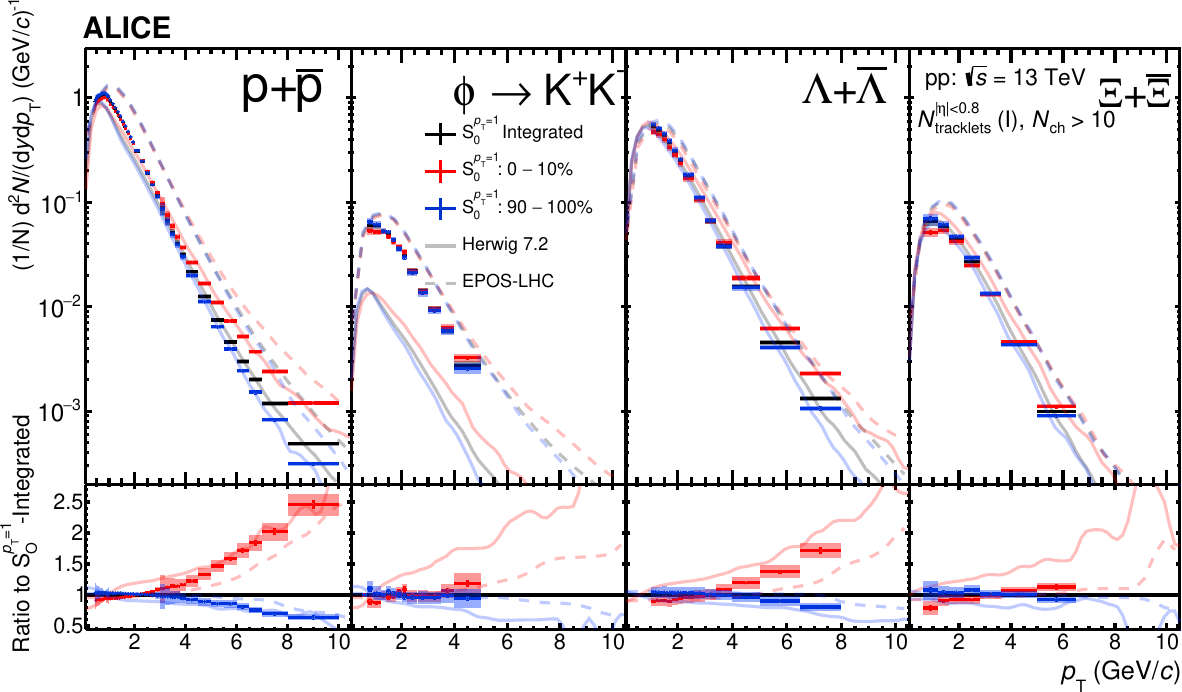}
    \includegraphics[scale=0.8]{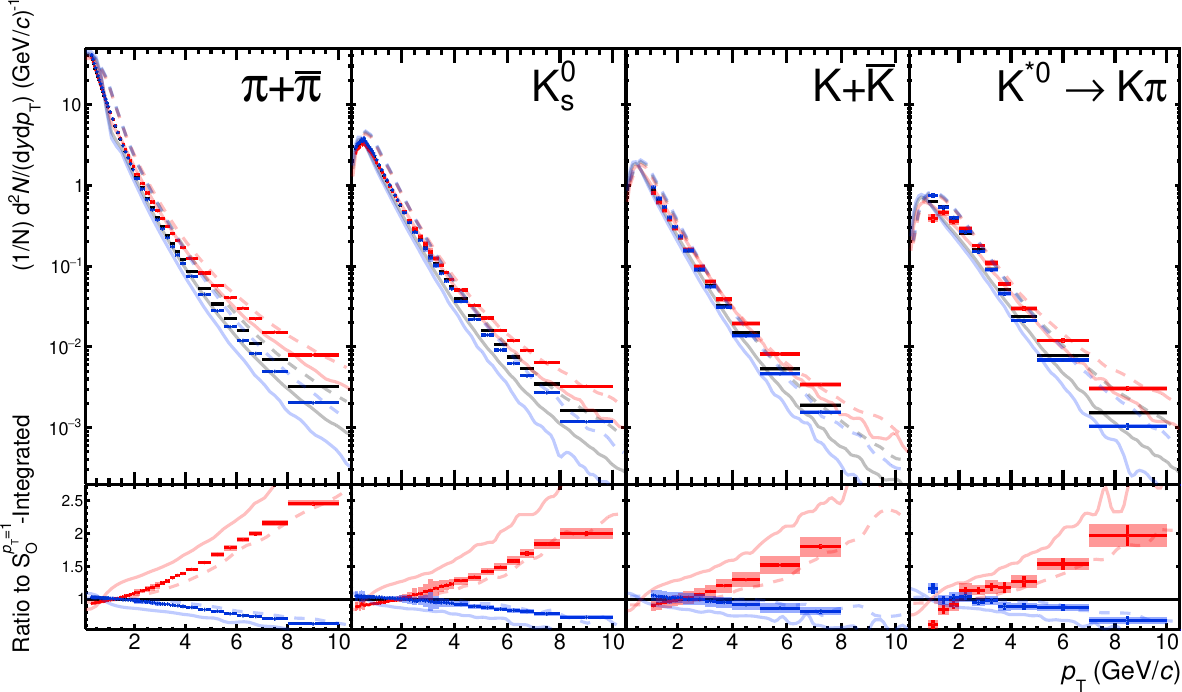}
    \caption{Transverse momentum distribution of \pikp, \KSTAR, \PHI, \KOs, \LA and \XI for \SOPT classes selected for events at high-multiplicity, determined by events in the top 1\% of \tracklet. The lower panels present the ratio between the \SOPT-integrated and \SOPT-differential events. Statistical and total systematic uncertainties are shown by error bars and boxes, respectively. Fig.~\ref{Fig:CombSpectraA} and Fig.~\ref{Fig:CombSpectraB} both show the same experimental data. The curves represent Herwig 7.2 and EPOS-LHC predictions of the same measurement. The average statistical uncertainties from the predictions across all particle species range in the order of 1--25 \% from low-to-high \pt.}
    \label{Fig:CombSpectraB}
\end{figure}

The \pt spectra for jet-like and isotropic events are presented in Figures~\ref{Fig:CombSpectraA} and~\ref{Fig:CombSpectraB} for \tracklet \topone{} together with the \SOPT unbiased reference. The trends of the spectral shapes are consistent between all observed particle species, showcasing a significant hardening (softening) of the \pt in the low (high) \SOPT selection, relative to the inclusive high-multiplicity event class, respectively. These trends are also well reflected in the model predictions. The PYTHIA 8.2 default Monash tune can describe the qualitative trends of the \SOPT selection. However, a large quantitative deviation from data is observed in the \pt differential production of light-flavor hadrons, in particular the non-strange hadrons. The PYTHIA 8.2 rope tune is able to describe the measured data for strange hadrons very well, but overestimates the total amount of produced non-strange hadrons. Similar observations can also be seen when contrasting Herwig 7.2 with EPOS-LHC, where EPOS-LHC overestimates the total yields, but is able to describe the \SOPT-differential interplay for the mesons quite well. These large deviations are well known from previous studies~\cite{13tevvsmultpapers}.

\begin{figure}[ht!]
    \centering
    \includegraphics[scale=0.84]{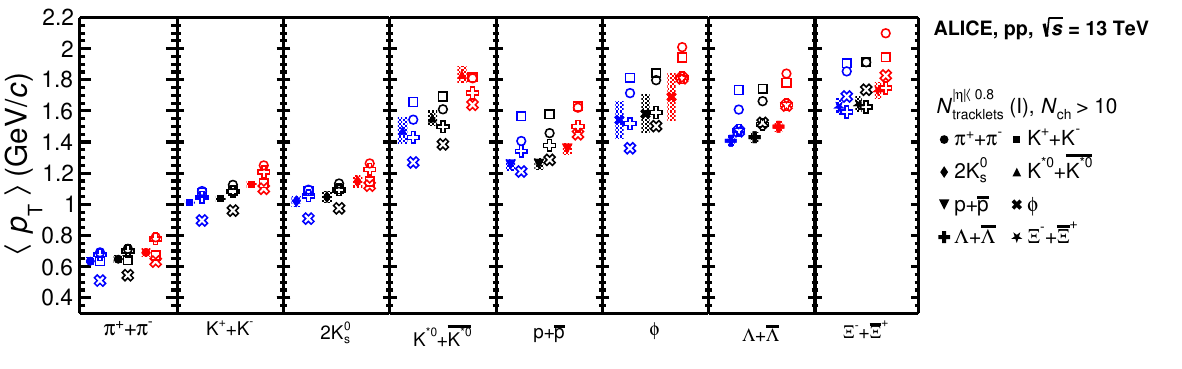}
    \includegraphics[scale=0.84]{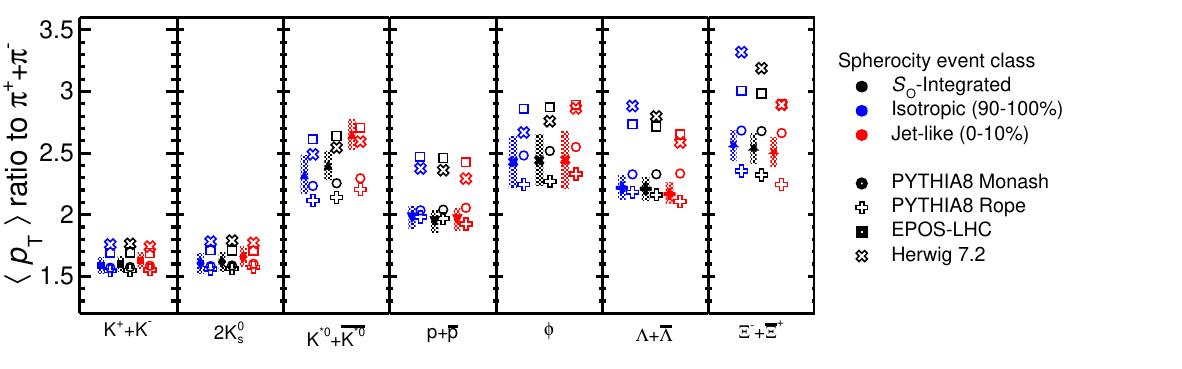}
    \hspace*{-2.45cm}\includegraphics[scale=0.84,trim={0 0 4.6cm 0},clip]{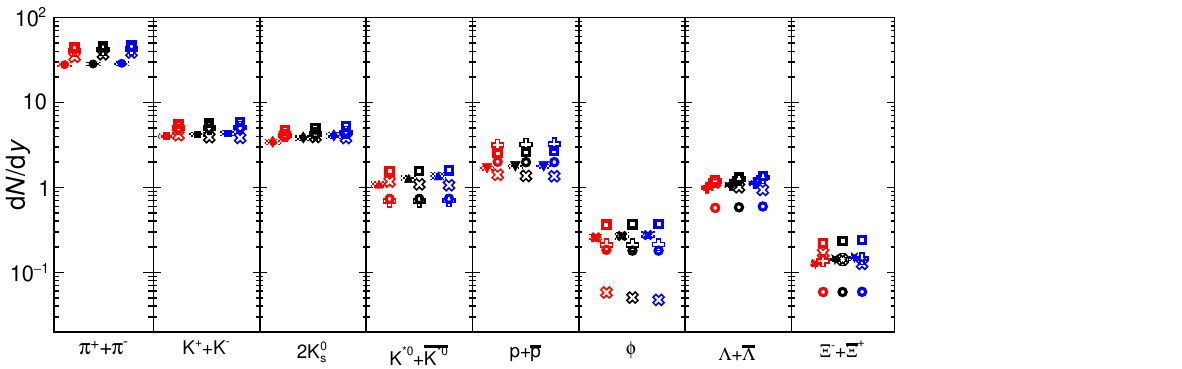}
    \hspace*{-1.3cm}\includegraphics[scale=0.84,trim={0 0 5.9cm 0},clip]{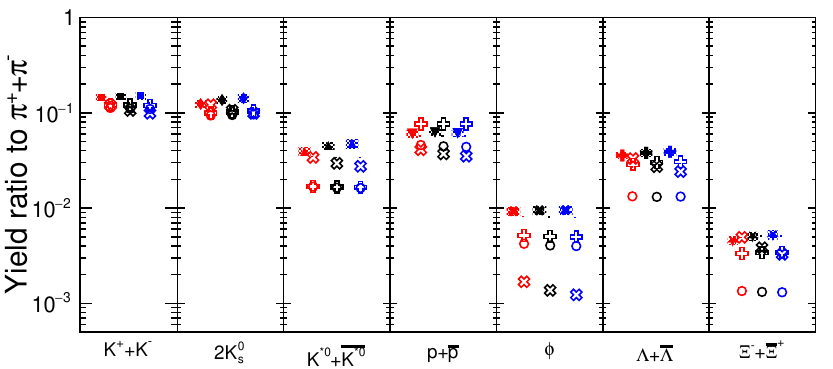}

\caption{The \mpt and \myield  as a function of particle masses obtained for the various particle species in \SOPT classes selected for  high-multiplicity events, determined by the events in the 0--1\% of \tracklet. Upper (lower) panels show the \mpt (\myield). The total systematic uncertainty is represented by the shaded regions. The measured data is compared to predictions from PYTHIA8 Monash, PYTHIA8 Rope, EPOS-LHC, and Herwig 7.2.}
    \label{Fig:Levy}
\end{figure}

 The \SOPT-differential average \pt (\mpt) and yield \myield are reported in Fig.~\ref{Fig:Levy} as a function of the extracted particle masses. The measured \mpt values confirms what is qualitatively observed in the \pt-differential spectra: there is a significant \pt-hardening in jet-like events, and this trend is consistent across all measured light-flavor particle species. Furthermore, the \mpt of the integrated \SOPT high-multiplicity events are consistent with the \mpt of the isotropic sample. This observation indicates that average high-multiplicity events and \SOPT selected isotropic events are dominated by similar underlying physics processes.
 This shows that the \SOPT-integrated event class is not the arithmetic average of the jet-like and isotropic subsamples, indicating that jet-like events are rare outliers of a much more homogeneous group of high-multiplicity events. This observation is of particular interest, and we will focus on exploring it in further detail in the following. 
 
The bottom panels of Fig.~\ref{Fig:Levy} show that the top-1\% \tracklet estimator constrains the variance in \myield between the different \SOPT classes, for all measured particles.  Hence, any observed deviations in comparisons between \tracklet 0--1\% jet-like and isotropic events are unlikely to be driven by a trivial difference in multiplicity. This is not only true for the \PI mesons as reported in Section~\ref{Sec:ResHM}, but across all measured particle masses as a function of \SOPT. 

Most of the presented models overestimate the \mpt for all particle species except \KSTAR, however, this effect can be partially explained by the fact that the models and data are compared at different midrapidity charged particle multiplicities~\cite{ALICE:2020nkc}. Moreover, the models give a reasonable description of the particle specie differences when one normalizes the \mpt relative to that of pions. Furthermore, one can note that the PYTHIA 8.2 predictions, in particular with the enabled color rope framework, overestimates the production of protons and pions, but qualitatively describes the production of most strange hadrons. Moreover, the color rope framework seems to primarily affect the baryons, as the resonance particles show only a small difference in the integrated yield between PYTHIA 8.2 color rope framework and the default PYTHIA 8.2 Monash, in particular for the \KSTAR. In contrast, EPOS-LHC describes the production rates of most particles within the uncertainties of the data, but is unable to capture the trend seen for charged kaons. Herwig 7.2 is unable to capture most of the trends observed in the measured data.

\subsection{Particle ratios for \tracklet 0--1\%}\label{Sec:ResRat}

QGP-like effects, such as collective flow and strangeness enhancement, affect hadrons of heavier species distinctly. Collective flow will boost heavier particles to larger \pt than lighter particles, and strangeness enhancement in previous measurements was found to scale with the number of strange quarks~\cite{Naturepaper}\cite{alice_newstrangeness}. Due to its low mass, abundant production and lack of strangeness, the pion constitutes a good reference for the QCD physics we are less interested in. Therefore, by studying the \pt-differential particle-to-pion ratios, we can potentially identify QGP-like features in the data.

To be able to observe the quantitative trends with the highest precision, we construct double ratio (DR), as defined in Eq.~\ref{Eq:DR}
\begin{equation}\label{Eq:DR}
\centering
    \left(\frac{{\rm d}N/{\rm d}\pt}{{\rm d}N_\PI/{\rm d}\pt}\right)_{\SOPT} \bigg/ \left(\frac{{\rm d}N/ {\rm d}\pt}{{\rm d}N_\PI/{\rm d}\pt}\right)_{\scaleto{\int}{15pt}\SOPT},
\end{equation}
where the denominator represents the \SOPT-integrated, high-multiplicity event sample. The advantage of the double ratio from an experimental standpoint is that most of the systematic uncertainties will cancel in the ratio between the \SOPT-differential and \SOPT-integrated spectra, for the same species. The double ratios therefore have the best precision in terms of systematic uncertainties of all the data presented in this section, where only the uncorrelated sources listed in Tab.~\ref{tab:systematics_pikp} --~\ref{Tab:SysLong} are applied. For the comparison with MC models, the DR means that focus can be shifted away from the large discrepancies between data and model that exists for most particle species, to test if the quantitative \textit{relative} trends are the same in data and MC. Therefore, we advocate that these comparisons are the ones that are most sensitive to the physics of \SOPT selection.

\begin{figure}[htbp!]
    \centering
	\includegraphics[scale=0.8]{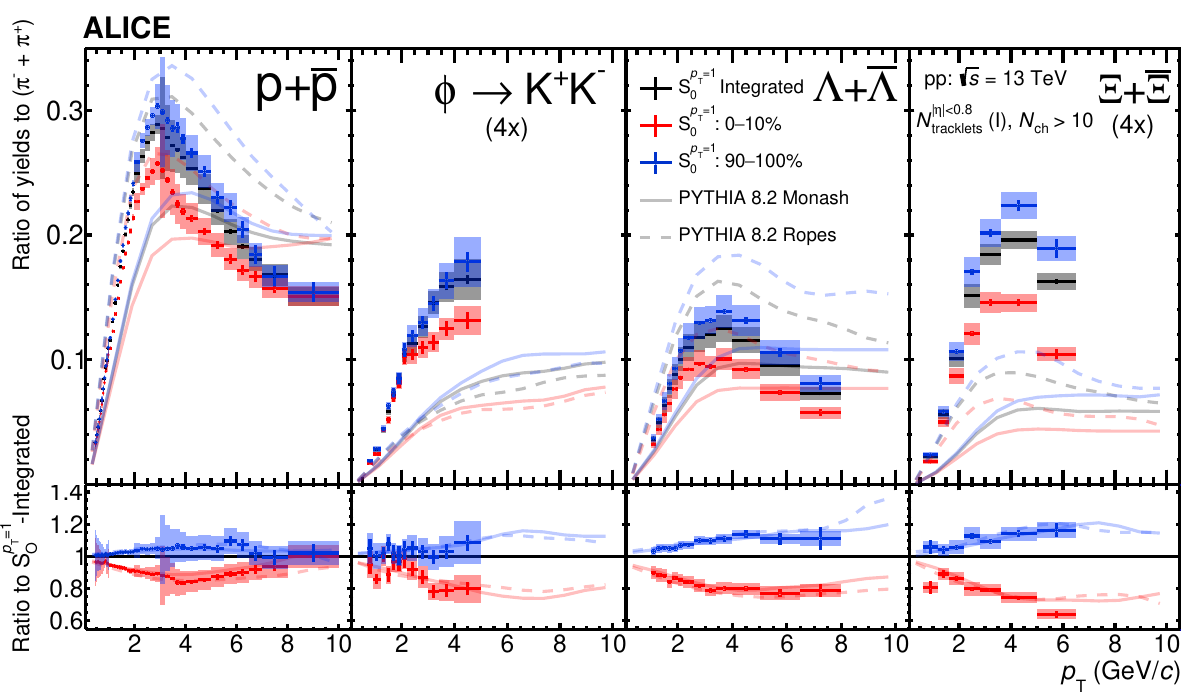}
	\newline
     	 \includegraphics[scale=0.8]{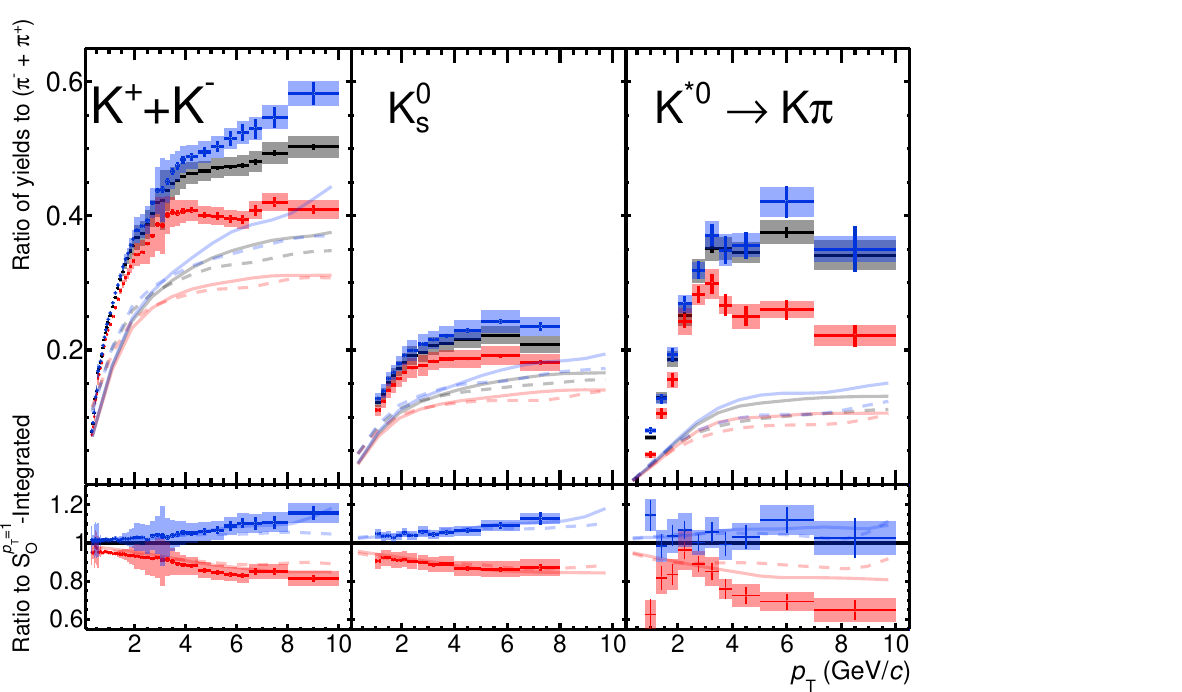}
  \caption{Top panels show hadron-to-\PI ratios for 0--10\% \SOPT classes selected for the 0--1\% \tracklet multiplicity events. Bottom panels present the hadron-to-\PI double ratios of \SOPT classes relative to \SOPT integrated high-multiplicity events. Statistical and systematic uncertainties are shown by  bars and boxes, respectively. Experimental results are compared with predictions from PYTHIA 8.2 Monash and Ropes.}   \label{Fig:CombRatioPy}
\end{figure}

\begin{figure}[htbp!]
    \centering
	\includegraphics[scale=0.8]{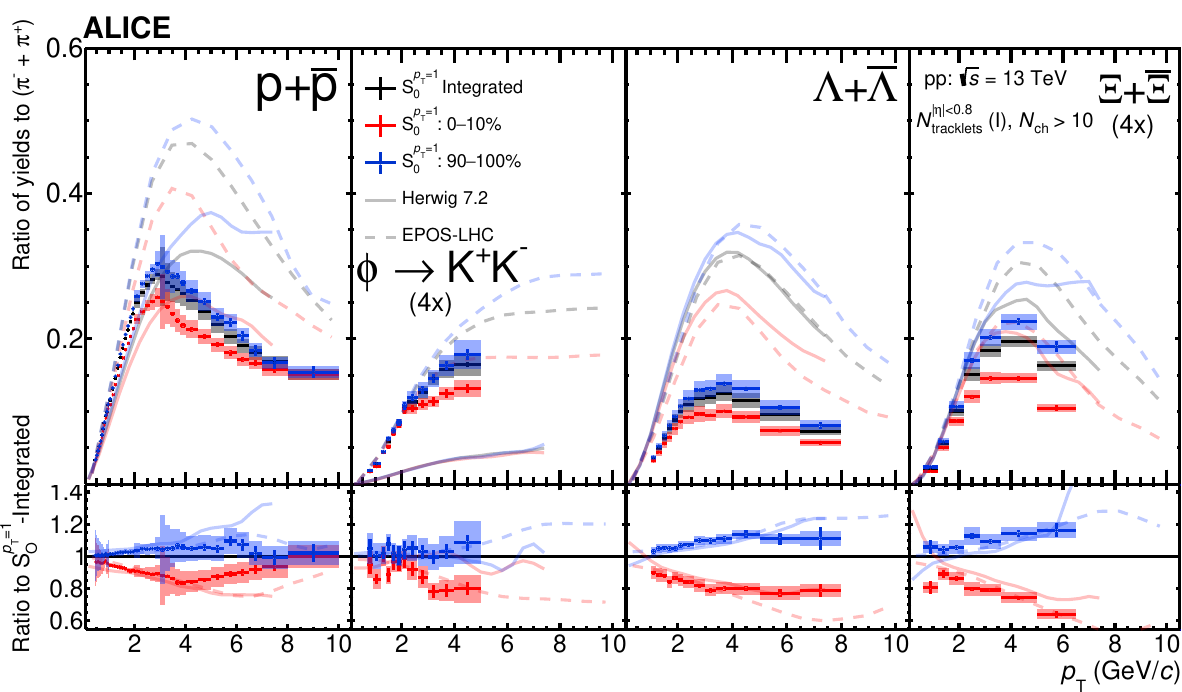}
	\newline
	 \includegraphics[scale=0.8]{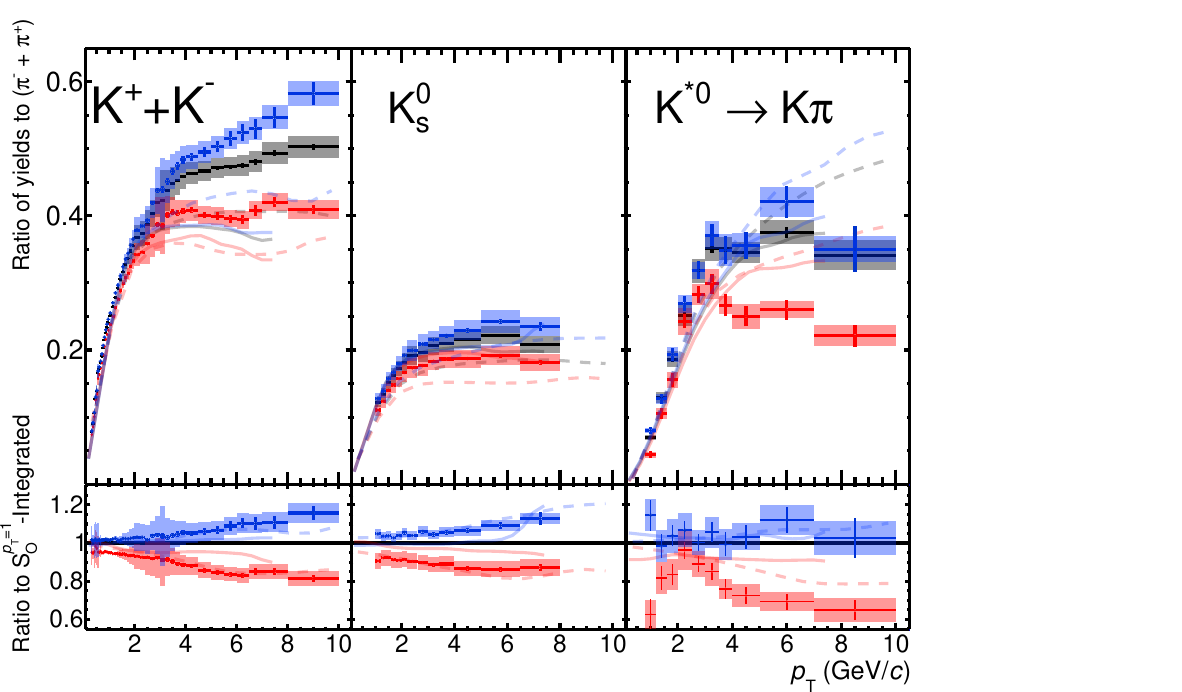}
  \caption{Top panels show hadron-to-\PI ratios for 0--10\% \SOPT classes selected for the 0--1\% \tracklet multiplicity events. Figure~\ref{Fig:CombRatioPy} and Fig.~\ref{Fig:CombRatioHer} both contain the same experimental data, but the vertical ranges are modified to accommodate the model predictions. Bottom panels present the hadron-to-\PI double ratios of \SOPT classes relative to \SOPT integrated high-multiplicity events. Statistical and systematic uncertainties are shown by  bars and boxes, respectively. Experimental results are compared with predictions from Herwig 7.2 and EPOS-LHC. The fluctuations present in the Herwig 7.2 predictions are due to statistical limitations.}   \label{Fig:CombRatioHer}
\end{figure}

The \pt-differential relative yield of identified light-flavor hadrons to \PI mesons, as well as the DR, for the \tracklet \topone~multiplicity events are presented for the 10\% \SOPT percentiles in Fig.~\ref{Fig:CombRatioPy} and Fig.~\ref{Fig:CombRatioHer}, compared with PYTHIA 8.2 predictions in the former, and EPOS-LHC and Herwig 7.2 predictions in the latter. 

The DR is presented in the lower panels, showcasing a clear enhancement of strangeness yield in isotropic topologies, and a suppression in jet-like topologies. Remarkably, the DR for all presented particle species decrease significantly for jet-like events. This can be interpreted as most particle ratios being shifted towards higher \pt in jet-like events, i.e. as the \pt of the partonic sub-collisions increase, the same hadrons will be produced with larger \pt. The hadron-to-\PI ratios exhibit a small enhancement of strange-hadron ratios in isotropic events, and a pronounced suppression in jet-like events. This suggests that strange particle production is favored in isotropic topologies. 

In Fig.~\ref{Fig:CombRatioPy}, the default Monash tune of PYTHIA 8.2 quantitatively predicts the interplay between the \SOPT-integrated high-multiplicity events and the \SOPT-differential event classes, but shows large quantitative deviation for most of the measured data. The rope tune does slightly better, in particular for baryons, but the deviation is for both tunes particularly large for the two resonance particles \PHI and \KSTAR. In contrast, neither Herwig 7.2 or EPOS-LHC are able to quantitatively describe the \pt-differential particle-to-\PI ratios, as seen in Fig.~\ref{Fig:CombRatioHer}.

The PYTHIA 8.2 Monash tune and the rope hadronization framework are both able to qualitatively capture the trend between isotropic and jet-like topologies in the DR for all presented light-flavor particles. Remarkably, there is no difference between the PYTHIA 8.2 Monash and the PYTHIA 8.2 Rope curves. Even though the production rates of light-flavor hadrons are different, both variations are able to describe the trends presented for the \SOPT characterized events. EPOS-LHC and Herwig 7.2 are able to qualitatively describe some trends, but are not able to describe the full evolution, mainly towards larger \pt.

\begin{figure}[htbp!]
    \centering
	\includegraphics[scale=0.8]{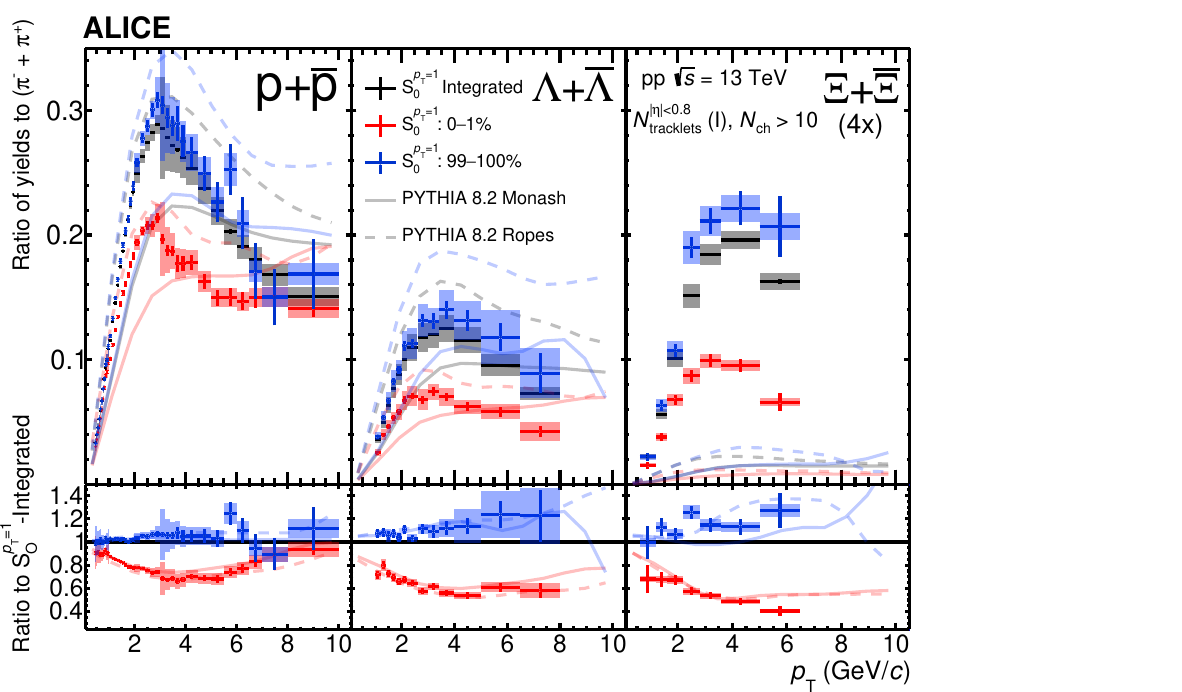}
	\newline
	\includegraphics[scale=0.8]{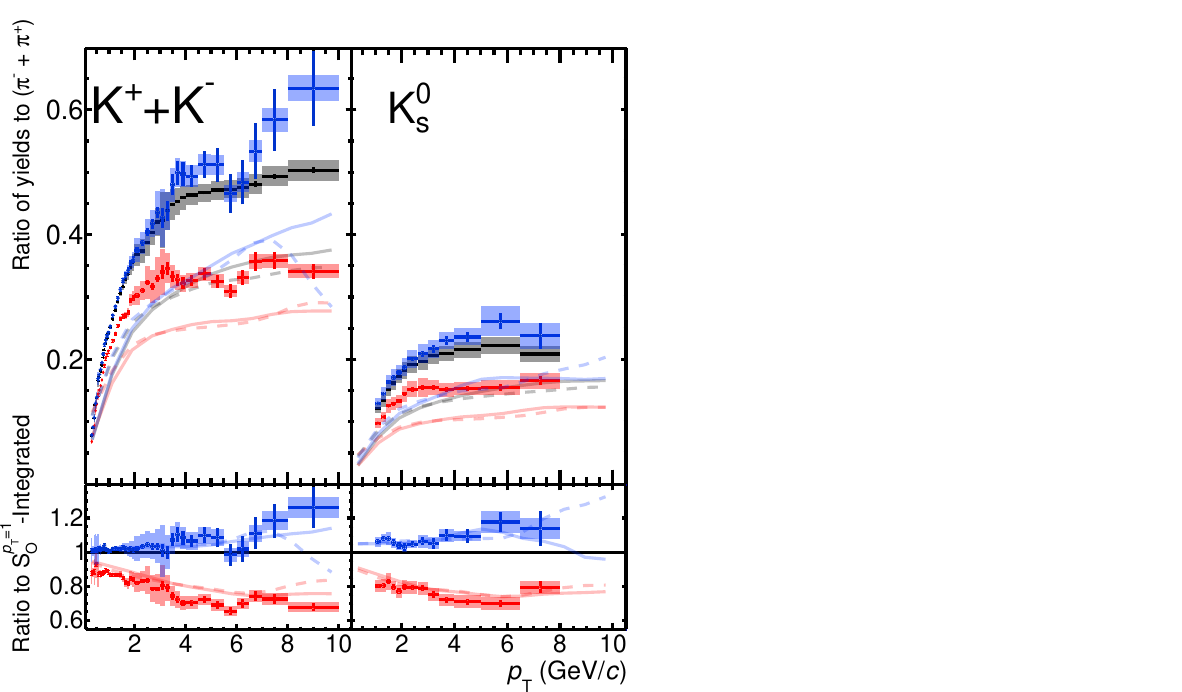}
  \caption{Top panels show hadron-to-\PI ratios for 0--1\% \SOPT classes selected for the 0--1\% \tracklet multiplicity events. Bottom panels present the hadron-to-\PI double ratios of \SOPT classes relative to \SOPT integrated high-multiplicity events. Statistical and systematic uncertainties are shown by  bars and boxes, respectively. Experimental results are compared with predictions from PYTHIA 8.2 Monash and Ropes.}
   \label{Fig:CombRatioPyEx}
\end{figure}

\begin{figure}[htbp!]
    \centering
	\includegraphics[scale=0.8]{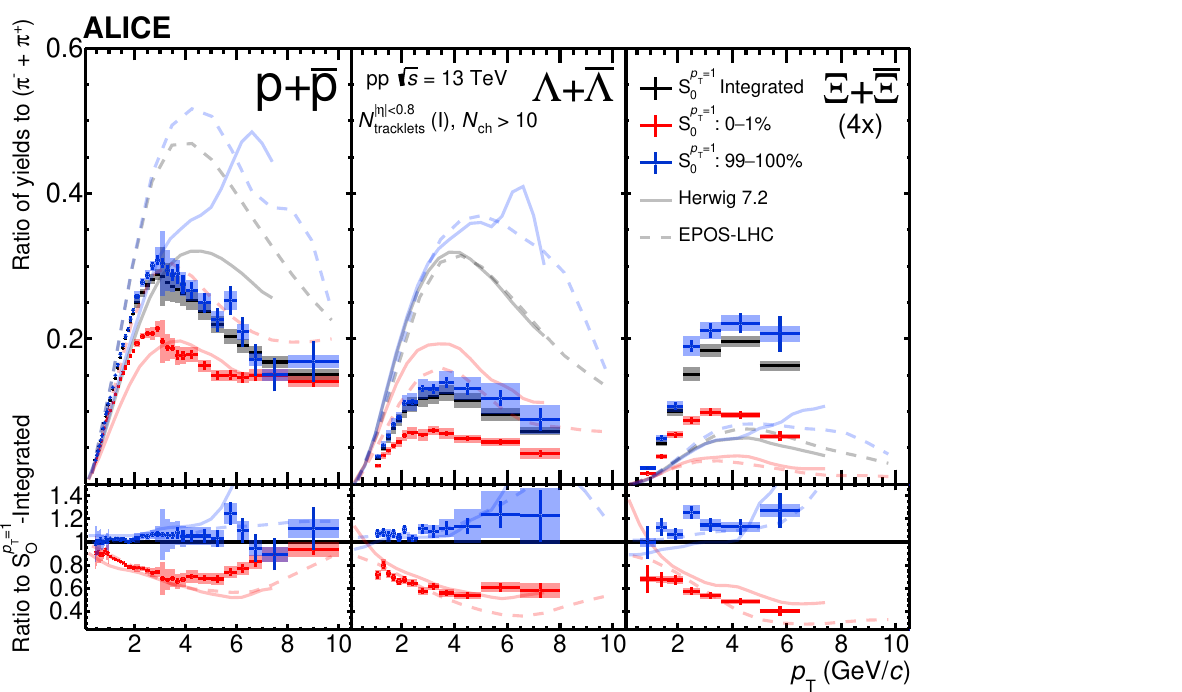}
	\newline
	\includegraphics[scale=0.8]{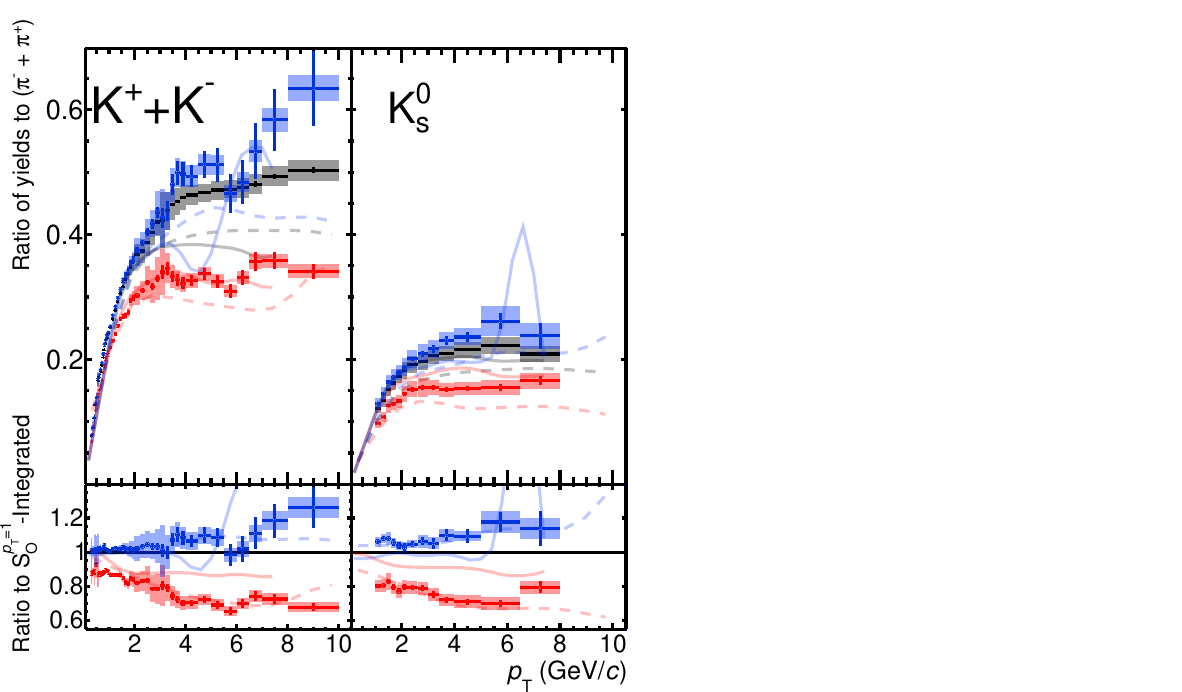}
  \caption{Top panels show hadron-to-\PI ratios for 0--1\% \SOPT classes selected for the 0--1\% \tracklet multiplicity events. Figure~\ref{Fig:CombRatioPyEx} and Fig.~\ref{Fig:CombRatioHerEx} both contain the same experimental data, but the vertical ranges are modified to accommodate the model predictions. Bottom panels present the hadron-to-\PI double ratios of \SOPT classes relative to \SOPT integrated high-multiplicity events. Statistical and systematic uncertainties are shown by  bars and boxes, respectively. Experimental results are compared with predictions from Herwig 7.2 and EPOS-LHC. The large fluctuations present in the Herwig 7.2 predictions are due to statistical limitations.}
   \label{Fig:CombRatioHerEx}
\end{figure}

By utilizing a narrower event selection, we present the \pt-differential relative yield of identified light-flavor hadrons to \PI mesons, as well as the DR, for the \tracklet \topone ~multiplicity with 1\% \SOPT percentiles in Fig.~\ref{Fig:CombRatioPyEx} and Fig.~\ref{Fig:CombRatioHerEx}. The resonances are excluded from this measurement due to the low number of events in the narrow \SOPT and multiplicity selection. This is an experimental challenge that particularly affects both \PHI and \KSTAR, as the signals are contaminated by large combinatorial backgrounds. Likewise, the large fluctuations in the MC predictions present in Fig.~\ref{Fig:CombRatioPyEx} and Fig.~\ref{Fig:CombRatioHerEx} are driven by an insufficient amount of events in the double-differential selection, compounded by the low production rate of the rare particle yields in the various models.

For all measured particles, the effect of the enhancement and suppression of light-flavor particles relative to pions is stronger when approaching more extreme topologies, in particular the suppression of yield in jet-like events. This is clearly seen both through the single particle-to-\PI ratio, and in the DR, contrasting Fig.~\ref{Fig:CombRatioPy} with Fig.~\ref{Fig:CombRatioPyEx}. In particular, one can note a large suppression of strange hadrons across the entire measured \pt range for events with jet-like topologies. This novel feature suggests that the abundance of strange hadrons in high-multiplicity events are produced in events that are associated to soft physics in terms of the azimuthal topology. This observation also implies that there is a significant amount of high-multiplicity events that reflect the same rates of strangeness production found in low-multiplicity events. ALICE has previously published studies of differential \LA/\kzero production in jets relative to the underlying event (UE), where it was found that the ratio in the jet was far below that of the ratio in the UE~\cite{alice_UE}. This is qualitatively similar to what we observe for the most extreme jet-like events in this study. One could therefore understand the results obtained here as a generalization to jet-dominated events.

The $\rm p/\pi$ peak present for the 99--100\% most isotropic events, as well as in the average 0--1\% high-multiplicity events, is significantly suppressed in 0--1\% jet-like event sample. In conjunction with strangeness suppression, this hints towards a decrease of QGP-like effects in events with extremely jetlike topologies. Furthermore, the discrepancy between the $\rm p/\pi$ and $\rm K/\pi$ ratios at high-\pt is interesting to note, where the isotropic/jet-like ratios meet for the $\rm p/\pi$ while diverging for the $\rm K/\pi$. The underlying mechanism of this effect is currently not fully understood.

In Fig.~\ref{Fig:CombRatioPyEx}, the two PYTHIA 8.2 predictions are qualitatively able to describe some of the particle-to-\PI ratios for the 1\% \SOPT percentiles, but remarkably underestimate the \pt-differential production of \XI baryons, as well as the isotropic production of both charged and neutral kaons. However, PYTHIA 8.2 is still able to qualitatively describe the interplay between high-multiplicity events, isotropic and jet-like topologies in the DR. Similar to what was observed in Fig.~\ref{Fig:CombRatioHer} for the broader \SOPT selection, EPOS-LHC and Herwig 7.2 are unable to accurately capture the interplay of the DR towards larger \pt. EPOS-LHC and Herwig 7.2 are both able to qualitatively describe most of the observed trends for the charged and neutral kaons in Fig.~\ref{Fig:CombRatioHerEx}, with a slight underestimation of the absolute production rates. However, both model predictions are unable to describe the observed trends for the presented baryons. 

\begin{figure}[thbp!]
    \centering
    \hspace{-1.2cm}
    \includegraphics[scale=0.85]{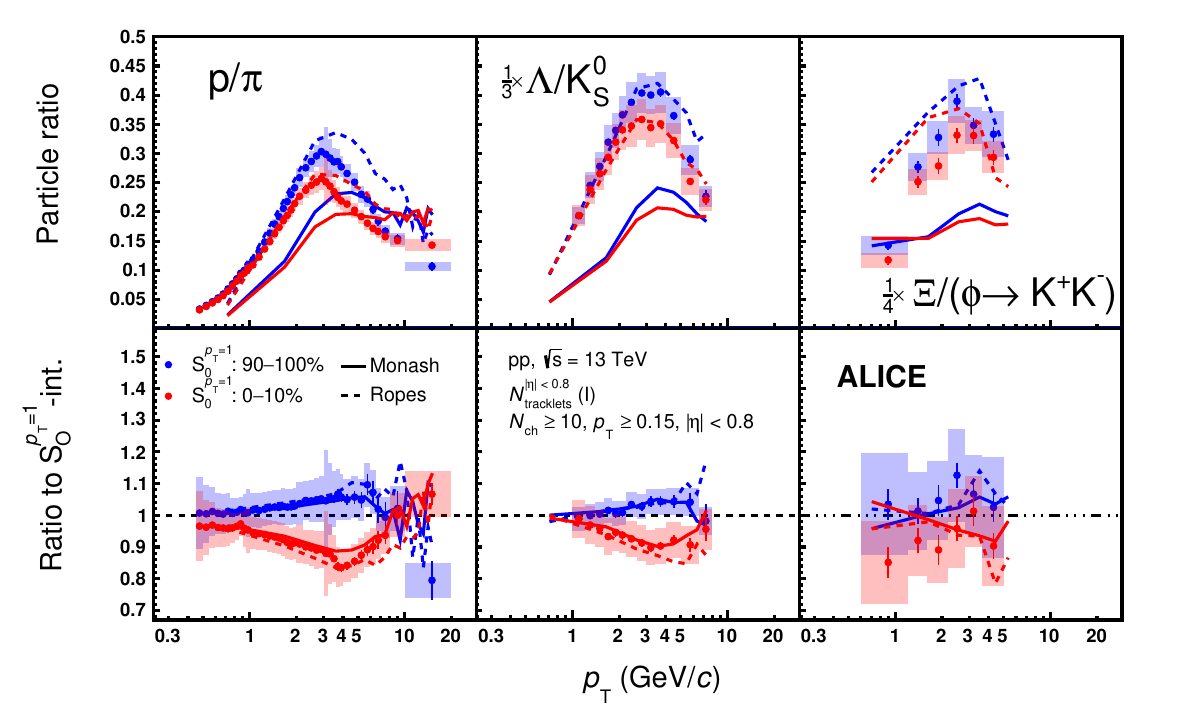}
    \caption{p/\PI,  \LA/$K$ and \XI/\PHI ratios for different \SOPT classes  are obtained for $0-1\%$ events measured by \tracklet.
    Lower panels show the ratio to \SOPT-integrated event selection.
    Statistical and total systematic uncertainties are shown by bars and boxes, respectively. The curves represent different model predictions of the same measurement. Note that for the p/\PI, the same data points are presented in~\ref{Fig:CombRatioPy}, with an extended \pt range now covering 10--20 \gev. }
    \label{Fig:Flow}
\end{figure}

In Fig.~\ref{Fig:Flow}, baryon-to-meson ratios are presented for p/\PI, \LA/\kzero and \XI/\PHI. These ratios are known to be interesting observables, able to highlight features of radial flow and possible recombination~\cite{ALICE:2019hno}. While the \XI/\PHI ratio is not usually associated to measurements of radial flow, phenomenological models have different views of the effective net-strangeness of the \PHI meson. In Lund-string-like models, such as PYTHIA 8.2, the \PHI meson is produced from the fragmentation of $\textrm{s}\bar{\textrm{s}}$ pairs, making it effectively double strange. On the other hand, statistical thermal models typically treats the \PHI meson as having no strangeness, where the production instead is driven by the hadron mass.

The overall trends among the three ratios are qualitatively similar, even as the strangeness content increases from p ($|S|=0$) to \XI ($|S|=2$). Furthermore, one can apply a traditional hypothesis of radial flow in larger collision systems, i.e., that a radial expansion of the system boosts heavier baryons out to high \pt, resulting in a depletion of baryons at low-\pt. In this context, Fig.~\ref{Fig:Flow} highlights an abundance (suppression) of isotropic (jet-like) protons at intermediate \pt in the $p/\PI$ ratio, without the depletion (enhancement) of isotropic (jet-like) protons at low \pt. The origin for the difference in the low-\pt behavior is still unclear, but we suspect that there is an interplay between soft radial flow and the hard suppression in the ratios to pions, observed in the jet-like events through Figs.~\ref{Fig:CombRatioPy}--~\ref{Fig:CombRatioHerEx}. One should keep in mind that the relative systematic uncertainties are smaller than the total systematic uncertainties reported in Fig.~\ref{Fig:Flow}, c.f., Sec.~\ref{Sec:Sys} for further discussion.

\begin{figure}[htbp!]
    \centering
    \includegraphics[scale=0.396]{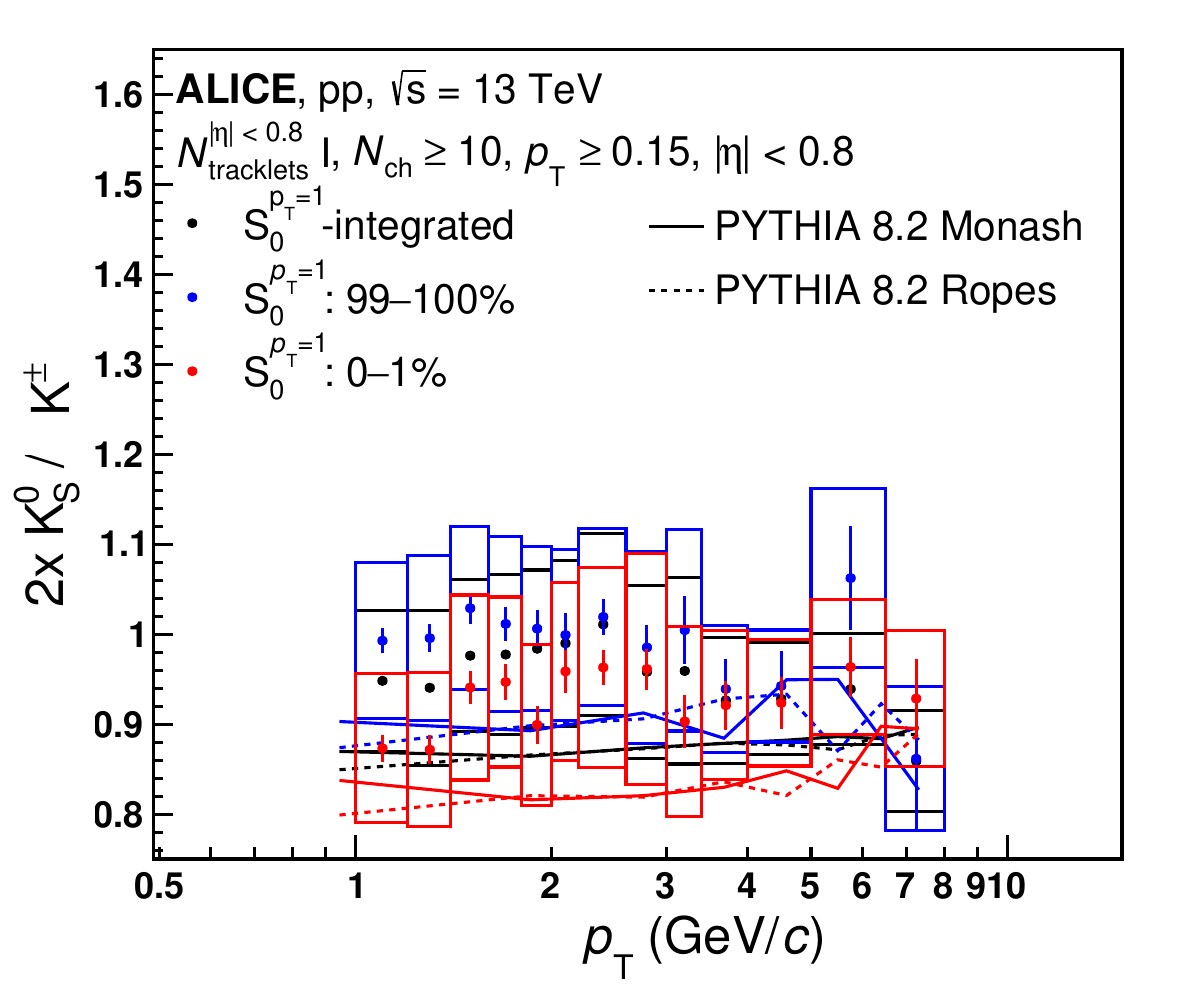}
    \includegraphics[scale=0.396]{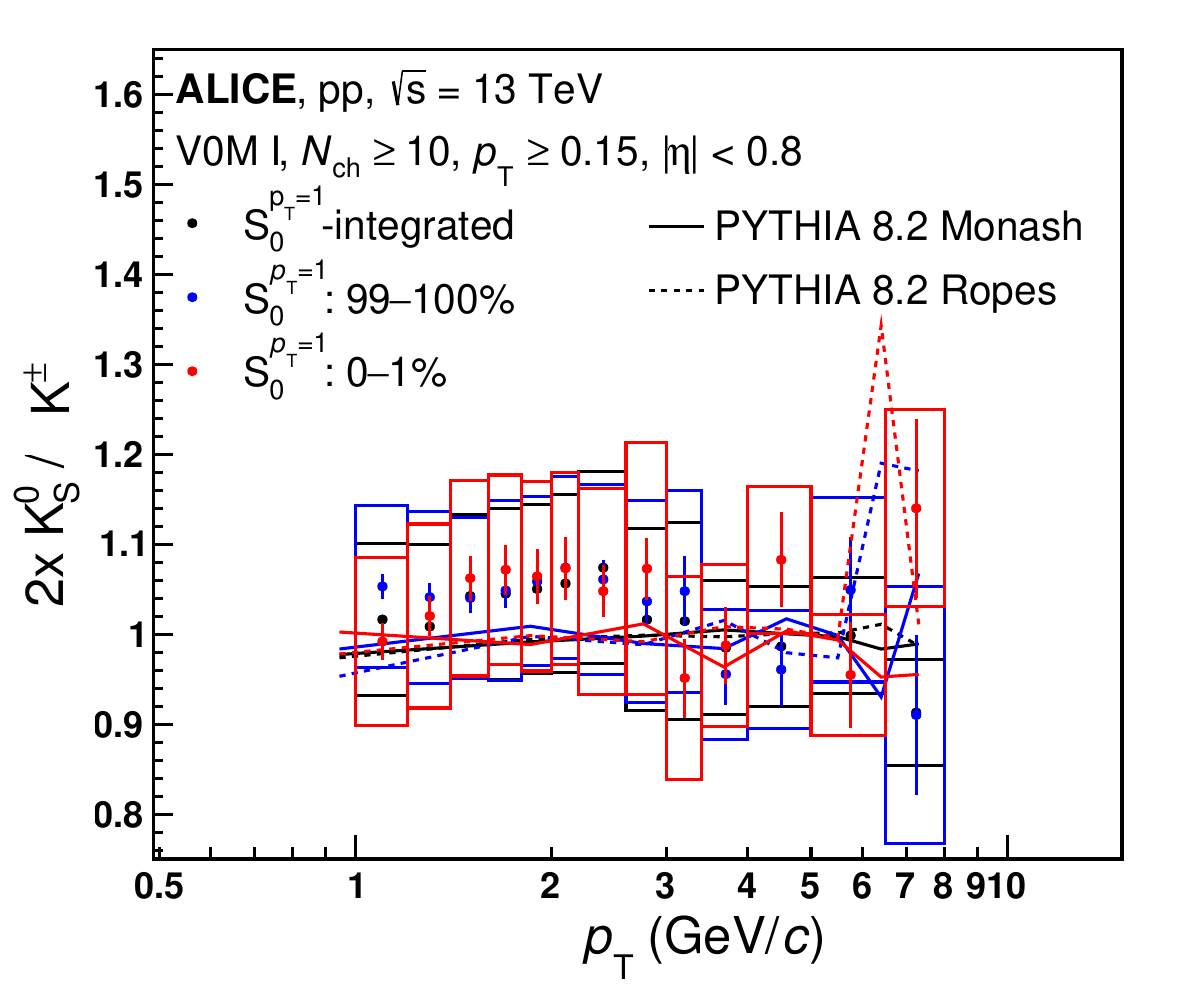}
    \caption{The neutral-to-charged \kzero/K ratios as a function of different multiplicity estimators and \SOPT. Statistical and total systematic uncertainties are shown by bars and boxes, respectively. The curves represent PYTHIA 8.2 model predictions of the same measurement. }
    \label{Fig:KtoK}
\end{figure}

Similar trends are observed in the \LA/\kzero ratio, although systematic uncertainties in the lower panel does not allow for a clear conclusion on the interplay between jet-like and isotropic events. The \XI/\PHI ratio suggests that there is a constant enhancement of \XI baryons relative to produced \PHI mesons, within systematic uncertainties.  
The rope hadronization framework in PYTHIA 8.2 predicts both the single and double ratios reasonably well. It is remarkable that even though PYTHIA 8.2 Monash fails to predict the single ratios, both PYTHIA 8.2 Monash and Rope predictions show no significant deviations for the double ratios.
The \kzero/K ratios for the different multiplicity estimators are presented in Fig.~\ref{Fig:KtoK}. The ratios highlight a consistency with unity, and showcase no significant \SOPT dependence. These results verify that the modified \SOPT estimator is robust through a data-driven approach, complementing the studies discussed in Sec.~\ref{Sec:SOPT}. 

\subsection{Integrated yields as a function of \SOPT}\label{Sec:IntYield}

\begin{figure}[h!]
    \vspace{0.1cm}
    \centering
    \includegraphics[scale=0.75]{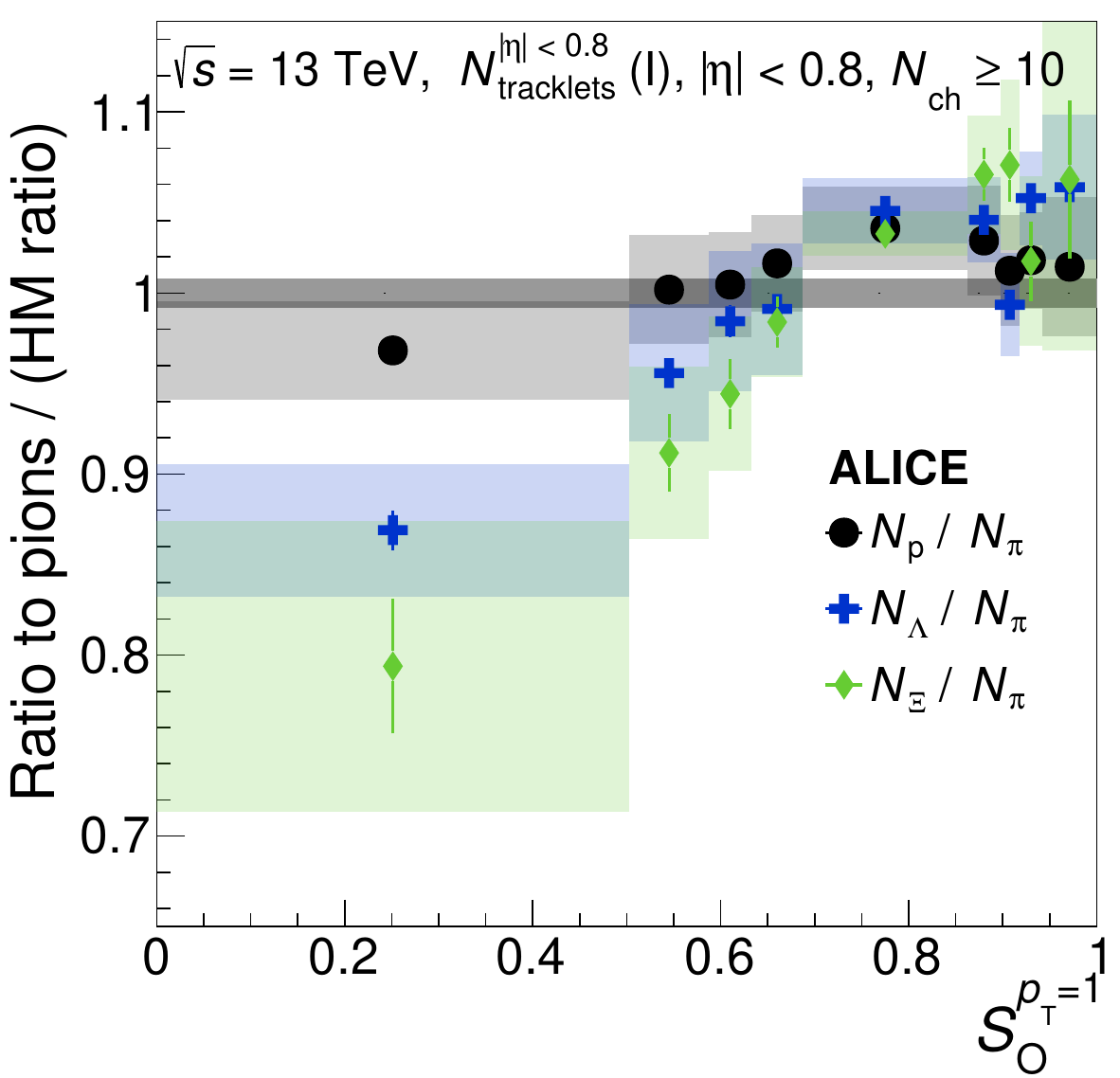}
    \caption{The double ratios of integrated yields as a function of \SOPT for the spectra of top-1\% \tracklet. The yield is estimated by extrapolating the spectra over the full \pt range. Statistical and systematic uncertainties are shown by bars and boxes, respectively. The grey band around unity represents the systematic uncertainty of the pion measurement.}
    \label{Fig:Extrapolation}
\end{figure}

The integrated double ratio over the full \pt range is presented in Fig.~\ref{Fig:Extrapolation}, for p,~\LA, and \XI. The fully integrated yields are obtained by extrapolating the measured \pt spectra for each particle species. Therefore, the systematic uncertainties shown in Fig.~\ref{Fig:Extrapolation} also account for the added uncertainty due to the extrapolation procedure. This added uncertainty is particle species dependent given the different measured \pt ranges for each particle species, particularly affecting the \LA yields, which, as shown in Tab.~\ref{Tab:S0bias}, can only be measured down to 1.0 \gev after \SOPT selection. 

The results demonstrate that the strange-hadron yield increases as a function of \SOPT, with indications of an ordering with strangeness content. In previous ALICE publications it was observed that in pp collisions at \s$= 13$ TeV the charge particle density, \dndeta, is a driving quantity for the enhancement of strange hadrons~\cite{Naturepaper}. For the results presented in Fig.~\ref{Fig:Extrapolation}, the \dndeta at midrapidity is restricted (see Fig.~\ref{Fig:CL1vsV0M}), which allows one to directly test if \SOPT is sensitive to strangeness enhancement. We find that the strangeness production is suppressed in events with jet-like topologies, and slightly enhanced in softer, isotropic event topologies.

\begin{figure}[th!]
    \vspace{0.1cm}
    \centering
    \includegraphics[scale=0.75]{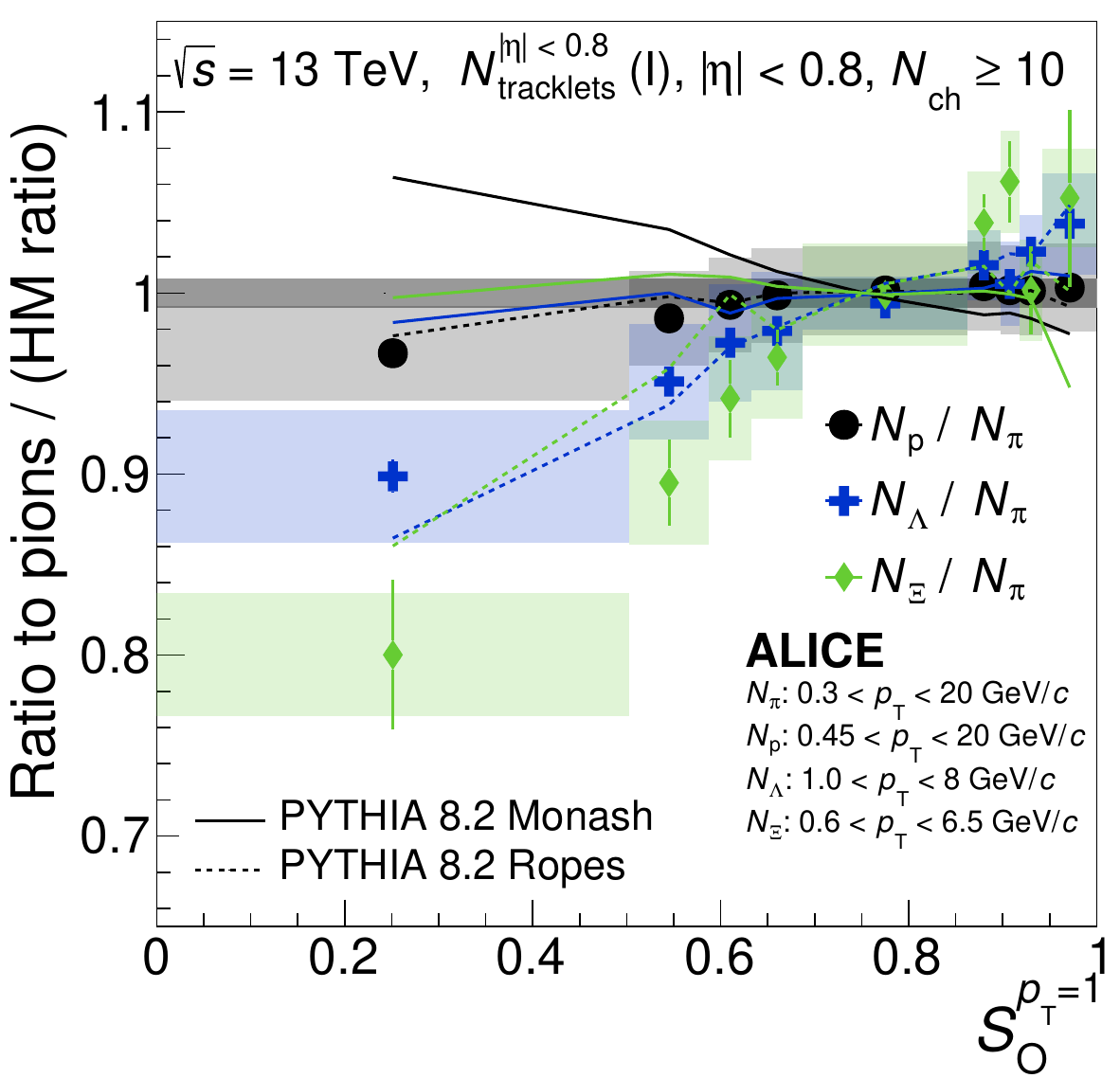}
    \caption{The double ratios of integrated yield as a function of \SOPT are represented in the top-1\% of \tracklet. The yields are integrated in the measured \pt ranges for each particle species. Statistical and systematic uncertainties are shown by bars and boxes, respectively. The curves represent different model predictions of the same measurement. The grey band around unity represents the systematic uncertainty of the pion measurement.}
    \label{Fig:CL1_1_Int_py}
\end{figure}

\begin{figure}[t!]
    \vspace{0.1cm}
    \centering
    \includegraphics[scale=0.75]{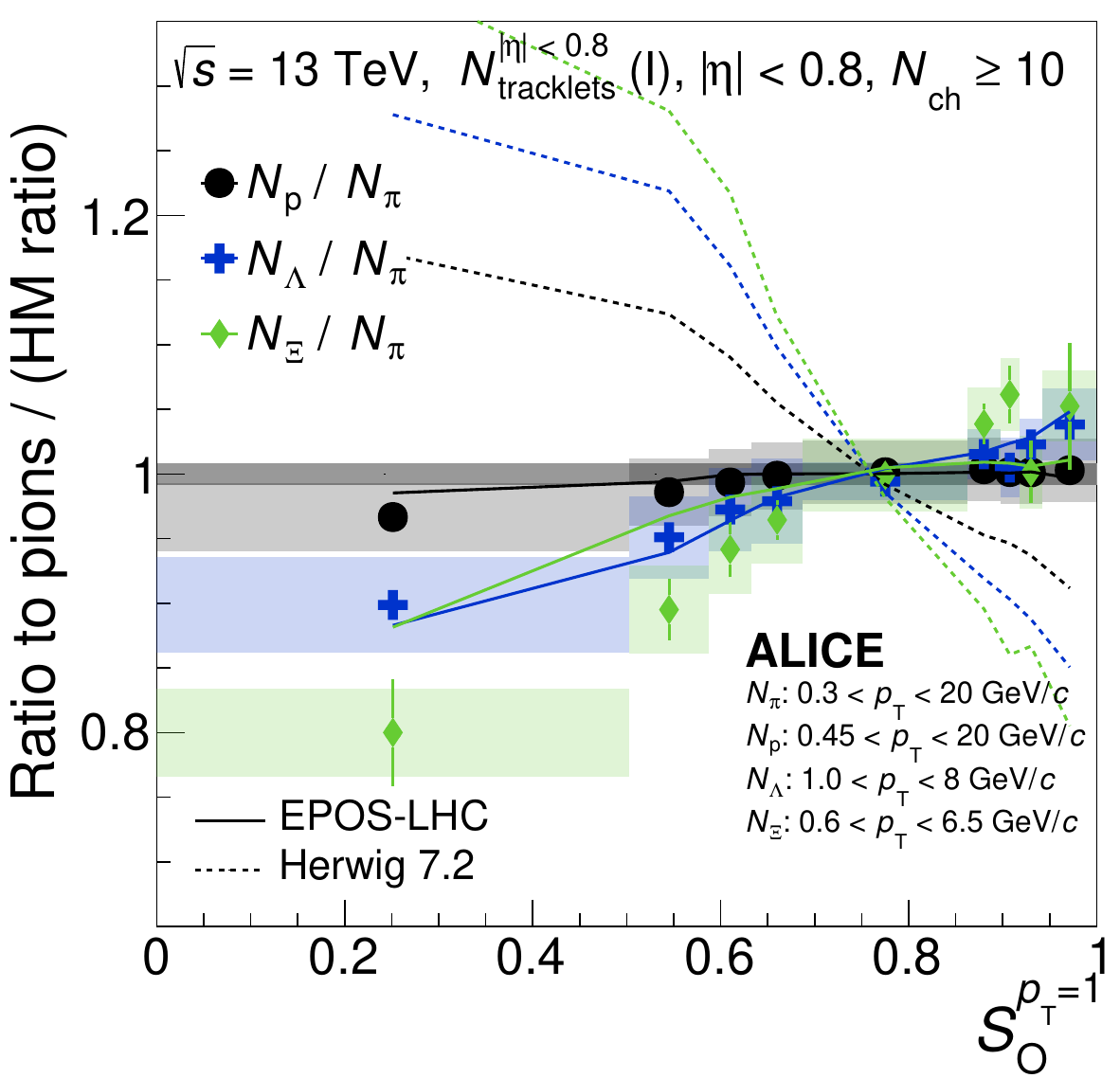}
    \caption{The double ratios of integrated yield as a function of \SOPT are represented in the top-1\% of \tracklet. The yields are integrated in measured \pt ranges for each particle species. Statistical and systematic uncertainties are shown by bars and boxes, respectively. Figure~\ref{Fig:CL1_1_Int_py} and Fig.~\ref{Fig:CL1_1_Int_hw} both contain the same experimental data, but the vertical ranges are modified to accommodate the model predictions. The curves represent different model predictions of the same measurement. The grey band around unity represents the systematic uncertainty of the pion measurement.}
    \label{Fig:CL1_1_Int_hw}
\end{figure}

In order to make the most precise comparison between the measured data and model predictions, the integrated double ratios are presented as a function of \SOPT for the measured \pt ranges in Fig.~\ref{Fig:CL1_1_Int_py} and Fig.~\ref{Fig:CL1_1_Int_hw}, for p,~\LA and \XI. The two figures contain the same data points, but use different ordinate ranges in the ratio to accommodate the MC-generator predictions. The yields for the measured \pt ranges are estimated by counting the bin entries in each \pt spectra, without the use of the Levy--Tsallis extrapolation. Therefore, the systematic uncertainties for the integrated yields utilizing the measured \pt ranges are significantly reduced compared to the extrapolated yields presented in Fig.~\ref{Fig:Extrapolation}. Moreover, while the fraction of yields gained from the extrapolation is species dependent, it was tested that the relative increase of yields is consistent across all spherocity classes and the high-multiplicity reference. As the integrated yields presented in Figs.~\ref{Fig:Extrapolation} --~\ref{Fig:CL1_1_Int_hw} are self-normalized, the relative yields obtained from the extrapolation therefore largely cancels. As such, one obtains the same physics conclusions by utilizing either extrapolated or measured \pt ranges, which is reflected in the comparison between Fig.~\ref{Fig:Extrapolation} and Fig.~\ref{Fig:CL1_1_Int_py}.

Figures ~\ref{Fig:CL1_1_Int_py} and ~\ref{Fig:CL1_1_Int_hw} highlight that the relative decrease of \XI production in the most jet-like events is of order 20\%. We estimate, based on Ref.~\cite{Naturepaper}, that to obtain a similar effect driven solely by multiplicity, one would have to decrease the multiplicity by approximately 60 to 70\%. Given that the difference in multiplicity between the spherocity event classes is roughly 10\%, this indicates a substantial lifting of the strangeness suppression due to the event topology selection.

This novel feature can help to further elucidate the underlying mechanism(s) that drives the strangeness enhancement. Remarkably, these findings suggest that charged particle production is not driven by a single source, but instead driven by several sources with varying strangeness-to-\PI production rates, with the jet-like events showcasing a level of strangeness production usually found at lower multiplicities. In combination with the \pt-differential ratios from Figs.~\ref{Fig:CombRatioPy} --~\ref{Fig:CombRatioHerEx}, as well as the baryon-to-meson ratios in Fig.~\ref{Fig:Flow}, one can characterize jet-like events as exhibiting a large decrease of relative production at intermediate \pt, with an overall high degree of strangeness suppression in the total yields. Isotropic events can be characterized completely opposite to jet-like events, containing a boost of particles at intermediate \pt, with enhanced strangeness production in the \pt-integrated yields. These findings suggest that one is able to control the degree of QGP-like effects in small systems by categorizing events based on the azimuthal topology. Furthermore, it demonstrates that \SOPT-integrated high-multiplicity events are dominated by soft processes, and provides an important input to understanding the ALICE observation of universal scaling of strangeness enhancement with multiplicity~\cite{Naturepaper}.

The PYTHIA 8.2 Rope hadronization framework and EPOS-LHC models, which incorporate two-component phenomenologies, are able to predict the qualitative trend of enhancement and suppression of strange particle production as a function of \SOPT, albeit with a different mass-ordering for \LA and \XI. In contrast, both the PYTHIA8 Monash and Herwig 7.2 predictions are unable to describe the reported experimental observation. Surprisingly, Herwig 7.2 predicts the opposite trend; enhancement of all three baryons in jet-like events and a suppression in isotropic events. If this is a generic feature of the new strangeness-enhancement process introduced in Herwig 7.2\cite{Herwig}, then the results presented in this article appear to rule out this mechanism. Furthermore, it might seem counterintuitive that there can be large differences between model predictions in Fig.~\ref{Fig:CL1_1_Int_py} and Fig.~\ref{Fig:CL1_1_Int_hw}, while those same models had similar trends for the \pt-differential double-ratios presented in Figs.~\ref{Fig:CombRatioPy}--\ref{Fig:CombRatioHerEx}. However, it is important to note that the integrated double-ratios are weighted by the relative yields in each \pt interval. Therefore, the \mpt of each particle species play a major part in the integrated particle yields, see Fig.~\ref{Fig:Levy}.

The comparison between model and data suggests that models without a universal hadronization scheme, either through the core--corona in EPOS-LHC, or through the color ropes in PYTHIA 8.2, are able to qualitatively reproduce the observed trends. In contrast, the default PYTHIA 8.2 Monash variation, based on the concept of jet universality, is unable to capture the feature presented in the data. 

\subsection{\SOPT results with a broadened multiplicity range}\label{Sec:ResExpMult}

In Sec.~\ref{Sec:ResRat}, it was shown that the 0--1\% topology selection produced the largest effects, seen in Fig.~\ref{Fig:CombRatioPyEx} and Fig.~\ref{Fig:CombRatioHerEx}. However, the resonance particles had to be excluded in those measurements due to statistical limitations related to the signal extraction.
Therefore, in this Section, we report on \SOPT measurements with a broader multiplicity selection for the different \SOPT classes. We implement the broadening of the multiplicity estimation in two different ways:

\begin{enumerate}
    \item First, we broaden the midrapidity multiplicity estimation to 0--10\%, while simultaneously constricting the \SOPT event selection to the top 1\% quantile. This allows for a broader multiplicity range, while retaining the extreme topology selection, gaining a factor 10 in the number of events.
    
    \item Secondly, we investigate top-1\% multiplicity at forward rapidity in 0--10\% \SOPT quantiles. This is  to study the impact of a broader $\langle \rm{d}\it{N}_{\pi}/\rm{d}\it{y} \rangle$, and to compare midrapidity to forward rapidity multiplicity estimation between roughly similar \dndeta, as is seen in Fig.~\ref{Fig:CL1vsV0M}.
\end{enumerate}

\begin{figure}[htbp!]
    \centering
        \hspace*{-1cm}
	\includegraphics[scale=0.87]{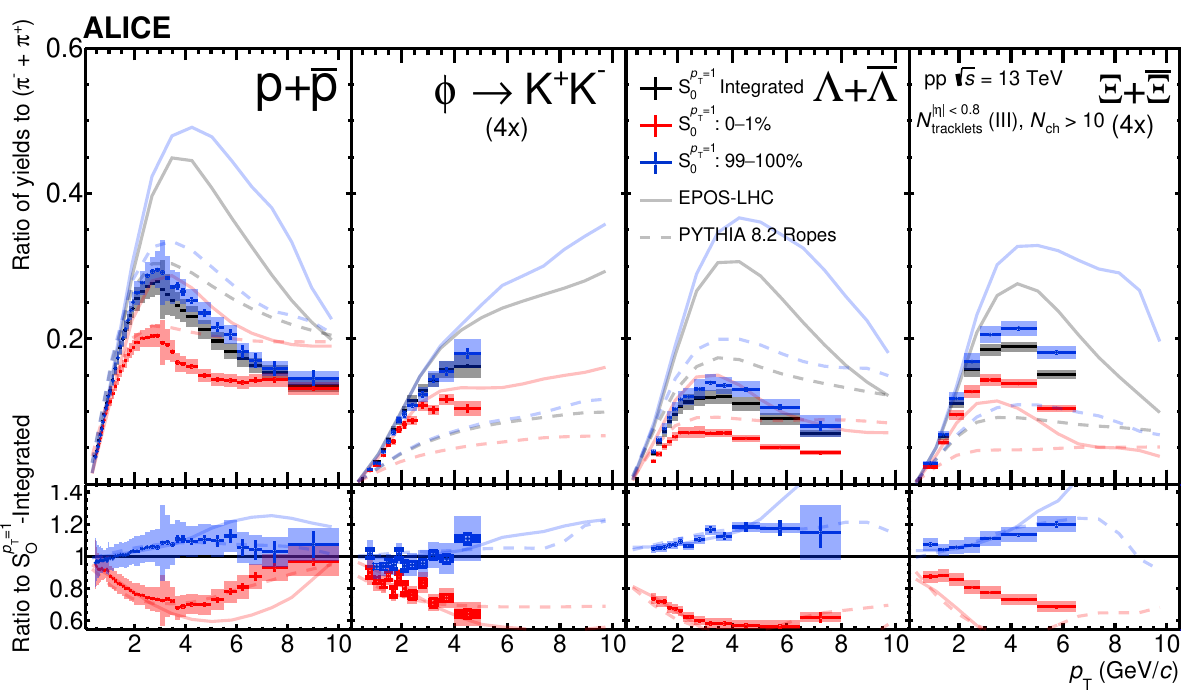}
	\newline
	\hspace*{-1cm}
	    \includegraphics[scale=0.87]{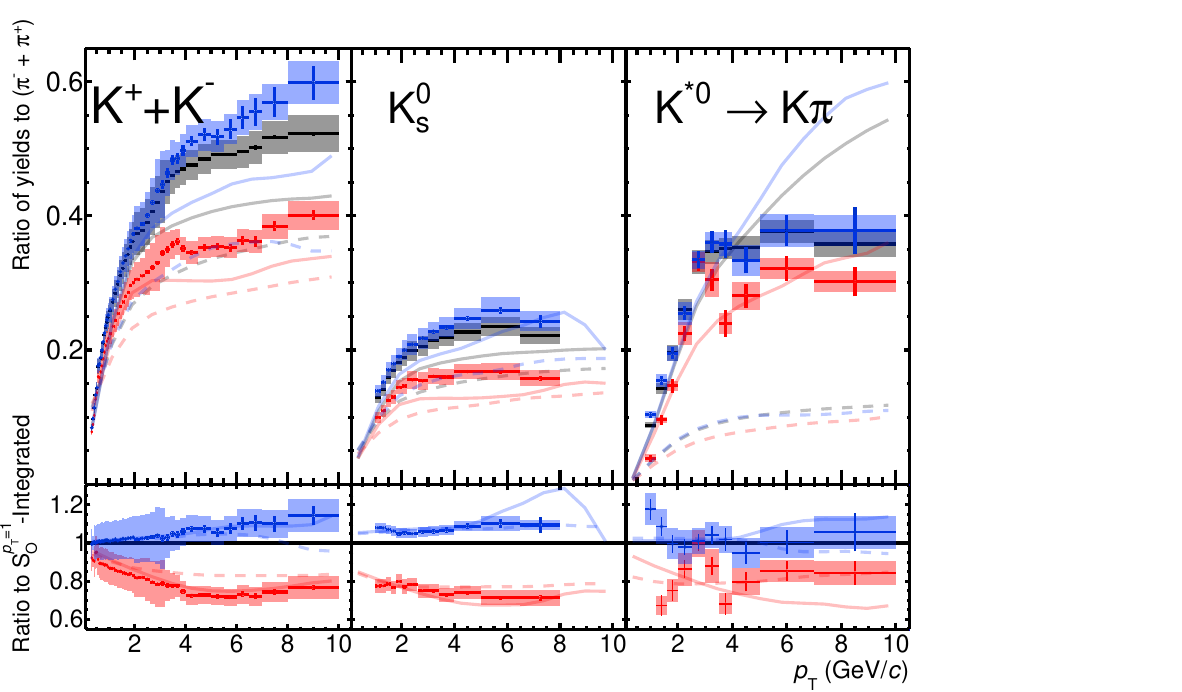}
  \caption{Top panels show hadron-to-\PI ratios for 0--1\% \SOPT classes selected for the 0--10\% \tracklet multiplicity events. Bottom panels present the hadron-to-\PI double-ratios of \SOPT classes relative to \SOPT integrated high-multiplicity events. Statistical and systematic uncertainties are shown by  bars and boxes, respectively. Experimental results are compared with predictions from EPOS-LHC and PYTHIA 8.2 Rope hadronization framework.}
   \label{Fig:DRCL1}
\end{figure}

Fig.~\ref{Fig:DRCL1} illustrates the \SOPT-differential ratios to $(\pi^+ + \pi^-)$ with multiplicity measured at midrapidity, with a simultaneous broadened multiplicity (\tracklet 0--10\%) and tightened \SOPT selection criteria, compared to the measurement presented in Sec.~\ref{Sec:ResRat}. The experimental data is compared with both PYTHIA 8.2 Ropes and EPOS-LHC,  which, as shown in previous sections of this article, give the most accurate predictions of the observed trends.

The double ratios for \KSTAR suggest that the production in isotropic topologies is similar to that of average high-multiplicity events. In contrast, there is a pronounced structure of the ratio for jet-like topologies, highlighting an overall suppression of \KSTAR production. The \PHI has similar features: suppression in jet-like events (qualitatively the same trend as for \XI), and consistent with unity for isotropic events. For all presented particle species, an overall narrower \SOPT selection highlights a large suppression of the \pt-differential yield of strange hadrons relative to pions in events with extreme jet-like topologies. Furthermore, there is a larger deviation among the four different models compared to the more constrained multiplicity quantile, in particular in jet-like topologies for protons, \XI, and \KSTAR. However, the overall trends are well predicted. Both resonance particles favor production in softer events containing QGP-like features, where the decrease of \KSTAR production in jet-like events could potentially be due to a rescattering effect. While the origin of the suppression of $\phi$ production in jet-like events is not fully understood, the behavior is consistent with the strange particles measured in this study.

\begin{figure}[htbp!]
    \centering
	\hspace*{-1cm}
	\includegraphics[scale=0.87]{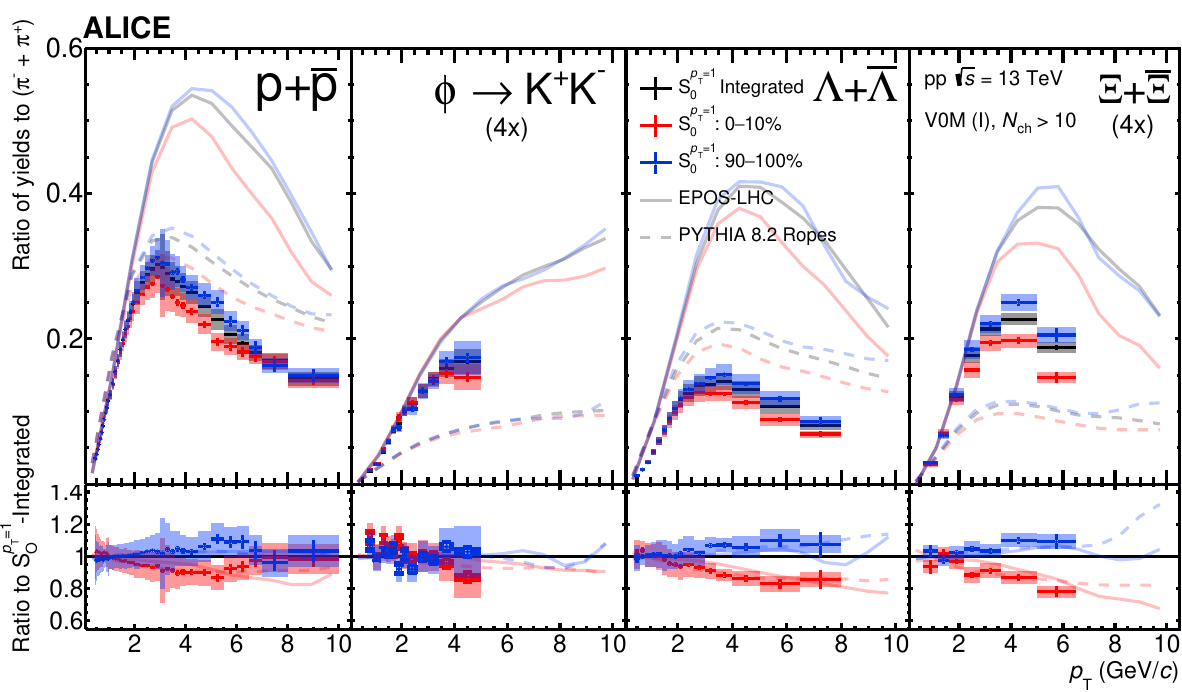}
	\newline
	\hspace*{-1cm}
	 \includegraphics[scale=0.87]{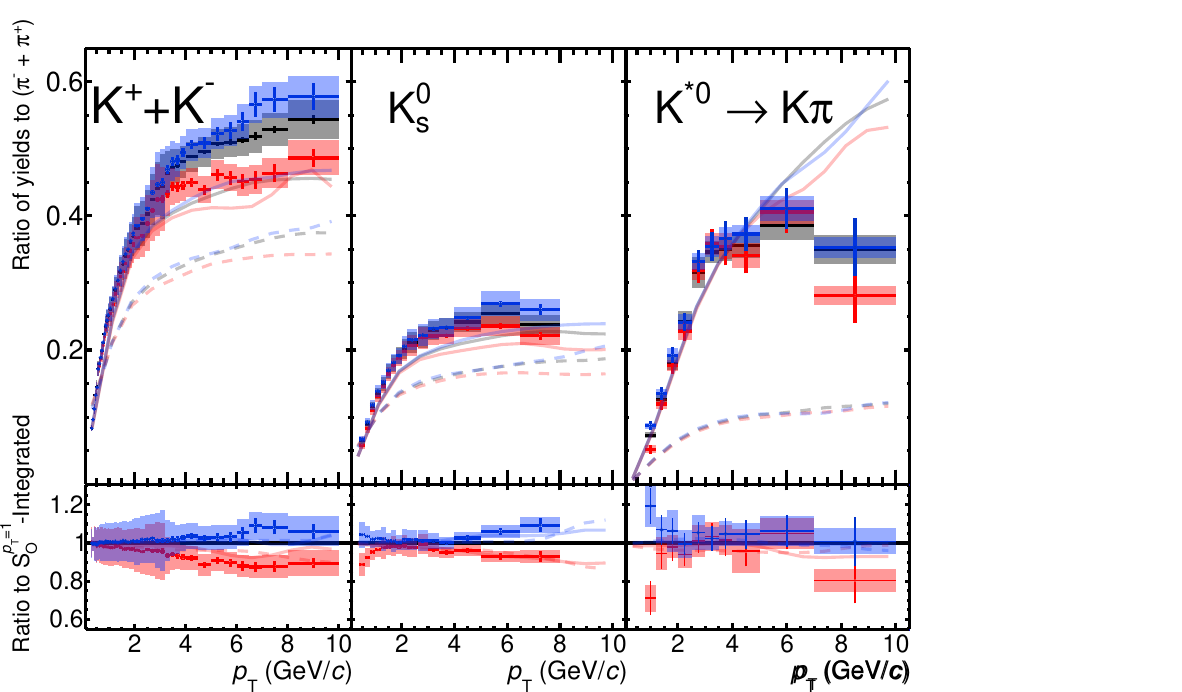}
  \caption{Top panels show hadron-to-\PI ratios for 0--10\% \SOPT classes selected for the 0--1\% V0M multiplicity events. Bottom panels present the hadron-to-\PI double-ratios of \SOPT classes relative to \SOPT integrated high-multiplicity events. Statistical and systematic uncertainties are shown by  bars and boxes, respectively. Experimental results are compared with predictions from EPOS-LHC and PYTHIA 8.2 Rope hadronization framework.}
   \label{Fig:DRV0M}
\end{figure}

The particle ratios to $(\pi^+ + \pi^-)$ for events with forward rapidity multiplicity estimations are presented in 0--10\% \SOPT event classes in Fig.~\ref{Fig:DRV0M}. The reported effects of \SOPT event-selection while estimating multiplicity through the V0M is comparatively weak to either percentile of tracklets, both for the looser and stricter \SOPT criteria. The PYTHIA 8.2 rope hadronization framework is able to accurately describe the interplay between different \SOPT event classes, while estimating multiplicity in the 
 forward rapidity region. In contrast, EPOS-LHC overestimates the relative enhancement and suppression of \XI and \LA in isotropic and jet-like topologies, respectively. The midrapidity multiplicity estimations (both 0--1\% and 0--10\%) showcase an increased sensitivity of strange hadron production as a function of \SOPT compared to the forward rapidity multiplicity estimation, even though the fractional $\langle \rm{d}\it{N}_{\pi}/\rm{d}\it{y} \rangle$ difference between the \SOPT event classes in this case are similar. This finding suggests that the effects of the relative enhancement (suppression) of light-flavor hadron production to \PI mesons in isotropic (jet-like) topologies are smaller while estimating multiplicity with V0M, and that one can study different physical properties relative to the rapidity ranges in which the multiplicity is estimated. 

\begin{figure}[ht!]
    \vspace{0.1cm}
    \centering
    \includegraphics[scale=0.36]{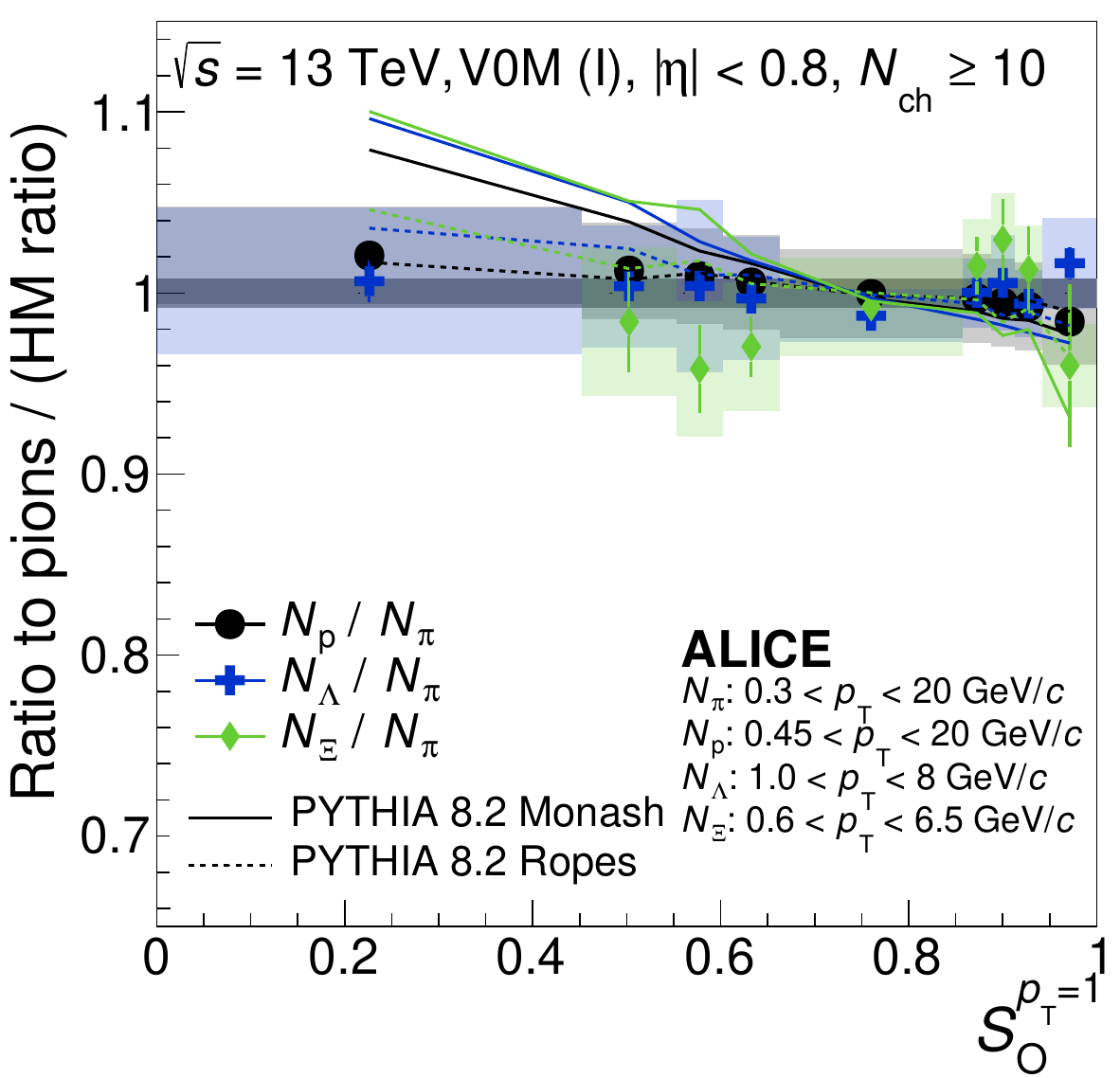}
    \includegraphics[scale=0.36]{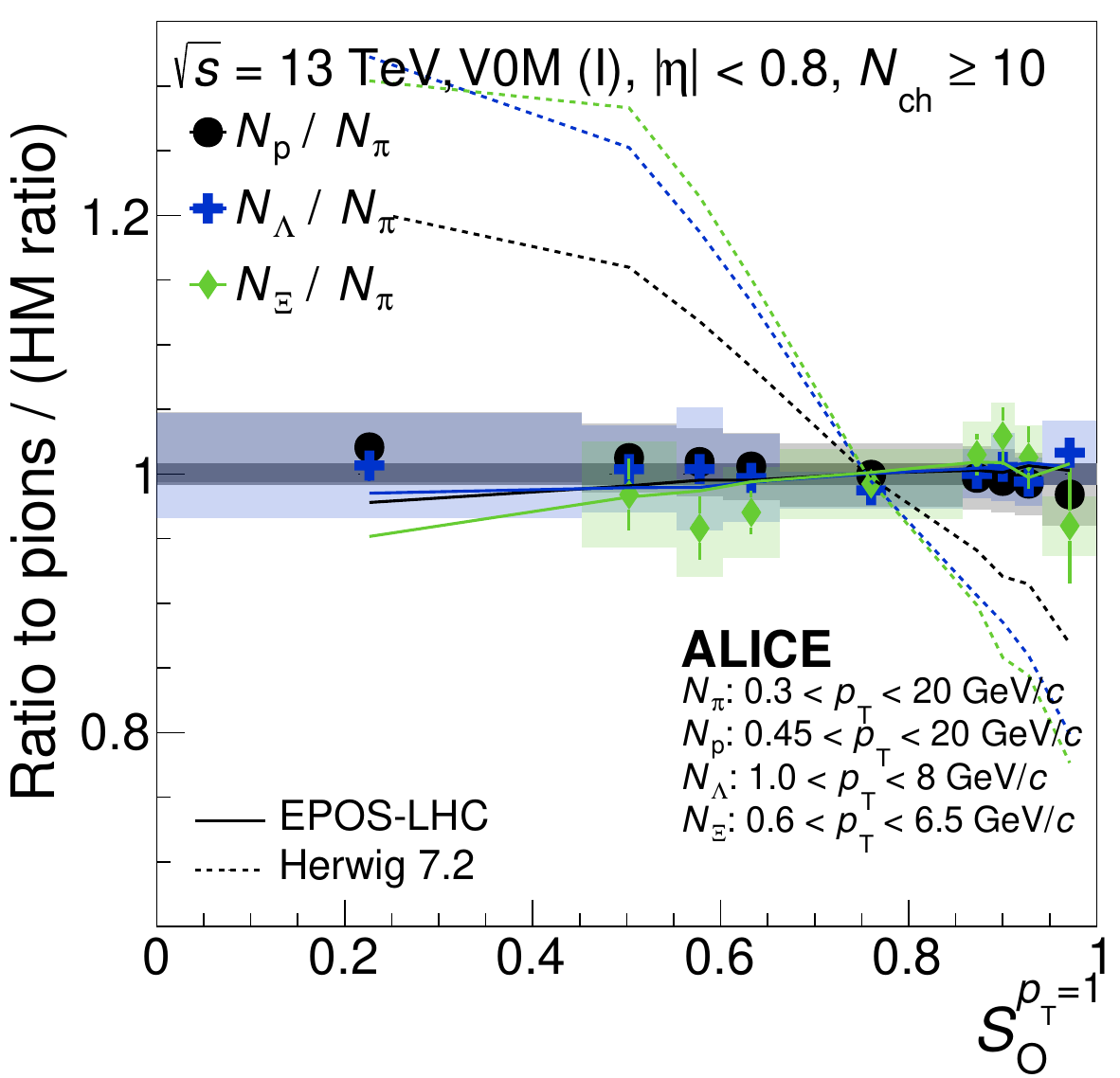}
    \includegraphics[scale=0.36]{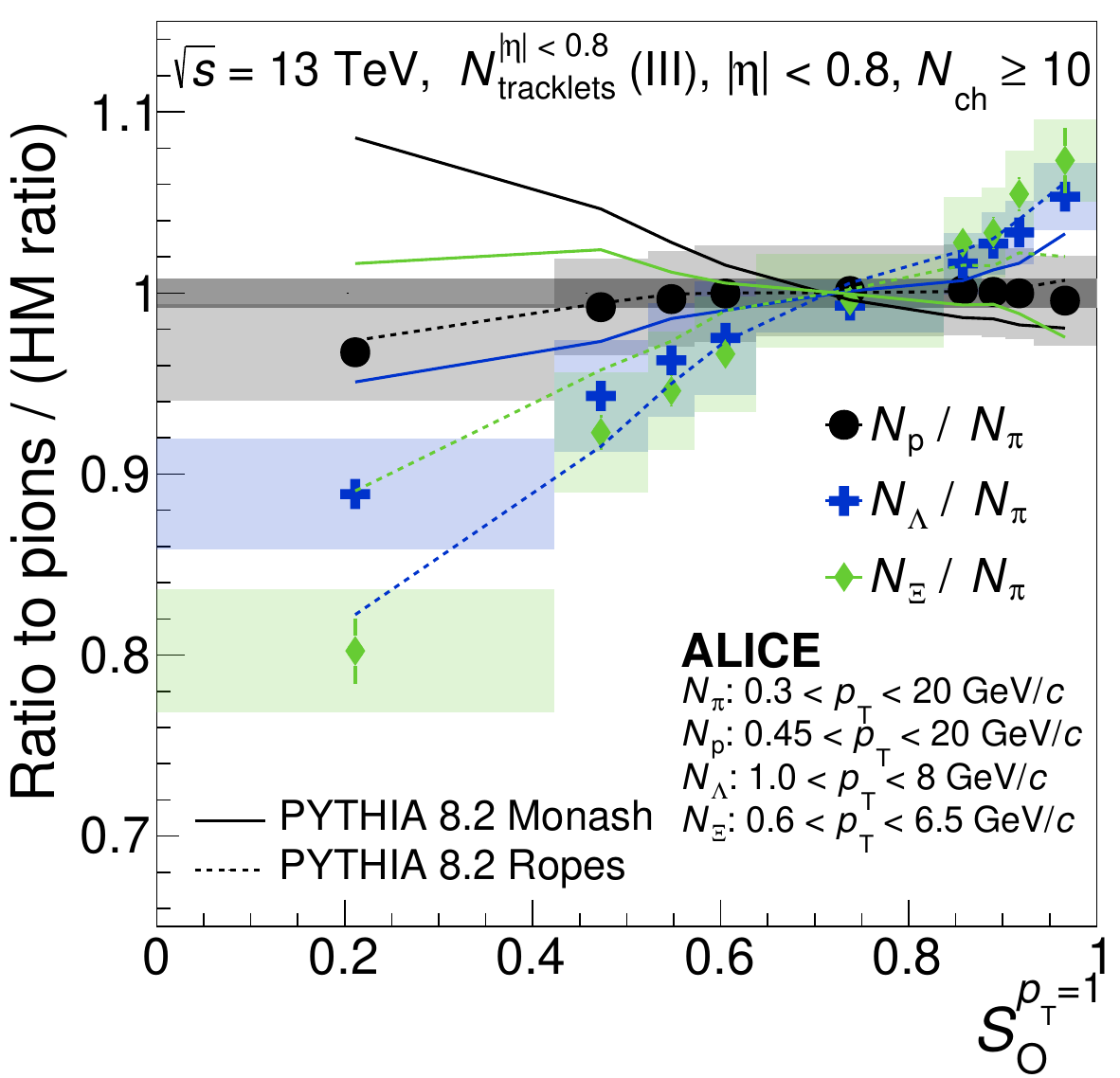}
    \includegraphics[scale=0.36]{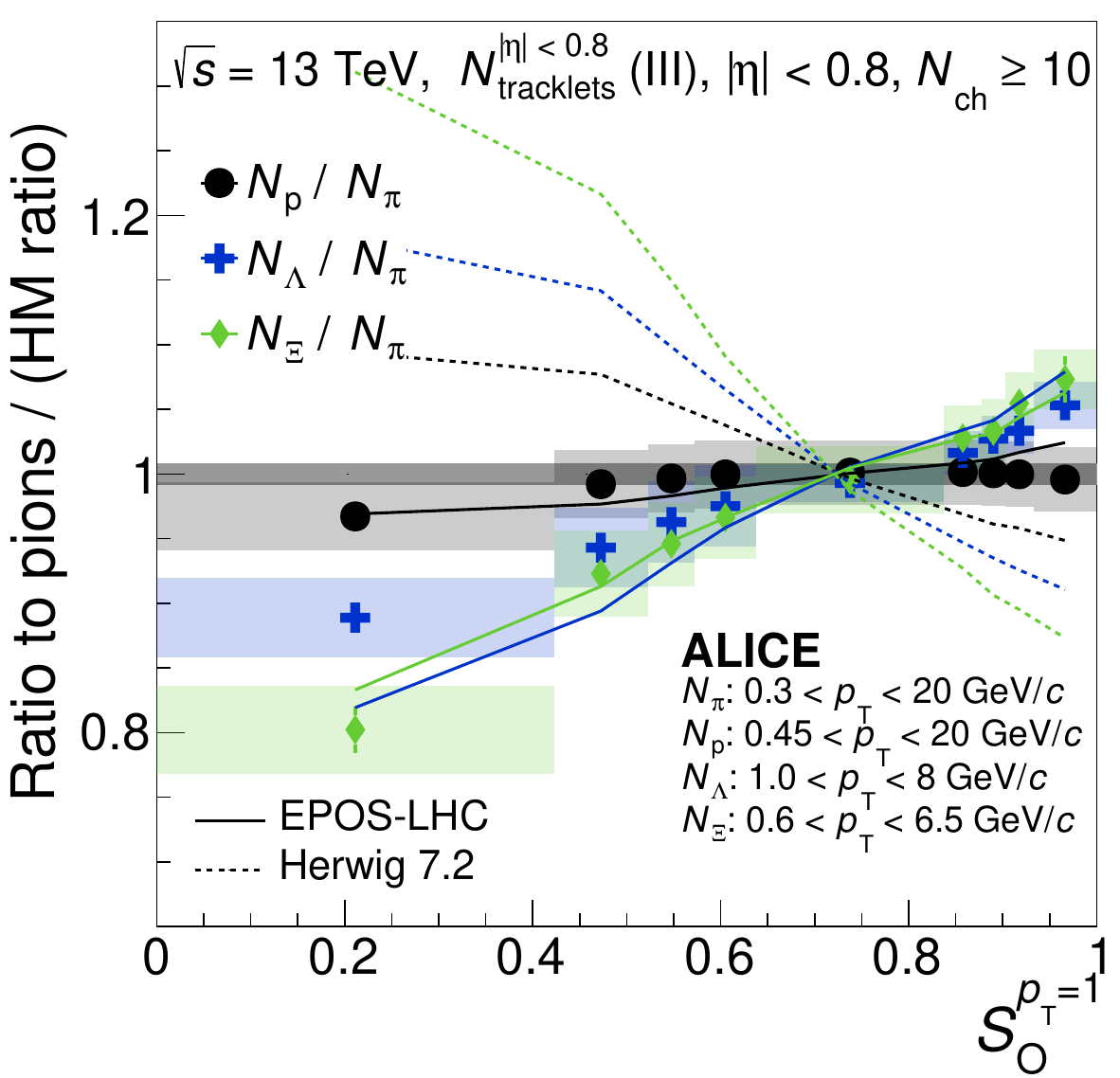}
    \caption{The double ratios of integrated yield as a function of \SOPT are presented for V0M 0--1\% (upper) and \tracklet 0--10\%  (lower). Left and right panels show the same data points, but with different model predictions. Statistical and total systematic uncertainties are shown by bars and boxes, respectively. The curves represent different model predictions of the same measurement. }
    \label{Fig:V0M_1_Int}
\end{figure}

Finally, in Fig.~\ref{Fig:V0M_1_Int} the \pt-integrated yields are presented as a function of \SOPT following the two expanded multiplicity estimations. It was shown in Sec.~\ref{Sec:IntYield} that the results utilizing the measured \pt ranges were consistent with yields obtained from integrating over the full, extrapolated \pt range. Therefore, for the sake of brevity, this section will only include yields integrated over the measured \pt ranges, to allow for maximal precision when compared with model predictions.  Remarkably, the \SOPT-dependent enhancement of strange hadrons seems to vanish once the multiplicity estimation is performed at forward rapidity. Similarly to Figs.~\ref{Fig:CL1_1_Int_py} and ~\ref{Fig:CL1_1_Int_hw}, the effect of strangeness enhancement is consistent between different ranges of the midrapidity multiplicity estimation, and the effect is suggested to be slightly stronger in the more extreme multiplicity case.

Both EPOS-LHC and the PYTHIA 8.2 rope hadronization framework are able to qualitatively describe the enhancement of \LA and \XI with increasing \SOPT, while simultaneously predicting the insensitivity of strange particle production as a function of \SOPT when estimating multiplicity at forward rapidity.

\begin{figure}[t]
    \vspace{0.1cm}
    \centering
    \includegraphics[scale=0.75]{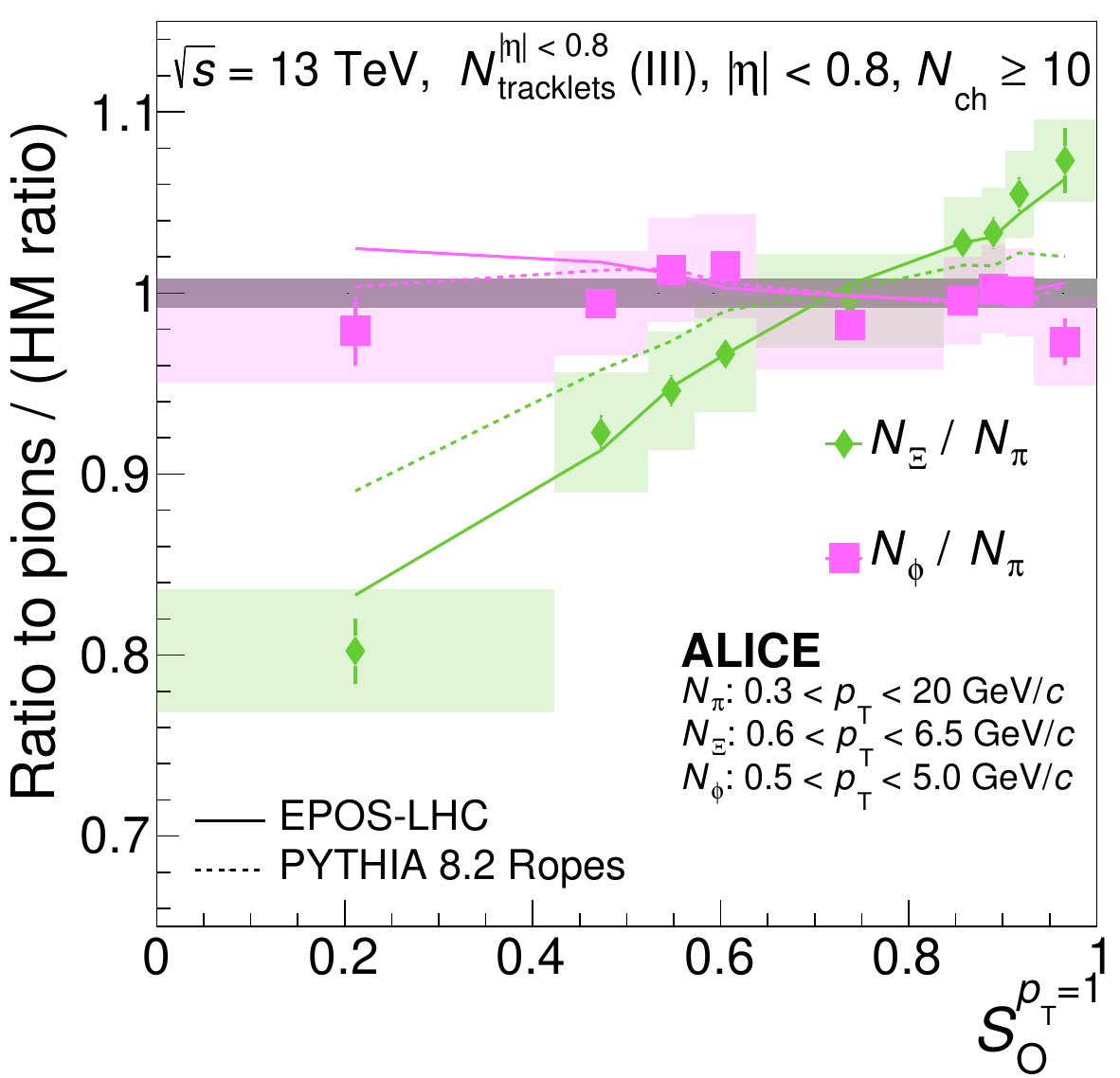}
    \caption{The double ratios of integrated yield as a function of \SOPT in \tracklet 0--10\% for $\phi$ and \XI. Statistical and total systematic uncertainties are shown by bars and boxes, respectively. The curves represent different model predictions of the same measurement}
    \label{Fig:CL1_10_Int_phi}
\end{figure}

Finally, due to the increase of the number of events from the broader multiplicity interval, it is possible to measure the integrated $\phi$ over the full \SOPT range. This is presented in Fig.~\ref{Fig:CL1_10_Int_phi}, where integrated \PHI yields are compared to \XI yields as function of \SOPT, for \tracklet 0--10\%, relative to the equivalent ($\pi^+\pi^-$) yields. It is observed that there is no significant modification of the relative \PHI meson yield as a function of \SOPT within systematic uncertainties. This is in clear contrast to the features exhibited when measuring the relative \XI production. Given the similarity of the proton curves in Fig.~\ref{Fig:V0M_1_Int}, these results imply that the $\phi$ meson has the properties of particles with net-zero strangeness content when measured as a function of \SOPT. The resulting suppression or enhancement of relative hadron yields as a function of \SOPT are therefore suggested to be driven by the hadron mass, rather than by the effective strangeness of the hadron.

The data is compared to PYTHIA 8.2 ropes and EPOS-LHC, both able to quantitatively predict the dynamics of the relative $\phi$ yields as a function of \SOPT. However, in PYTHIA 8.2, one would naively expect that the \PHI dynamics should follow the same trend as \XI, as both are effectively double-strange. This observed difference suggests that the \XI enhancement in the PYTHIA 8.2 rope model can mainly be attributed to addition of Junction formations, which only enhances baryon production.

\section{Summary and conclusions}\label{Sec:Conc}

In this article, we have presented the production of strange and non-strange light-flavor particles at midrapidity (\etarange 0.8) in high-multiplicity pp collisions at \s $=$ 13 TeV, as a function of the unweighted transverse spherocity \SOPT. In order to account for the biases between weakly-decaying and charged primary hadrons, we have elaborated on how the observable was updated from the traditional transverse spherocity.

The multiplicity was estimated in two different kinematic regions: at midrapidity (\etarange 0.8) by  measuring the activity in the SPD (\tracklet), and at forward rapidities ($2.8 < \eta < 5.1$ and $-3.7 < \eta < -1.7$) by measuring the activity in both sides of the V0 forward detector (V0M amplitude). In Fig.~\ref{Fig:CL1vsV0M}, we report that the different \SOPT  event classes have a large variance in local charged particle density when estimating the multiplicity through the V0M, while being roughly equal in terms of \mpt. In contrast, the \tracklet-selected \SOPT events are constrained in terms of $\langle \rm{d}\it{N}_{\pi}/\rm{d}\it{y} \rangle$ within $\sim 10\%$, while having a large variance in \mpt. We conclude that estimating multiplicity at midrapidity allows one to isolate and study the dynamics of particle production in events that are driven by either soft or hard-QCD physics.

The main features of this analysis are reported in the particle-to-pion ratios in Figs.~\ref{Fig:CombRatioPy}--~\ref{Fig:CombRatioHerEx}, highlighting an enhancement of strange hadrons in events with an isotropic topology, and a strong suppression in events with a jet-like topology. This indicates that events with an isotropic topology describe the average high-multiplicity event fairly well, while events heavily influenced by jet-like physics are outliers. For the baryon-to-meson ratios presented in Fig.~\ref{Fig:Flow}, the \SOPT selection also highlights an enhancement at intermediate \pt in the p/\PI, \LA/\kzero and \XI/\PHI ratios, but without a depletion of low-\pt particles for jet-like events. The effects are consistent among the three ratios, and the \PHI exhibits a behavior similar to the other strange particles. However, the \pt-integrated yield of \PHI as a function of \SOPT presented in Fig.~\ref{Fig:CL1_10_Int_phi} highlights features similar to other non-strange particles, reporting no observed modification of relative \PHI meson production as a function of \SOPT.

The presented models are not able to describe the quantitative observations found in data for the particle ratios, but are overall able to describe the interplay between \SOPT classified events at high-multiplicity with good accuracy in the double ratios. Remarkably, even though the production mechanisms for the PYTHIA 8.2 Monash and PYTHIA 8.2 Ropes variations (ropes formed by layers of overlapping strings) are qualitatively different, they both predict the interplay with only minor differences between them. 

Finally, we report the relative integrated strange particle yield to pions as function of \SOPT in Fig.~\ref{Fig:CL1_1_Int_py}. It is found that one can achieve a similar strangeness enhancement found in multiplicity differential analyses, by fixing the local charged particle density and varying the azimuthal topology, instead of only varying the charged-particle density. This indicates that particle production at high multiplicities is driven by more than a single source, with different strangeness-to-pion production rates. The same measurement is then performed in high-multiplicity events selected at forward rapidities. Fig.~\ref{Fig:CL1vsV0M} highlights that the differential-\SOPT selection in this event-class primarily varies the local charged particle density approximately a factor of two. Therefore, when contrasted to the multiplicity-dependent ALICE measurements in Ref.~\cite{Naturepaper}, one could naively expect to observe strangeness suppression and enhancement due to the difference in charged particle density alone. Surprisingly, we show in Fig.~\ref{Fig:V0M_1_Int} that this effect is negligible when estimating multiplicity in the forward-rapidity region, with no significant strangeness enhancement or suppression within systematic uncertainties. This goes contrary to expectations based on previous ALICE publications, and the effect is currently not well understood. PYTHIA 8.2 Monash and Herwig 7.2 are unable to describe this effect, but PYTHIA 8.2 Ropes and EPOS-LHC qualitatively capture the trend, showcasing a large suppression of strangeness production in jet-like events, although with a mass ordering that is incompatible with the experimental data. 

We conclude that with \SOPT, one is able to categorize events into classes based on characteristic azimuthal topologies, from jet-like events associated with hard-QCD physics, to isotropic events associated with soft-QCD physics. Furthermore, the findings presented in this article suggest that average high-multiplicity events, even in the most extreme cases, are dominated by soft processes, where rare hard processes play little or no role for bulk observables.


\newenvironment{acknowledgement}{\relax}{\relax}
\begin{acknowledgement}
\section*{Acknowledgements}

The ALICE Collaboration would like to thank all its engineers and technicians for their invaluable contributions to the construction of the experiment and the CERN accelerator teams for the outstanding performance of the LHC complex.
The ALICE Collaboration gratefully acknowledges the resources and support provided by all Grid centres and the Worldwide LHC Computing Grid (WLCG) collaboration.
The ALICE Collaboration acknowledges the following funding agencies for their support in building and running the ALICE detector:
A. I. Alikhanyan National Science Laboratory (Yerevan Physics Institute) Foundation (ANSL), State Committee of Science and World Federation of Scientists (WFS), Armenia;
Austrian Academy of Sciences, Austrian Science Fund (FWF): [M 2467-N36] and Nationalstiftung f\"{u}r Forschung, Technologie und Entwicklung, Austria;
Ministry of Communications and High Technologies, National Nuclear Research Center, Azerbaijan;
Conselho Nacional de Desenvolvimento Cient\'{\i}fico e Tecnol\'{o}gico (CNPq), Financiadora de Estudos e Projetos (Finep), Funda\c{c}\~{a}o de Amparo \`{a} Pesquisa do Estado de S\~{a}o Paulo (FAPESP) and Universidade Federal do Rio Grande do Sul (UFRGS), Brazil;
Bulgarian Ministry of Education and Science, within the National Roadmap for Research Infrastructures 2020-2027 (object CERN), Bulgaria;
Ministry of Education of China (MOEC) , Ministry of Science \& Technology of China (MSTC) and National Natural Science Foundation of China (NSFC), China;
Ministry of Science and Education and Croatian Science Foundation, Croatia;
Centro de Aplicaciones Tecnol\'{o}gicas y Desarrollo Nuclear (CEADEN), Cubaenerg\'{\i}a, Cuba;
Ministry of Education, Youth and Sports of the Czech Republic, Czech Republic;
The Danish Council for Independent Research | Natural Sciences, the VILLUM FONDEN and Danish National Research Foundation (DNRF), Denmark;
Helsinki Institute of Physics (HIP), Finland;
Commissariat \`{a} l'Energie Atomique (CEA) and Institut National de Physique Nucl\'{e}aire et de Physique des Particules (IN2P3) and Centre National de la Recherche Scientifique (CNRS), France;
Bundesministerium f\"{u}r Bildung und Forschung (BMBF) and GSI Helmholtzzentrum f\"{u}r Schwerionenforschung GmbH, Germany;
General Secretariat for Research and Technology, Ministry of Education, Research and Religions, Greece;
National Research, Development and Innovation Office, Hungary;
Department of Atomic Energy Government of India (DAE), Department of Science and Technology, Government of India (DST), University Grants Commission, Government of India (UGC) and Council of Scientific and Industrial Research (CSIR), India;
National Research and Innovation Agency - BRIN, Indonesia;
Istituto Nazionale di Fisica Nucleare (INFN), Italy;
Japanese Ministry of Education, Culture, Sports, Science and Technology (MEXT) and Japan Society for the Promotion of Science (JSPS) KAKENHI, Japan;
Consejo Nacional de Ciencia (CONACYT) y Tecnolog\'{i}a, through Fondo de Cooperaci\'{o}n Internacional en Ciencia y Tecnolog\'{i}a (FONCICYT) and Direcci\'{o}n General de Asuntos del Personal Academico (DGAPA), Mexico;
Nederlandse Organisatie voor Wetenschappelijk Onderzoek (NWO), Netherlands;
The Research Council of Norway, Norway;
Commission on Science and Technology for Sustainable Development in the South (COMSATS), Pakistan;
Pontificia Universidad Cat\'{o}lica del Per\'{u}, Peru;
Ministry of Education and Science, National Science Centre and WUT ID-UB, Poland;
Korea Institute of Science and Technology Information and National Research Foundation of Korea (NRF), Republic of Korea;
Ministry of Education and Scientific Research, Institute of Atomic Physics, Ministry of Research and Innovation and Institute of Atomic Physics and Universitatea Nationala de Stiinta si Tehnologie Politehnica Bucuresti, Romania;
Ministry of Education, Science, Research and Sport of the Slovak Republic, Slovakia;
National Research Foundation of South Africa, South Africa;
Swedish Research Council (VR) and Knut \& Alice Wallenberg Foundation (KAW), Sweden;
European Organization for Nuclear Research, Switzerland;
Suranaree University of Technology (SUT), National Science and Technology Development Agency (NSTDA) and National Science, Research and Innovation Fund (NSRF via PMU-B B05F650021), Thailand;
Turkish Energy, Nuclear and Mineral Research Agency (TENMAK), Turkey;
National Academy of  Sciences of Ukraine, Ukraine;
Science and Technology Facilities Council (STFC), United Kingdom;
National Science Foundation of the United States of America (NSF) and United States Department of Energy, Office of Nuclear Physics (DOE NP), United States of America.
In addition, individual groups or members have received support from:
Czech Science Foundation (grant no. 23-07499S), Czech Republic;
European Research Council, Strong 2020 - Horizon 2020 (grant nos. 950692, 824093), European Union;
ICSC - Centro Nazionale di Ricerca in High Performance Computing, Big Data and Quantum Computing, European Union - NextGenerationEU;
Academy of Finland (Center of Excellence in Quark Matter) (grant nos. 346327, 346328), Finland.

\end{acknowledgement}

\bibliographystyle{utphys}   
\bibliography{bibliography}

\newpage
\appendix

%
%

\section{The ALICE Collaboration}
\label{app:collab}
\begin{flushleft} 
\small

S.~Acharya\,\orcidlink{0000-0002-9213-5329}\,$^{\rm 128}$, 
D.~Adamov\'{a}\,\orcidlink{0000-0002-0504-7428}\,$^{\rm 87}$, 
G.~Aglieri Rinella\,\orcidlink{0000-0002-9611-3696}\,$^{\rm 33}$, 
M.~Agnello\,\orcidlink{0000-0002-0760-5075}\,$^{\rm 30}$, 
N.~Agrawal\,\orcidlink{0000-0003-0348-9836}\,$^{\rm 52}$, 
Z.~Ahammed\,\orcidlink{0000-0001-5241-7412}\,$^{\rm 136}$, 
S.~Ahmad\,\orcidlink{0000-0003-0497-5705}\,$^{\rm 16}$, 
S.U.~Ahn\,\orcidlink{0000-0001-8847-489X}\,$^{\rm 72}$, 
I.~Ahuja\,\orcidlink{0000-0002-4417-1392}\,$^{\rm 38}$, 
A.~Akindinov\,\orcidlink{0000-0002-7388-3022}\,$^{\rm 142}$
% ...

\section*{Affiliation Notes}

$^{\rm I}$ Deceased\\
$^{\rm II}$ Also at: Max-Planck-Institut fur Physik, Munich, Germany\\
$^{\rm III}$ Also at: Italian National Agency for New Technologies, Energy and Sustainable Economic Development (ENEA), Bologna, Italy\\
$^{\rm IV}$ Also at: Dipartimento DET del Politecnico di Torino, Turin, Italy\\
$^{\rm V}$ Also at: Yildiz Technical University, Istanbul, T\"{u}rkiye\\
$^{\rm VI}$ Also at: Department of Applied Physics, Aligarh Muslim University, Aligarh, India\\
$^{\rm VII}$ Also at: Institute of Theoretical Physics, University of Wroclaw, Poland\\
$^{\rm VIII}$ Also at: An institution covered by a cooperation agreement with CERN\\

\section*{Collaboration Institutes}

$^{1}$ A.I. Alikhanyan National Science Laboratory (Yerevan Physics Institute) Foundation, Yerevan, Armenia\\
$^{2}$ AGH University of Krakow, Cracow, Poland\\
$^{3}$ Bogolyubov Institute for Theoretical Physics, National Academy of Sciences of Ukraine, Kiev, Ukraine\\
$^{4}$ Bose Institute, Department of Physics  and Centre for Astroparticle Physics and Space Science (CAPSS), Kolkata, India\\
$^{5}$ California Polytechnic State University, San Luis Obispo, California, United States\\
$^{6}$ Central China Normal University, Wuhan, China\\
$^{7}$ Centro de Aplicaciones Tecnol\'{o}gicas y Desarrollo Nuclear (CEADEN), Havana, Cuba\\
$^{8}$ Centro de Investigaci\'{o}n y de Estudios Avanzados (CINVESTAV), Mexico City and M\'{e}rida, Mexico\\
$^{9}$ Chicago State University, Chicago, Illinois, United States\\
$^{10}$ China Institute of Atomic Energy, Beijing, China\\
$^{11}$ China University of Geosciences, Wuhan, China\\
$^{12}$ Chungbuk National University, Cheongju, Republic of Korea\\
$^{13}$ Comenius University Bratislava, Faculty of Mathematics, Physics and Informatics, Bratislava, Slovak Republic\\
$^{14}$ COMSATS University Islamabad, Islamabad, Pakistan\\
$^{15}$ Creighton University, Omaha, Nebraska, United States\\
$^{16}$ Department of Physics, Aligarh Muslim University, Aligarh, India\\
$^{17}$ Department of Physics, Pusan National University, Pusan, Republic of Korea\\
$^{18}$ Department of Physics, Sejong University, Seoul, Republic of Korea\\
$^{19}$ Department of Physics, University of California, Berkeley, California, United States\\
$^{20}$ Department of Physics, University of Oslo, Oslo, Norway\\
$^{21}$ Department of Physics and Technology, University of Bergen, Bergen, Norway\\
$^{22}$ Dipartimento di Fisica, Universit\`{a} di Pavia, Pavia, Italy\\
$^{23}$ Dipartimento di Fisica dell'Universit\`{a} and Sezione INFN, Cagliari, Italy\\
$^{24}$ Dipartimento di Fisica dell'Universit\`{a} and Sezione INFN, Trieste, Italy\\
$^{25}$ Dipartimento di Fisica dell'Universit\`{a} and Sezione INFN, Turin, Italy\\
$^{26}$ Dipartimento di Fisica e Astronomia dell'Universit\`{a} and Sezione INFN, Bologna, Italy\\
$^{27}$ Dipartimento di Fisica e Astronomia dell'Universit\`{a} and Sezione INFN, Catania, Italy\\
$^{28}$ Dipartimento di Fisica e Astronomia dell'Universit\`{a} and Sezione INFN, Padova, Italy\\
$^{29}$ Dipartimento di Fisica `E.R.~Caianiello' dell'Universit\`{a} and Gruppo Collegato INFN, Salerno, Italy\\
$^{30}$ Dipartimento DISAT del Politecnico and Sezione INFN, Turin, Italy\\
$^{31}$ Dipartimento di Scienze MIFT, Universit\`{a} di Messina, Messina, Italy\\
$^{32}$ Dipartimento Interateneo di Fisica `M.~Merlin' and Sezione INFN, Bari, Italy\\
$^{33}$ European Organization for Nuclear Research (CERN), Geneva, Switzerland\\
$^{34}$ Faculty of Electrical Engineering, Mechanical Engineering and Naval Architecture, University of Split, Split, Croatia\\
$^{35}$ Faculty of Engineering and Science, Western Norway University of Applied Sciences, Bergen, Norway\\
$^{36}$ Faculty of Nuclear Sciences and Physical Engineering, Czech Technical University in Prague, Prague, Czech Republic\\
$^{37}$ Faculty of Physics, Sofia University, Sofia, Bulgaria\\
$^{38}$ Faculty of Science, P.J.~\v{S}af\'{a}rik University, Ko\v{s}ice, Slovak Republic\\
$^{39}$ Frankfurt Institute for Advanced Studies, Johann Wolfgang Goethe-Universit\"{a}t Frankfurt, Frankfurt, Germany\\
$^{40}$ Fudan University, Shanghai, China\\
$^{41}$ Gangneung-Wonju National University, Gangneung, Republic of Korea\\
$^{42}$ Gauhati University, Department of Physics, Guwahati, India\\
$^{43}$ Helmholtz-Institut f\"{u}r Strahlen- und Kernphysik, Rheinische Friedrich-Wilhelms-Universit\"{a}t Bonn, Bonn, Germany\\
$^{44}$ Helsinki Institute of Physics (HIP), Helsinki, Finland\\
$^{45}$ High Energy Physics Group,  Universidad Aut\'{o}noma de Puebla, Puebla, Mexico\\
$^{46}$ Horia Hulubei National Institute of Physics and Nuclear Engineering, Bucharest, Romania\\
$^{47}$ HUN-REN Wigner Research Centre for Physics, Budapest, Hungary\\
$^{48}$ Indian Institute of Technology Bombay (IIT), Mumbai, India\\
$^{49}$ Indian Institute of Technology Indore, Indore, India\\
$^{50}$ INFN, Laboratori Nazionali di Frascati, Frascati, Italy\\
$^{51}$ INFN, Sezione di Bari, Bari, Italy\\
$^{52}$ INFN, Sezione di Bologna, Bologna, Italy\\
$^{53}$ INFN, Sezione di Cagliari, Cagliari, Italy\\
$^{54}$ INFN, Sezione di Catania, Catania, Italy\\
$^{55}$ INFN, Sezione di Padova, Padova, Italy\\
$^{56}$ INFN, Sezione di Pavia, Pavia, Italy\\
$^{57}$ INFN, Sezione di Torino, Turin, Italy\\
$^{58}$ INFN, Sezione di Trieste, Trieste, Italy\\
$^{59}$ Inha University, Incheon, Republic of Korea\\
$^{60}$ Institute for Gravitational and Subatomic Physics (GRASP), Utrecht University/Nikhef, Utrecht, Netherlands\\
$^{61}$ Institute of Experimental Physics, Slovak Academy of Sciences, Ko\v{s}ice, Slovak Republic\\
$^{62}$ Institute of Physics, Homi Bhabha National Institute, Bhubaneswar, India\\
$^{63}$ Institute of Physics of the Czech Academy of Sciences, Prague, Czech Republic\\
$^{64}$ Institute of Space Science (ISS), Bucharest, Romania\\
$^{65}$ Institut f\"{u}r Kernphysik, Johann Wolfgang Goethe-Universit\"{a}t Frankfurt, Frankfurt, Germany\\
$^{66}$ Instituto de Ciencias Nucleares, Universidad Nacional Aut\'{o}noma de M\'{e}xico, Mexico City, Mexico\\
$^{67}$ Instituto de F\'{i}sica, Universidade Federal do Rio Grande do Sul (UFRGS), Porto Alegre, Brazil\\
$^{68}$ Instituto de F\'{\i}sica, Universidad Nacional Aut\'{o}noma de M\'{e}xico, Mexico City, Mexico\\
$^{69}$ iThemba LABS, National Research Foundation, Somerset West, South Africa\\
$^{70}$ Jeonbuk National University, Jeonju, Republic of Korea\\
$^{71}$ Johann-Wolfgang-Goethe Universit\"{a}t Frankfurt Institut f\"{u}r Informatik, Fachbereich Informatik und Mathematik, Frankfurt, Germany\\
$^{72}$ Korea Institute of Science and Technology Information, Daejeon, Republic of Korea\\
$^{73}$ KTO Karatay University, Konya, Turkey\\
$^{74}$ Laboratoire de Physique Subatomique et de Cosmologie, Universit\'{e} Grenoble-Alpes, CNRS-IN2P3, Grenoble, France\\
$^{75}$ Lawrence Berkeley National Laboratory, Berkeley, California, United States\\
$^{76}$ Lund University Department of Physics, Division of Particle Physics, Lund, Sweden\\
$^{77}$ Nagasaki Institute of Applied Science, Nagasaki, Japan\\
$^{78}$ Nara Women{'}s University (NWU), Nara, Japan\\
$^{79}$ National and Kapodistrian University of Athens, School of Science, Department of Physics , Athens, Greece\\
$^{80}$ National Centre for Nuclear Research, Warsaw, Poland\\
$^{81}$ National Institute of Science Education and Research, Homi Bhabha National Institute, Jatni, India\\
$^{82}$ National Nuclear Research Center, Baku, Azerbaijan\\
$^{83}$ National Research and Innovation Agency - BRIN, Jakarta, Indonesia\\
$^{84}$ Niels Bohr Institute, University of Copenhagen, Copenhagen, Denmark\\
$^{85}$ Nikhef, National institute for subatomic physics, Amsterdam, Netherlands\\
$^{86}$ Nuclear Physics Group, STFC Daresbury Laboratory, Daresbury, United Kingdom\\
$^{87}$ Nuclear Physics Institute of the Czech Academy of Sciences, Husinec-\v{R}e\v{z}, Czech Republic\\
$^{88}$ Oak Ridge National Laboratory, Oak Ridge, Tennessee, United States\\
$^{89}$ Ohio State University, Columbus, Ohio, United States\\
$^{90}$ Physics department, Faculty of science, University of Zagreb, Zagreb, Croatia\\
$^{91}$ Physics Department, Panjab University, Chandigarh, India\\
$^{92}$ Physics Department, University of Jammu, Jammu, India\\
$^{93}$ Physics Program and International Institute for Sustainability with Knotted Chiral Meta Matter (SKCM2), Hiroshima University, Hiroshima, Japan\\
$^{94}$ Physikalisches Institut, Eberhard-Karls-Universit\"{a}t T\"{u}bingen, T\"{u}bingen, Germany\\
$^{95}$ Physikalisches Institut, Ruprecht-Karls-Universit\"{a}t Heidelberg, Heidelberg, Germany\\
$^{96}$ Physik Department, Technische Universit\"{a}t M\"{u}nchen, Munich, Germany\\
$^{97}$ Politecnico di Bari and Sezione INFN, Bari, Italy\\
$^{98}$ Research Division and ExtreMe Matter Institute EMMI, GSI Helmholtzzentrum f\"ur Schwerionenforschung GmbH, Darmstadt, Germany\\
$^{99}$ Saga University, Saga, Japan\\
$^{100}$ Saha Institute of Nuclear Physics, Homi Bhabha National Institute, Kolkata, India\\
$^{101}$ School of Physics and Astronomy, University of Birmingham, Birmingham, United Kingdom\\
$^{102}$ Secci\'{o}n F\'{\i}sica, Departamento de Ciencias, Pontificia Universidad Cat\'{o}lica del Per\'{u}, Lima, Peru\\
$^{103}$ Stefan Meyer Institut f\"{u}r Subatomare Physik (SMI), Vienna, Austria\\
$^{104}$ SUBATECH, IMT Atlantique, Nantes Universit\'{e}, CNRS-IN2P3, Nantes, France\\
$^{105}$ Sungkyunkwan University, Suwon City, Republic of Korea\\
$^{106}$ Suranaree University of Technology, Nakhon Ratchasima, Thailand\\
$^{107}$ Technical University of Ko\v{s}ice, Ko\v{s}ice, Slovak Republic\\
$^{108}$ The Henryk Niewodniczanski Institute of Nuclear Physics, Polish Academy of Sciences, Cracow, Poland\\
$^{109}$ The University of Texas at Austin, Austin, Texas, United States\\
$^{110}$ Universidad Aut\'{o}noma de Sinaloa, Culiac\'{a}n, Mexico\\
$^{111}$ Universidade de S\~{a}o Paulo (USP), S\~{a}o Paulo, Brazil\\
$^{112}$ Universidade Estadual de Campinas (UNICAMP), Campinas, Brazil\\
$^{113}$ Universidade Federal do ABC, Santo Andre, Brazil\\
$^{114}$ Universitatea Nationala de Stiinta si Tehnologie Politehnica Bucuresti, Bucharest, Romania\\
$^{115}$ University of Cape Town, Cape Town, South Africa\\
$^{116}$ University of Derby, Derby, United Kingdom\\
$^{117}$ University of Houston, Houston, Texas, United States\\
$^{118}$ University of Jyv\"{a}skyl\"{a}, Jyv\"{a}skyl\"{a}, Finland\\
$^{119}$ University of Kansas, Lawrence, Kansas, United States\\
$^{120}$ University of Liverpool, Liverpool, United Kingdom\\
$^{121}$ University of Science and Technology of China, Hefei, China\\
$^{122}$ University of South-Eastern Norway, Kongsberg, Norway\\
$^{123}$ University of Tennessee, Knoxville, Tennessee, United States\\
$^{124}$ University of the Witwatersrand, Johannesburg, South Africa\\
$^{125}$ University of Tokyo, Tokyo, Japan\\
$^{126}$ University of Tsukuba, Tsukuba, Japan\\
$^{127}$ Universit\"{a}t M\"{u}nster, Institut f\"{u}r Kernphysik, M\"{u}nster, Germany\\
$^{128}$ Universit\'{e} Clermont Auvergne, CNRS/IN2P3, LPC, Clermont-Ferrand, France\\
$^{129}$ Universit\'{e} de Lyon, CNRS/IN2P3, Institut de Physique des 2 Infinis de Lyon, Lyon, France\\
$^{130}$ Universit\'{e} de Strasbourg, CNRS, IPHC UMR 7178, F-67000 Strasbourg, France, Strasbourg, France\\
$^{131}$ Universit\'{e} Paris-Saclay, Centre d'Etudes de Saclay (CEA), IRFU, D\'{e}partment de Physique Nucl\'{e}aire (DPhN), Saclay, France\\
$^{132}$ Universit\'{e}  Paris-Saclay, CNRS/IN2P3, IJCLab, Orsay, France\\
$^{133}$ Universit\`{a} degli Studi di Foggia, Foggia, Italy\\
$^{134}$ Universit\`{a} del Piemonte Orientale, Vercelli, Italy\\
$^{135}$ Universit\`{a} di Brescia, Brescia, Italy\\
$^{136}$ Variable Energy Cyclotron Centre, Homi Bhabha National Institute, Kolkata, India\\
$^{137}$ Warsaw University of Technology, Warsaw, Poland\\
$^{138}$ Wayne State University, Detroit, Michigan, United States\\
$^{139}$ Yale University, New Haven, Connecticut, United States\\
$^{140}$ Yonsei University, Seoul, Republic of Korea\\
$^{141}$  Zentrum  f\"{u}r Technologie und Transfer (ZTT), Worms, Germany\\
$^{142}$ Affiliated with an institute covered by a cooperation agreement with CERN\\
$^{143}$ Affiliated with an international laboratory covered by a cooperation agreement with CERN.\\

\end{flushleft} 

\end{document}